\documentclass[preprintnumbers, superscriptaddress, showpacs, nofootinbib, amsfonts, amsmath, amssymb, aps, prd, notitlepage, floatfix]{revtex4-2}
\usepackage[colorlinks=true, linkcolor=blue, citecolor=blue, urlcolor=blue]{hyperref}
\usepackage{xcolor}
\usepackage{graphicx}
\usepackage{sub
figure}
\usepackage{multirow}
\usepackage{slashed}
\usepackage{physics}
\usepackage{dsfont}
\usepackage{lineno}

\newcommand{\dhd}{{\textstyle d} \lower.03ex\hbox{\kern-0.38em$^{\scriptstyle-}$}\kern-0.05em{}}

\begin{document}

\title{{Perturbative Corrections to Quark TMDPDFs in the Background-Field Method: Gauge Invariance, Equations of Motion, and Multiple Interactions}}

\author{Swagato~Mukherjee}
\affiliation{Physics Department, Brookhaven National Laboratory, Upton, New York 11973, USA}
\author{Vladimir~V.~Skokov}
\affiliation{Department of Physics, North Carolina State University, Raleigh, NC 27695, USA}
\author{Andrey~Tarasov}
\email{ataraso@ncsu.edu}
\affiliation{Department of Physics, North Carolina State University, Raleigh, NC 27695, USA}
\affiliation{CFNS, Department of Physics and Astronomy, Stony Brook University, Stony Brook, NY 11794, USA}
\author{Shaswat~Tiwari}
\email{sstiwari@ncsu.edu}
\affiliation{Department of Physics, North Carolina State University, Raleigh, NC 27695, USA}
\begin{abstract}
 We calculate the perturbative corrections in the strong coupling to the unpolarized quark transverse-momentum dependent parton distribution function (TMDPDF) operator within a background-field framework, extending the approach of Ref.~\cite{Mukherjee:2023snp}. We focus on ensuring gauge invariance, identifying two key components needed: a gauge-invariant TMDPDF operator with a transverse gauge link at spatial infinity, and accounting of the equations of motion (EoM) of the background fields. We go beyond next-to-leading order in strong coupling expansion, considering multiple interactions with the background field at all orders of strong coupling. By examining the interplay between quark and gluon contributions, we show that spurious singularities, proportional to EoM, can be misinterpreted as genuine divergences in QCD factorization unless properly identified and removed.
\end{abstract}
\maketitle
\section{Introduction}
The QCD factorization approach~\cite{Collins:2011zzd} is a powerful method of describing the dynamics of the QCD medium in the high-energy scattering reaction. Nowadays, among a few other techniques, it is a state-of-the-art approach that is commonly applied to study observables at the current and future experimental facilities, e.g. the Electron-Ion Collider (EIC).

The QCD factorization applies when the scattering reaction involves a wide separation of kinematic scales, so that the different types of QCD dynamics can be disentangled. The essence of the approach is that the part of dynamics characterized by a hard scale can be handled by means of the perturbative methods, while the components characterized by a non-perturbative scale can be encoded in terms of matrix elements of the QCD operators. The matrix elements between the hadron states give rise to the various types of parton distribution functions (PDFs) defining distribution of quarks and gluons inside the hadron.

The structure of these operators is by no means arbitrary. Instead, the operators appear within a formal mathematical procedure summarized in the QCD factorization theorems. The operators, obtained within a particular factorization scheme, rigorously define the structure of separated dynamical modes in the QCD factorization. The precise knowledge of these operators allows us to systematically study the properties of QCD in the high-energy scattering, and, in particular, its non-perturbative dynamics, which generates the fundamental properties of hadrons. In this sense, the definition of operators has an intrinsic physical content.

The defining feature of the QCD factorization is emergence of large kinematic logarithms, that is due to the presence of widely separated scales. These logarithms dominate experimental observables, and the possibility of their resumation is the foundation of predictive power of the QCD factorization approach. The nature of logarithms is specific to a particular structure of dynamical modes in the QCD factorization regime, for this reason, it can be obtained from analysis of the QCD operators specific to a given factorization scheme.

For example, the deeply inelastic scattering (DIS), which is a process $l(l) + N(P,S)\to l(l') + X$, where the lepton $l$ interacts with the hadron target $N$ of momenta $P$ and spin $S$ via the virtual photon $\gamma^\ast$ exchange of momenta $q = l - l'$, develops two factorization regimes depending on the values of the photon virtuality $Q^2 = - q^2$ and the Bjorken $x_B = Q^2/2P\cdot q$ variable. 

In the Bjorken limit of large $Q^2\to\infty$ and fixed Bjorken-$x_B$ QCD scatterings are determined by the wide separation in transverse momenta, which can be described within the collinear factorization scheme. This separation leads to the appearance of large transverse logarithms. The collinear factorization prevails in the region of large Bjorken-$x_B$. In this factorization regime, the QCD scatterings are described by the light-cone string QCD operators depending on the longitudinal momentum fraction variable $x$. The dominant transverse logarithms of these operators can be separated using, for instance, the light-cone expansion technique, see e.g. Refs.~\cite{Anikin:1978tj,Shuryak:1981kj,Karchev:1983ai,Geyer:1985vw,Braunschweig:1985nr,Balitsky:1987bk}, and resumed with the Dokshitzer–Gribov–Lipatov–Altarelli–Parisi  (DGLAP) evolution equations~\cite{Dokshitzer:1977sg,Gribov:1972ri,Altarelli:1977zs}.

At the same time, in the Regge limit of small $x_B\to 0$ and fixed virtuality $Q^2$, the QCD medium develops the strong separation in rapidity. Such a QCD factorization regime dominates the kinematic region of small Bjorken-$x_B$, and can be described using, for instance, the high-energy rapidity factorization scheme \cite{Balitsky:1995ub}. In this regime, with no ordering in the transverse momenta, the QCD operators have the dipole structure, see e.g. Refs.~\cite{Dominguez:2010xd,Dominguez:2011wm,Xiao:2017yya}, and depend on the momentum variable $p_\perp$, or a conjugate variable $b_\perp$. The ordering in rapidity leads to dominance of the rapidity logarithms. These logarithms can be separated using the high-energy expansion techniques, see e.g. Ref.~\cite{Balitsky:1995ub}, and resumed using the Balitsky-Fadin-Kuraev-Lipatov (BFKL) evolution equation~\cite{Fadin:1975cb,Kuraev:1976ge,Kuraev:1977fs,Balitsky:1978ic,Ioffe:2010zz}, which is the linearization of the Balitsky-Kovchegov (BK) ~\cite{Balitsky:1995ub,Balitsky:1997mk,Kovchegov:1999yj} and the Jalilian-Marian-Iancu-McLerran-Weigert-Leonidov-Kovner ~\cite{Jalilian-Marian:1997qno,Jalilian-Marian:1997ubg,Kovner:2000pt,Iancu:2000hn,Ferreiro:2001qy,Kovner:2013ona,Kovner:2014lca,Lublinsky:2016meo,vanHameren:2025hyo} evolution.

Though these two factorization regimes have been extensively studied and employed to describe various observables in the corresponding kinematic regions, there is however growing interest in expanding the limits of applicability of these two factorization schemes \cite{Kutak:2004ym,Motyka:2009gi,Mueller:2013wwa,Mueller:2012uf,Balitsky:2015qba,Iancu:2015vea,Iancu:2015joa,Balitsky:2016dgz,Boussarie:2020fpb,Caucal:2021ent,Taels:2022tza,Caucal:2022ulg,Caucal:2023fsf,Altinoluk:2023hfz,Duan:2024qck,Duan:2024qev}. This interest is motivated by the need to describe experimental data in the region of moderate values of Bjorken-$x_B$. Such data will form a substantial fraction at the future EIC~\cite{Accardi:2012qut,AbdulKhalek:2021gbh,Aschenauer:2017jsk} and will be instrumental in various studies, including the search for signatures of gluon saturation~\cite{Gribov:1972ri,Iancu:2003xm,Weigert:2005us,Gelis:2010nm,Albacete:2014fwa,Kovchegov:2012mbw}.

The intricacy of this region is that it is not necessarily dominated by kinematic logarithms of only one kind. Dynamical modes in this region become significantly more involved due to various effects.

To account for these effects, a novel factorization scheme has been recently proposed in Ref. \cite{Mukherjee:2023snp}. In the following discussion we will refer to this scheme as the MSTT factorization approach. The MSTT approach aims to bridge the gap between the regions of large and small Bjorken-$x_B$, and faithfully describe the transition between two limits.

The MSTT scheme involves separation in both rapidity and transverse momenta, and, for this reason, belongs to the class of transverse-momentum dependent (TMD) factorization~\cite{Collins:1981uk,Collins:1984kg,Collins:1987pm,Collins:1989gx,Meng:1995yn,Ji:2004wu,Ji:2004xq,Boussarie:2023izj}. As a result, the non-perturbative part of QCD in this scheme is described by the TMDPDF operators that depend on both $x$ and $p_\perp$ variables.

The main difference of the MSTT approach from other TMD factorization schemes is that it genuinely describes the flow of transverse momenta between different dynamical modes in the high-energy scattering, in particular in the region of small $p_\perp \gtrsim \Lambda_{\rm QCD}$, or equivalently large $b_\perp \lesssim \Lambda^{-1}_{\rm QCD}$. This leads to a different structure of the IR logarithms of the TMD distributions compared to the standard results obtained using the collinear matching procedure~\cite{Collins:1984kg,Collins:1981uw,Kang:2012em,Sun:2013hua,Dai:2014ala, Braun:2009mi,Echevarria:2015byo,Scimemi:2019gge}. In the latter case, the IR structure of distributions is dominated by transverse logarithms of the DGLAP type. This reflects an underlying assumption about strict ordering of the transverse momenta flow between different dynamical modes in the QCD factorization.\footnote{This is valid in the region of small $b_\perp \ll \Lambda^{-1}_{\rm QCD}$.}

Meanwhile, in the MSTT approach the DGLAP logarithms dominate the IR structure of TMDPDFs only in the region of large $p_\perp \gg \Lambda_{\rm QCD}$, or equivalently small $b_\perp \ll \Lambda^{-1}_{\rm QCD}$, where the collinear expansion of the TMD operators is valid. Outside this region, the logarithmic DGLAP contribution decays. Instead, at small $x$, the IR structure of TMDPDFs in the MSTT scheme is dominated by large logarithms of the BFKL type.

In Ref. \cite{Mukherjee:2023snp} the analysis of structure of TMDPDFs in the MSTT scheme has been done for the unpolarized gluon TMD distribution. In this paper, we turn our attention to the quark TMD operators aiming to address an important issue of the gauge invariance of calculations in the MSTT approach. However, the results of our analysis are general and can be applied to perturbative calculations of PDFs in any QCD factorization scheme.

Specifically, we calculate the NLO correction to the quark TMD operator in a dilute quark background field. In the spirit of the background field approach to the QCD factorization, see Refs.~\cite{Abbott:1980hw,Abbott:1981ke}, we refer to the perturbative fields as the ``quantum" fields, and call the fields of the target the ``background" fields.

First, it's important to recognize that any perturbative calculation is always {\it gauge specific}, since one has to fix the gauge of gluon fields in the calculation. By this choice, one can cancel the certain types of quantum emission, e.g. the emission of gauge links of the QCD operators etc. However, as we emphasized at the beginning of this section, the structure of operators in the definition of PDFs has an intrinsic physical content. This raises the question of whether the gauge fixing causes a certain loss of physics by dropping some types of the quantum emission.

It is very well known, however, that this is not the case. Though by choosing a specific gauge one can eliminate a particular type of emission, the physical content is never lost. It simply gets redistributed between other non-trivial types of the quantum emission in Feynman diagrams. This reflects the {\it gauge invariance} of the perturbative calculation. Even though a particular diagram contribution is gauge dependent, the sum of all types of diagrams (emissions) is gauge independent and fully represents the physical content of PDFs.

However, this is not always easy to observe, especially in the case of a dense background field. To illustrate this, we first perform a ``naive" calculation of the NLO correction to the quark TMD operator in the MSTT scheme using two gauges: the Feynman gauge and the axial gauge with the gauge fixing condition for the quantum fields $A_+ = 0$. In this ``naive" calculation we do not include contribution of the transverse gauge link at infinity in the definition of the quark TMD operator following the common lore that in non-singular gauges, like the Feynman gauge, the fields nullify at infinities, i.e. $A_\mu(+\infty n)=0$, where $n$ is a light-like vector, and in the axial gauge the gauge fixing condition can be supplemented, see e.g. Ref.~\cite{Chirilli:2015fza}, with an advanced boundary condition $A_\perp(+\infty n)=0$, so that the contribution of the transverse link at infinity can be neglected as well.

One can expect that due to the gauge invariance, the results of our calculation in two gauges should coincide. However, after performing the calculation in two gauges in the MSTT scheme, we find that two results are drastically different. Yet, performing the collinear expansion of the results, i.e. the collinear matching procedure that is valid at small values of $b_\perp \ll \Lambda^{-1}_{\rm QCD}$, we observe that, at least in the leading order of the expansion, the difference does disappear and two results perfectly agree with each other. So what do we miss in our ``naive" calculation that brings a nontrivial difference at large values of $b_\perp \lesssim \Lambda^{-1}_{\rm QCD}$? 

To study the difference between the two gauges, we analyze the Feynman diagrams in our calculation and follow the known strategy, see Refs. \cite{Brodsky:2002ue,Ji:2002aa}, applying the Ward identity to the contraction of the gluon momenta $p_\mu$ with the quark-gluon interaction vertex:
\begin{eqnarray}
&&\frac{i}{\slashed{k}}~ i\slashed{p} ~ \frac{i}{\slashed{k} + \slashed{p}}=\frac{i}{\slashed{k} + \slashed{p}} - \frac{i}{\slashed{k}}\,,
\label{eq:Wid-dilute}
\end{eqnarray}
which in our formalism corresponds to the integration by parts procedure. Here $p_\mu$ is the gluon momentum, while $k_\mu+p_\mu$ and $k_\mu$ are quark momenta before and after interaction. Eq. (\ref{eq:Wid-dilute}) is in fact a well-known trick, which is commonly used for proving the Ward identity in QED, and can be found in standard textbooks, see e.g. \cite{Peskin:1995ev}. In the following discussion we will refer to this relation as ``the vertex identity". Though, in our analysis we go beyond this leading order relation and study the diagrams in presence of an {\it infinite number of interactions} with the dense background field, which constitutes one of the main results of this paper.

Applying the vertex identity to all interaction vertices we immediately find that the difference between two gauges reduces to contributions of two types.

The first type, quite well studied before, see e.g. Refs.~\cite{Ji:2002aa,Ebert:2019okf}, can be interpreted as terms containing some kind of quantum emission from the spatial infinity. The usual approach to such terms is to eliminate them by choosing an appropriate prescription for the light-cone singularity in the axial gauge, or equivalently introducing a supplemental boundary condition for the gluon fields at the infinity. However, we argue that such an approach is by no means general, and is very much specific to a particular choice of the gauge fixing condition $e^\mu A_\mu = 0$, where $e_\mu$ is the gauge fixing vector.

Meanwhile, following the general principle, the gauge invariance should not depend on this choice. In other words, the results of calculations performed in different gauges should coincide regardless of the choice of either the gauge fixing vector or the prescription for the light-cone singularity.

This motivates us to drop the prescription argument and include the contribution of the transverse gauge link at the spatial infinity in the definition of the quark TMD operator in its full generality. In fact, this is quite anticipated since one can talk about the gauge invariance only in the context of the gauge invariant operators. For the TMD operators, the transverse gauge link at the spatial infinity is indeed necessary for the  gauge invariance of the operator.

The inclusion of the transverse gauge link is necessary to ensure that the result of calculation does not depend on the choice of the gauge fixing vector $e_\mu$ in the axial gauge. We demonstrate that dependence on the gauge fixing vector in the terms associated with the quantum emission at the spatial infinity cancels in the sum of all Feynman diagrams at any given order of the perturbative expansion. The requirement to consider QCD operators that are explicitly gauge invariant constitutes the first condition to ensure the gauge invariance of the result. In the context of our problem, this implies calculation of the TMD operators with the transverse gauge link at the spatial infinity.

However, we find that while the above condition is necessary, it is not sufficient to obtain a fully gauge invariant result. In the MSTT scheme the Feynman diagrams are calculated in the background fields of the target. Hence, the application of the Ward identity leads to the second type of terms in the difference between results calculated in the Feynman and the axial gauges. Our analysis shows that these terms cannot be eliminated by the contribution of the transverse gauge link at the spatial infinity. In other words, such terms are going to appear even in a calculation of the gauge invariant operators. 
However, we show that these terms can be transformed to explicitly show their proportionality to the QCD equations of motion for the background fields of the target. We demonstrate this in full generality for an arbitrary background field and to all orders of perturbation theory.

We thus conclude that in order to satisfy the gauge invariance one has to enforce the QCD equations of motion for the background fields. This constitutes the second condition necessary for the gauge invariance of the result.

We'd like to emphasize that this requirement is not specific to calculations in the MSTT approach. In fact, it has to be imposed in any QCD factorization scheme as this condition reflects the gauge invariance of the hadron state. Indeed, while the gauge of the target fields can be rotated, the rotation should not lead to any physical effect. The role of the equations of motion is exactly to ensure this invariance.

In the bulk of available calculations concerning the large Bjorken-$x_B$ region, the equations of motion are usually implicitly satisfied since the dilute limit of the background fields is  commonly taken and the background partons are set to be on the mass-shell. However, we believe the situation is not as trivial in the field of small-$x_B$ physics since one has to deal with dense background fields.

 For instance, as we conclude from our analysis, the result of a perturbative calculation in the QCD factorization approach is never unique. Regardless of the kinematic region, i.e. large or small Bjorken-$x_B$, there is an inevitable ambiguity related to the equations of motion for the background fields: one can always add an arbitrary contribution proportional to the equations of motion and not change any physics since such contribution is to be trivial. We demonstrate that such ``addition" effectively corresponds to a transition from one choice of the gauge fixing vector $e_{1\mu}$ to another $e_{2\mu}$. We find that the limiting case, when the explicit dependence on the $e_\mu$ vector is eliminated, corresponds to the background-Feynman gauge, which coincides with the standard Feynman gauge only in the absence of the background gluon fields, e.g. a dilute quark target. We argue that the background-Feynman gauge, which in general does not coincide with the standard Feynman gauge, is a proper generalization of the latter to the case of dense background fields of the target.

The ambiguity due to the equations of motion leads to a non-trivial interplay between quark and gluon background field contributions. Essentially, the quark and gluon sectors are not independent and a certain physical content can be transferred between various quark and gluon field configurations. We provide a detailed description of this mechanism which inevitably plays a crucial role in calculations with the dense target.

Apart from the ambiguity in results, the equations of motion for the background fields do not pose any issues, in general. This is correct, however, except for the case when the terms proportional to the equations of motion contain singularities. We argue that such singularities are non-physical since they originate in terms that are to be removed using the equations of motion. However, in practice, it's not always straightforward to identify whether a particular singularity is non-physical and corresponds to a contribution of the equations of motion, or reflects a genuine divergence originating in the factorization procedure.

We provide an example of this effect by observing that the result of our ``naive" calculation in the axial gauge contains a rapidity divergence $\int dz/z$, which on the face of it can be interpreted as an IR divergence defining the IR structure of the TMDPDFs. Yet, we do not find such a divergence in our calculation performed in the Feynman gauge, which means that this divergence is non-physical and can -- and must -- be removed by applying the equations of motion. 
One of the main goals of this paper is to highlight this non-trivial effect, as neglecting it could lead to fundamentally incorrect results.

The paper is organized as follows. Starting with some basic definitions in Sec. \ref{sec:defI}, we calculate the NLO correction to the quark TMD opertor in the dilute limit in the Feynman gauge in Sec. \ref{sec:Feynmancalc} folowing with a similar calculation in the axial gauge in Sec. \ref{sec:ax-calc}. In Sec. \ref{sec:mapping} we compare two results and show that the difference between them should be attributed to the contribution of the transverse gauge link at the spatial infinity in the TMDPDF operator and the equations of motion for the background fields of the target. While the analysis in Sec. \ref{sec:mapping} is conducted in the dilute approximation with only two background quarks fields, in Sec. \ref{sec:beyondDL} we generalize it to the dense case with an arbitrary number of interactions with the background fields.

\section{Definitions\label{sec:defI}}
In the subsequent sections, we will consider calculation of the NLO correction to the quark TMD operator. We will perform this calculation in the dilute limit by computing the corresponding Feynman diagrams in the background of two quarks. For the background quarks we will assume general kinematics, in particular, keeping finite value of the corresponding transverse momenta $p_\perp$, and assuming it to be small $p_\perp \gtrsim \Lambda_{QCD}$, which corresponds to large $b_\perp \lesssim \Lambda^{-1}_{QCD}$. This latter makes our calculation different from the collinear matching procedure, valid in the region of small $b_\perp \ll \Lambda^{-1}_{\rm QCD}$, in which the background partons are essentially collinear and carry only the longitudinal momenta. In our MSTT discussion, we will refer to such approximations as the collinear limit or the collinear expansion.

The non-zero value of the transverse momenta $p_\perp$ is essential for our analysis and by no means trivial. There are various physical consequences, both of conceptual and practical nature, but in this paper we mainly focus on only one of them, which is the gauge invariance of the perturbative corrections to the TMD operator. To demonstrate the issue, in the next two sections we will perform a ``naive" calculation in two gauges for the quantum gluon fields - the Feynman gauge and the axial gauge fixed as $A_+ = 0$. By inspecting the results, we will find that in the general kinematics the results calculated in two gauges do not agree. We will see that the results coincide only in a special case of the collinear expansion when the background quarks are effectively on the mass-shell. The latter approximation is commonly used in the calculations of the TMDPDFs at small $b_\perp \ll \Lambda^{-1}_{\rm QCD}$. For this reason, the aforementioned discrepancy between the results calculated in different gauges has not been previously observed.

Let us start with a definition of the TMDPDF matrix element
\begin{eqnarray}
&&\Phi^{[\Gamma]}(x, b_\perp) = \frac{1}{2}{\rm tr}\Big(\Phi(x, b_\perp)\Gamma\Big)\,,
\label{TMDshort:def}
\end{eqnarray}
where
\begin{eqnarray}
&&\Phi_{ij}(x, b_\perp) = \int \frac{dz^-}{\pi} e^{-2ixp^+z^-} \langle P, S|\bar{q}_j(z^-, \frac{b_\perp}{2})[z^-, \pm\infty]_{\frac{b}{2}}\mathcal{L}_{\frac{b}{2}}\mathcal{L}^\dag_{-\frac{b}{2}}[\pm\infty, -z^- ]_{- \frac{b}{2}}q_i(-z^-, - \frac{b_\perp}{2})|P, S\rangle\,
\label{TMDmat:def}
\end{eqnarray}
is a matrix element for a quark $q$ with the longitudinal momentum fraction $x$ at the transverse parameter $b_\perp$.
Here $\Gamma$ is a $\gamma$-matrix projector, and $i$, $j$ are Dirac indexes. The light-cone Wilson lines are defined as
\begin{eqnarray}
&&[x^-, y^-]_b = \exp\Big(ig\int^{x^-}_{y^-} dz^-A_-(z^-, b_\perp)\Big)\,.
\label{wl:def}
\end{eqnarray}
The light-cone Wilson lines in Eq. (\ref{TMDmat:def}) are connected at infinity by semi-infinite transverse gauge links $\mathcal{L}_{\frac{b}{2}}$ and $\mathcal{L}^\dag_{-\frac{b}{2}}$, which correspondingly start at transverse positions $b/2$ and $-b/2$.

The matrix element in Eq. (\ref{TMDmat:def}) is calculated between the hadron states of momenta $P$, spin $S$ and mass $M$. To define the kinematic variables, we introduce two light-cone vectors $n_\mu$ and $\bar{n}_\mu$, such that $(n\bar{n}) = 1$. With this vectors we can decompose an arbitrary vector $v_\mu$ as
\begin{eqnarray}
&&v^\mu = v^+ \bar{n}^\mu + v^- n^\mu + v^\mu_T\,,
\end{eqnarray}
where the light-cone components are defined as
\begin{eqnarray}
n^\nu v_\nu = v^+\,;\ \ \ \bar{n}^\nu v_\nu = v^-\,.
\end{eqnarray}
In particular, we have
\begin{eqnarray}
&&n^+ = 0\,,\ \ \  n^- = 1\,,\ \ \  n_T = 0\,,
\nonumber\\
&&\bar{n}^+ = 1\,,\ \ \  \bar{n}^- = 0\,,\ \ \  \bar{n}_T = 0\,.
\end{eqnarray}

While vector $n$ can be chosen in an arbitrary way, which corresponds to the choice of the reference frame, vector $\bar{n}$ can be introduced using a light-cone momenta
\begin{eqnarray}
&&p^\mu = P^\mu - \frac{n^\mu}{2}\frac{M^2}{(n\cdot P)}\,,
\end{eqnarray}
where $P_\mu$ is the momentum of the target, $P^2 = M^2$. Using $p_\mu$ we can construct the light-cone momenta $\bar{n}_\mu$ as $\bar{n}_\mu = p_\mu/(n\cdot p)$.

The TMDPDF matrix element (\ref{TMDmat:def}) is a bare matrix element which has to be renormalized. In particular, this procedure involves a soft factor $\sqrt{\mathcal{S}(b_\perp)}$ which is a vaccum matrix element, see e.g. Ref. \cite{Collins:2011zzd}, so the full TMDPDF matrix element has a form
\begin{eqnarray}
&&f_{ij}(x, b_\perp) = \sqrt{\mathcal{S}(b_\perp)} \Phi_{ij}(x, b_\perp)\,.
\end{eqnarray}
The full matrix element can be parametrized in terms of TMDPDFs. This parameterized depends on the projection matrix, which in our calculation we choose to be $\Gamma = \gamma^+$. We also choose the forward-pointing Wilson lines, which corresponds to the TMDPDFs appearing in the SIDIS process. In this case the matrix element can be parameterized in terms of two independent TMDPDFs:
\begin{eqnarray}
&&f^{[\gamma^+]}(x, b_\perp) = f_1(x, b_\perp) + i \epsilon^{\mu\nu}_\perp b_\mu s_{T\nu} M f^\perp_{qT}(x, b_\perp)\,,
\label{TMD:gplus}
\end{eqnarray}
where $s_{T\nu}$ is the transverse part of the hadron spin vector $S_\mu$. The function $f_1$ is the unpolarized TMDPDF, which describes distribution of the unpolarized quarks in the unpolarized hadron. The Sivers functions $f^\perp_{qT}$ is a TMDPDF, which describes the unpolarized quark distribution in the polarized hadron.


Since the separation between any pair of operators in Eq. (\ref{TMDmat:def}) is space-like, the operators commute and we can rewrite the matrix element as

\begin{eqnarray}
\Phi^{[\gamma^+]}(x, b_\perp) = \int \frac{dz^-}{2\pi} e^{-2ix p^+ z^-} \langle P, S|\mathcal{U}^{[\gamma^+]}(z^- , -z^-, \frac{b_\perp}{2})|P, S\rangle\,,
\label{def:matrix-fourier1}
\end{eqnarray}
with the TMDPDF operator
\begin{eqnarray}
\mathcal{U}^{[\gamma^+]}(z^-_1, z^-_2, b_\perp) = T\{\bar{q}(z^-_1, b_\perp)[z^-_1, \infty]_{b}\mathcal{L}_{b}\gamma^+\mathcal{L}^\dag_{-b}[\infty, z^-_2]_{-b}q(z^-_2, -b_\perp)\}\,,
\label{def:op}
\end{eqnarray}
where $T$ indicates the time ordering of operators. 

Using the translational invariance of the matrix element (\ref{TMDmat:def}), we can write another form of Eq. (\ref{def:matrix-fourier1}):
\begin{eqnarray}
\Phi^{[\gamma^+]}(x, b_\perp) = \frac{1}{4\pi \delta(0)} \int dz^-_1 e^{-ix p^+ z^-_1 } \int dz^-_2 e^{ix p^+ z^-_2 } \langle P, S|\mathcal{U}^{[\gamma^+]}(z^-_1 , z^-_2, \frac{b_\perp}{2})|P, S\rangle\,,
\label{def:matrix-fourier2}
\end{eqnarray}
where $\delta(0) \equiv \int^\infty_{-\infty} dz^- $ is an infinite longitudinal length.

\begin{figure}[tb]
 \begin{center}
\includegraphics[width=0.5\textwidth]{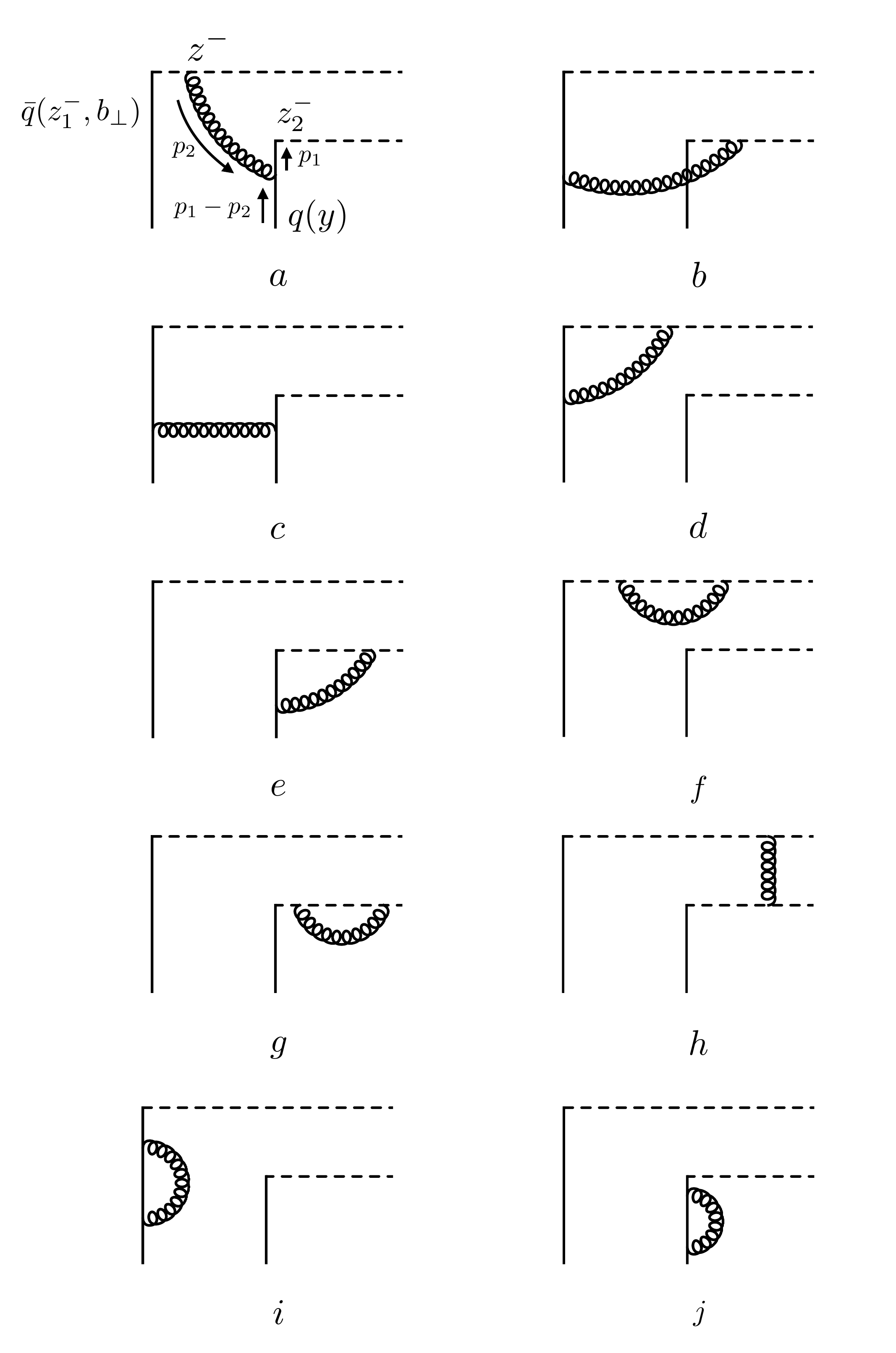}
 \end{center}
\caption{\label{fig:Fdiag}The Feynman diagrams contributing to the TMDPDF operator (\ref{def:op}) at the NLO order in the dilute approximation with two background quark fields.}
 \end{figure}

We will compute the operator (\ref{def:op}) at the next-to-leading order (NLO) in strong coupling, assuming the background field associated with the target hadron consists of only two quarks, i.e. in the so-called dilute limit. As we mentioned above, we will start with a ``naive" calculation in the Feynman and axial gauges. In this ``naive" calculation we will drop the contribution of the transverse gauge links at the spatial infinity $\mathcal{L}$, though in our subsequent analysis we will demonstrate that these links are essential, but not sufficient, to demonstrate the gauge invariance of the final result. The contributing Feynman diagrams are presented in Fig. \ref{fig:Fdiag}.

\section{Calculation of the NLO corrections to the quark TMDPDF operator in the Feynman gauge\label{sec:Feynmancalc}}
In this section we will present calculation of the diagrams in Fig. \ref{fig:Fdiag} using the gluon propagator in the Feynman gauge. In our discussion we will use the Schwinger’s notation, see Ref. \cite{Schwinger:1951nm}, which utilizes the idea of coherent states $|x)$ and $|p)$, which are eigenvectors of the position and momentum operators:
\begin{eqnarray}
&&\hat{p}_\mu|p) = p_\mu|p);\ \ \ \hat{x}_\mu|x) = x_\mu|x)
\end{eqnarray}
and satisfy the following completeness and orthogonality relations:
\begin{eqnarray}
\int d^4x\, |x)( x| = 1\,;\ \ \ \int \dhd^4p \, |p) (p| = 1\,;\ \ \ ( x|p) = e^{ipx}\,,
\label{bosonort}
\end{eqnarray}
where
\begin{align}
    \int \dhd^d p \equiv \int \frac{d^d p}{(2\pi)^d}\,.
\end{align}

Using this relations it is straightforward to obtain the following equation for an arbitrary function of the momentum operator
\begin{eqnarray}
&&(x|f(\hat{p})|y) = \int \dhd ^4p e^{-ip(x-y)}f(p)\,.
\label{def:fopsw}
\end{eqnarray}
Consequently, the gluon propagator in the Feynman gauge can be written in the following compact form
\begin{eqnarray}
&&\langle A^{a}_\mu(x) A^{b}_\nu(y) \rangle = (x|\frac{-ig_{\mu\nu}(\hat{p})\delta^{ab}}{\hat{p}^2} |y) \,.
\end{eqnarray}

For brevity, in the following discussion, we  will drop the ``hat" notation for the operators. However, one should always remember the commutation properties of the operators. For example, the momentum operator commutes with an arbitrary coordinate operator as
\begin{eqnarray}
&&\hat{p}_\mu\mathcal{O}(\hat{x}) = \mathcal{O}(\hat{x}) \hat{p}_\mu + i\partial_\mu \mathcal{O}(\hat{x})\,.
\label{eq:op-com}
\end{eqnarray}

Let us start our calculation of the diagrams in Fig. \ref{fig:Fdiag}. Though this calculation is straightforward, we will provide some details of computation for the diagram in Fig. \ref{fig:Fdiag}a. Starting with the initial expression for this diagram in the Feynman gauge, using the Schwinger’s notation we write
\begin{eqnarray}
&&\mathcal{U}^{[\gamma^+]}(z^-_1, z^-_2, b_\perp)\Big|^{Feyn.}_{Fig.~\ref{fig:Fdiag}a}
\label{eq:f1ainit}
\\
&&= -g^2 \int^{z^-_1}_{\infty}dz^- \int d^4y \bar{q}(z^-_1, b_\perp) t^a \gamma^+ (z^-_2, - b_\perp|\frac{i\slashed{p}}{p^2+i\epsilon}|y) \gamma^\nu t^b q(y) n^\mu (z^-, b_\perp|\frac{-ig_{\mu\nu}\delta^{ab}}{p^2+i\epsilon}|y)\,,
\nonumber
\end{eqnarray}
where it is easy to identify quark and gluon propagators, as well as quark background fields $q$ and $\bar{q}$.

At this point, it is convenient to project the product of background fields onto the basis of $\gamma$-matrices using the following general decomposition
\begin{eqnarray}
&&q_i(z)\bar{q}_j(y) = - \frac{1}{4}{\rm tr}\{\bar{q}(y) q(z)\} \mathds{1}_{ij} - \frac{1}{4}{\rm tr}\{\bar{q}(y) \gamma_\mu q(z)\}\gamma^\mu_{ij} - \frac{1}{8}{\rm tr}\{\bar{q}(y) \sigma_{\mu\nu}q(z)\} \sigma^{\mu\nu}_{ij} 
\nonumber\\
&&+ \frac{1}{4}{\rm tr}\{\bar{q}(y) \gamma_\mu\gamma^5q(z)\}(\gamma^\mu\gamma^5)_{ij} - \frac{1}{4}{\rm tr}\{\bar{q}(y) \gamma^5q(z)\}\gamma^5_{ij}\,,
\label{eq:gamma-decomp}
\end{eqnarray}
where ${\rm tr}\{\bar{q}(y) \Gamma q(z)\}\equiv \bar{q}_i(y) \Gamma_{ij} q_j(z)$ with $\Gamma$ being an arbitrary product of $\gamma$-matrices. 

Note that, while our initial operator (\ref{def:op}) contains a bilocal quark operator projected onto $\gamma^+$, decomposition (\ref{eq:gamma-decomp}) for the background fields introduces mixing with other projections in the diagrams in Fig. \ref{fig:Fdiag}. However, for our discussion it is sufficient to neglect the mixing terms and keep only $\gamma^+$ projection in Eq. (\ref{eq:gamma-decomp}) using the following replacement in Eq. (\ref{eq:f1ainit}):
\begin{eqnarray}
&&q_i(z)\bar{q}_j(y) \to - \frac{1}{4}{\rm tr}\{\bar{q}(y) \gamma^+ q(z)\}\gamma^-_{ij}\,.
\label{eq:gamma-subst}
\end{eqnarray}

Making this replacement and calculating the trace of $\gamma$-matrices we obtain
\begin{eqnarray}
&&\mathcal{U}^{[\gamma^+]}(z^-_1, z^-_2, b_\perp)\Big|^{Feyn.}_{Fig.~\ref{fig:Fdiag}a}
 \label{eq:after-proj}
\\
&&= - 2 g^2 C_F \int^{z^-_1}_{\infty}dz^- \int d^4y~ {\rm tr}\{\bar{q}(z^-_1,  b_\perp) \gamma^+ q(y) \} (z^-_2, - b_\perp|\frac{ p^+ }{p^2+i\epsilon}|y) (y|\frac{1}{p^2+i\epsilon}|z^-, b_\perp)\,.
\nonumber
\end{eqnarray}

Now, let us use the standard approximation and assume that the background fields is independent of $y^+$: $q(y) = q(y^-, y_\perp)$. This approximation is a consequence of the wide separation of the longitudinal momentum 
between the target and projectile in the high-energy scattering. In particular, the longitudinal momentum 
of the boosted target satisfies $P^+ \gg P^-$, so the $y^+$-dependence of the background fields can be neglected. This assumption is essentially an assumption about factorization between different kinematic modes that takes place in the high-energy scattering, see e.g. Ref. \cite{Collins:2011zzd}.

As a result, we can perform some integrals in Eq. (\ref{eq:after-proj}) and rewrite it as
\begin{eqnarray}
&&\mathcal{U}^{[\gamma^+]}(z^-_1, z^-_2, b_\perp)\Big|^{Feyn.}_{Fig.~\ref{fig:Fdiag}a} = - 2 g^2 C_F \int^{z^-_1}_{\infty}dz^- \int dy^- \int d^2y_\perp~ {\rm tr}\{\bar{q}(z^-_1,  b_\perp) \gamma^+ q(y^-, y_\perp) \}
\\
&&\times \int \frac{dp^-}{2\pi} \int \frac{dp^+_1}{2\pi} e^{-ip^+_1(z^-_2 - y^-)}  ( - b_\perp|\frac{ p^+_1 }{2p^+_1 p^- - p^2_\perp+i\epsilon}|y_\perp) 
 \int \frac{dp^+_2}{2\pi} e^{-ip^+_2(y^- - z^-)} (y_\perp|\frac{1}{2p^+_2p^- - p^2_\perp+i\epsilon}|b_\perp) \,.
\nonumber
\end{eqnarray}

Changing the sign of $p^+_2$ and evaluating the $p^-$ integral we get
\begin{eqnarray}
&&\mathcal{U}^{[\gamma^+]}(z^-_1, z^-_2, b_\perp)\Big|^{Feyn.}_{Fig.~\ref{fig:Fdiag}a} = - ig^2 C_F \int^{z^-_1}_{\infty}dz^- \int dy^- \int d^2y_\perp {\rm tr}\{\bar{q}(z^-_1,  b_\perp) \gamma^+ q(y^-, y_\perp) \} \int \dhd^2 p_{1\perp} e^{ip_{1\perp} (-b_\perp - y_\perp) }
\nonumber\\
&&\times \int \dhd^2 p_{2\perp} e^{ip_{2\perp} (y_\perp - b_\perp)} \Big\{ \int^\infty_0 \frac{dp^+_1}{2\pi}  \int_0^{\infty} \frac{dp^+_2}{2\pi} - \int^0_{-\infty} \frac{dp^+_1}{2\pi}  \int_{-\infty}^0 \frac{dp^+_2}{2\pi} \Big\} \frac{1}{2p^+_2} e^{-ip^+_1 (z^-_2 - y^-)} e^{ip^+_2 (y^- - z^-)} \frac{1}{ \frac{p^2_{1\perp}}{2p^+_1} + \frac{p^2_{2\perp}}{2p^+_2}}\,.
 \label{eq:after-yplus}
\end{eqnarray}

Now, let us introduce a variable
\begin{eqnarray}
&&z = \frac{p^+_1}{p^+_1 + p^+_2}\,,
\label{eq:zdef}
\end{eqnarray}
which is a ratio of the plus momentum components of the quark propagator and the background quark field, see Fig.~\ref{fig:Fdiag}a. Replacing $p^+_2$ component in favor of the $z$ variable we rewrite Eq. (\ref{eq:after-yplus}) as
\begin{eqnarray}
&&\mathcal{U}^{[\gamma^+]}(z^-_1, z^-_2, b_\perp)\Big|^{Feyn.}_{Fig.~\ref{fig:Fdiag}a} = - \frac{ig^2 C_F}{2\pi}  \int^{z^-_1}_{\infty}dz^- \int dy^- \int d^2y_\perp~ {\rm tr}\{\bar{q}(z^-_1,  b_\perp) \gamma^+ q(y^-, y_\perp) \} \int \dhd^2 p_{1\perp} e^{ip_{1\perp} (-b_\perp - y_\perp) } 
\nonumber\\
&&\times \int \dhd^2 p_{2\perp} e^{ip_{2\perp} (y_\perp - b_\perp)} \int_0^1 \frac{dz}{z} \Big\{ \int^\infty_0 \frac{dp^+_1}{2\pi}  + \int^0_{-\infty} \frac{dp^+_1}{2\pi} \Big\} e^{-i p^+_1 z^-_2} e^{i \frac{1}{z} p^+_1 y^- } e^{-i \frac{1-z}{z} p^+_1 z^- } \frac{p^+_1}{ (1-z) p^2_{1\perp} + z p^2_{2\perp} }\,.
\label{eq:withz}
\end{eqnarray}

Subsequently integrating over $z^-$, $p^+_1$ and $y^-$ variables we get
\begin{eqnarray}
&&\mathcal{U}^{[\gamma^+]}(z^-_1, z^-_2, b_\perp)\Big|^{Feyn.}_{Fig.~\ref{fig:Fdiag}a} = \frac{g^2 C_F}{2\pi} \int_0^1 dz \frac{z}{1-z} \int d^2y_\perp~ {\rm tr}\{\bar{q}(z^-_1,  b_\perp) \gamma^+ q((1-z) z^-_1 + zz^-_2, y_\perp) \}  
\nonumber\\
&&\times \int \dhd^2 p_{1\perp} e^{ip_{1\perp} (-b_\perp - y_\perp) } \int \dhd^2 p_{2\perp} e^{ip_{2\perp} (y_\perp - b_\perp)}  \frac{1}{ (1-z) p^2_{1\perp} + z p^2_{2\perp} }\,.
\label{eq:fin1a}
\end{eqnarray}

Using this result for the diagram in Fig. \ref{fig:Fdiag}a, we can easily calculate the form of the corresponding matrix element (\ref{def:matrix-fourier1}). Substituting Eq. (\ref{eq:fin1a}) into Eq. (\ref{def:matrix-fourier2}), using translational invariance of the matrix element, and redefining the integration variables as $p_{1\perp} \to -p_\perp$, $p_{2\perp} \to -k_\perp$, $b_\perp - y_\perp \to z_\perp$, and $z (z^-_1 - z^-_2) \to z^-$, we find
\begin{eqnarray}
&&\Phi^{[\gamma^+]}(x, b_\perp)\Big|^{Feyn.}_{Fig.~\ref{fig:Fdiag}a} = \frac{g^2 C_F}{8\pi^2} \int \dhd^2 p_{\perp} e^{ip_{\perp} b_\perp } \int_0^1 \frac{dz}{1-z} \int \dhd^2 k_\perp \frac{1}{ (1-z) p^2_{\perp} + z k^2_\perp }
\nonumber\\
&&\times
  \int d^2z_\perp ~ e^{i(k - p)_\perp z_\perp } \int d z^- e^{-i\frac{x}{z} p^+ z^- } \langle P, S|\bar{q}( z^- , z_\perp) \gamma^+ q(0) |P, S\rangle\,,
  \label{eq:mel1a}
\end{eqnarray}
which is our final result for the diagram in Fig. \ref{fig:Fdiag}a.

Calculation of the diagram in Fig. \ref{fig:Fdiag}b can be done in the same way leading to exactly the same result as the diagram in Fig. \ref{fig:Fdiag}a.

Similarly, for the diagrams in Figs. \ref{fig:Fdiag}d and \ref{fig:Fdiag}e we obtain
\begin{eqnarray}
&&\Phi^{[\gamma^+]}(x, b_\perp)\Big|^{Feyn.}_{Fig.~\ref{fig:Fdiag}d,~ \ref{fig:Fdiag}e} = -\frac{g^2 C_F}{8\pi^2 } \int \dhd^2 p_\perp e^{i p_\perp b_\perp }
 \int_0^1 dz \frac{z}{1-z} \int \dhd^2 k_\perp e^{- i k_\perp b_\perp } \frac{1}{ (1-z) p^2_\perp + z k^2_\perp }
\nonumber\\
&&\times   \int d^2z_\perp e^{i (k - p)_\perp z_\perp } \int dz^- e^{-ix p^+ z^- } \langle P, S| \bar{q}(z^-,  z_\perp) \gamma^+ q(0 ) |P, S\rangle\,.
\label{eq1dfn}
\end{eqnarray}

The diagram in Fig. \ref{fig:Fdiag}c can be easily calculated as well, which yields
\begin{eqnarray}
&&\Phi^{[\gamma^+]}(x, b_\perp)\Big|^{Feyn.}_{Fig.~\ref{fig:Fdiag}c} = \frac{ g^2 C_F}{8\pi^2} \int \dhd^2 p_{\perp} e^{ ip_{\perp} b_\perp } \int^1_0  \frac{dz}{z} (1-z) \int \dhd^2 k_{\perp} \frac{ p^2_{\perp} }{ \big( (1-z)p^2_{\perp} + z k^2_{\perp} \big)^2 }   
\nonumber\\
&&\times  \int d^2z_\perp e^{i (k -p )_\perp  z_\perp } 
  \int dz^- e^{-i \frac{ x }{z} p^+ z^- } \langle P, S| \bar{q}(z^- , z_\perp) \gamma^+ q(0) |P, S\rangle\,.
  \label{eq:Feynmat1c}
\end{eqnarray}

Note that the diagrams in Figs. \ref{fig:Fdiag}f, \ref{fig:Fdiag}g, and \ref{fig:Fdiag}h are trivial in the Feynman gauge since in this gauge the corresponding gluon propagators involve only trivial components of the metric tensor, however these diagrams are non-zero in the axial gauge.

Finally, for diagrams in Figs. \ref{fig:Fdiag}i and \ref{fig:Fdiag}j we have
\begin{eqnarray}
&&\Phi^{[\gamma^+]}(x, b_\perp)\Big|^{Feyn.}_{Fig.~\ref{fig:Fdiag}i,~\ref{fig:Fdiag}j} = -\frac{g^2C_F}{16\pi^2} \int \dhd^2p_{\perp} e^{ip_{\perp} b_\perp } \int^1_0 dz \int \dhd^2k_{\perp} e^{-i k_{\perp} b_\perp } \frac{ 1 }{ (1-z) p^2_{\perp} + zk^2_{\perp} }
\nonumber\\
&&\times \int d^2z_\perp e^{i(k_{\perp} - p_{\perp} ) z_\perp } \int^\infty_{-\infty} dz^- e^{-ix p^+ z^- } \langle P, S| \bar{q}(z^-, z_\perp) \gamma^+ q(0)  |P, S\rangle \,.
\end{eqnarray}

Combining all results together we obtain
\begin{eqnarray}
&&\Phi^{[\gamma^+]}(x, b_\perp)\Big|^{Feyn.}_{Fig.~\ref{fig:Fdiag}} 
= \frac{g^2 C_F}{4\pi^2} \int \dhd^2 p_{\perp} e^{ip_{\perp} b_\perp } \int \dhd^2 k_\perp \int d^2z_\perp e^{i (k -p )_\perp  z_\perp } \int_0^1 \frac{dz}{z} \Big(  \Big[ \frac{z}{1-z}\Big]_+ + \frac{1-z}{2} \frac{ p^2_{\perp} }{(1-z)p^2_{\perp} + z k^2_{\perp} } \Big)
\nonumber\\
&&\times \frac{1}{ (1-z) p^2_{\perp} + z k^2_\perp } 
  \int dz^- e^{-i \frac{ x }{z} p^+ z^- } \langle P, S| \bar{q}(z^- , z_\perp) \gamma^+ q(0) |P, S\rangle
\nonumber\\
&&- \frac{g^2 C_F}{4\pi^2} \int \dhd^2 p_{\perp} e^{ip_{\perp} b_\perp } \int \dhd^2 k_\perp \int d^2z_\perp e^{i (k - p)_\perp z_\perp } \int_0^1 dz \Big(  \Big[ \frac{z}{1-z}\Big]_+ \frac{1}{ (1-z) p^2_\perp + z k^2_\perp } e^{- i k_\perp b_\perp }
\nonumber\\
&&- \frac{z}{1-z} \frac{1}{ k^2_\perp } \Big( 1 - e^{- i k_\perp b_\perp } \Big) + \frac{1}{2}\frac{ 1 }{ (1-z) p^2_{\perp} + zk^2_{\perp} } e^{-i k_{\perp} b_\perp } \Big) \int dz^- e^{-ix p^+ z^- } \langle P, S| \bar{q}(z^-,  z_\perp) \gamma^+ q(0 ) |P, S\rangle\,,
\label{eq:fig1-plus}
\end{eqnarray}
where the plus prescription is defined in the usual way\footnote{To make expressions lighter, we introduced $\bar{z} \equiv 1 - z$.}
\begin{eqnarray}
&&[f(z)]_+ \equiv f(z) - \delta(\bar{z})\int^1_0 dz' f(z')\,.
\end{eqnarray}

Using the relation
\begin{eqnarray}
&&\Big[\frac{2z}{1 - z}\Big]_+ = \Big[\frac{1 + z^2}{1 - z}\Big]_+ - (1 - z)  + \frac{1}{2}\delta(\bar{z})\,
\label{eq:relplus}
\end{eqnarray}
we can transform Eq. (\ref{eq:fig1-plus}) to
\begin{eqnarray}
&&\Phi^{[\gamma^+]}(x, b_\perp)\Big|^{Feyn.}_{Fig.~\ref{fig:Fdiag}} 
= \frac{g^2 C_F}{8\pi^2} \int \dhd^2 p_{\perp} e^{ip_{\perp} b_\perp } \int \dhd^2 k_\perp \int_0^1 \frac{dz}{z} \Big(  \Big[\frac{1 + z^2}{1 - z}\Big]_+ + \frac{1}{2}\delta(\bar{z}) + z(1-z) \frac{ p^2_{\perp} -  k^2_{\perp}}{(1-z)p^2_{\perp} + z k^2_{\perp} } \Big)
\nonumber\\
&&\times \frac{1}{ (1-z) p^2_{\perp} + z k^2_\perp } \int d^2z_\perp e^{i (k -p )_\perp  z_\perp }  
  \int dz^- e^{-i \frac{ x }{z} p^+ z^- } \langle P, S| \bar{q}(z^- , z_\perp) \gamma^+ q(0) |P, S\rangle
\nonumber\\
&&- \frac{g^2 C_F}{4\pi^2} \int \dhd^2 p_{\perp} e^{ip_{\perp} b_\perp } \int \dhd^2 k_\perp \int d^2z_\perp e^{i (k - p)_\perp z_\perp } \int_0^1 dz \Big(  \Big[ \frac{z}{1-z}\Big]_+ \frac{1}{ (1-z) p^2_\perp + z k^2_\perp } e^{- i k_\perp b_\perp }
\nonumber\\
&&-  \frac{z}{1-z} \frac{1}{ k^2_\perp } \Big( 1 - e^{- i k_\perp b_\perp } \Big) + \frac{1}{2}\frac{ 1 }{ (1-z) p^2_{\perp} + zk^2_{\perp} } e^{-i k_{\perp} b_\perp } \Big)  \int dz^- e^{-ix p^+ z^- } \langle P, S| \bar{q}(z^-,  z_\perp) \gamma^+ q(0 ) |P, S\rangle\,.
\label{eq:fig1-fin}
\end{eqnarray}

Note that this result was obtained in assumption that the transverse momenta, $k_\perp - p_\perp$, of the background quarks are nonzero. However, it is instructive to consider the small transverse-momentum limit  and construct the collinear expansion of the r.h.s. of Eq. (\ref{eq:fig1-fin}):
\begin{eqnarray}
&&\Phi^{[\gamma^+]}(x, b_\perp)\Big|^{Feyn.}_{Fig.~\ref{fig:Fdiag}} 
\nonumber\\
&&= \frac{g^2 C_F}{8\pi^2} \int \frac{\dhd^2 p_{\perp}}{ p^2_{\perp} } e^{ip_{\perp} b_\perp } \int_0^1 \frac{dz}{z} \Big(  \Big[\frac{1 + z^2}{1 - z}\Big]_+ + \frac{1}{2}\delta(\bar{z}) \Big) \int dz^- e^{-i \frac{ x }{z} p^+ z^- } \langle P, S| \bar{q}(z^- , 0_\perp) \gamma^+ q(0) |P, S\rangle
\nonumber\\
&&+ \frac{g^2 C_F}{4\pi^2} \int \frac{\dhd^2 p_{\perp}}{ p^2_\perp  }e^{ip_{\perp} b_\perp } \int_0^1 dz \frac{z}{1-z} \Big( 1 - e^{- i p_\perp b_\perp } \Big)
 \int dz^- e^{-ix p^+ z^- } \langle P, S| \bar{q}(z^-,  0_\perp) \gamma^+ q(0 ) |P, S\rangle + \dots\,,
\label{eq:fig1-col}
\end{eqnarray}
where the ellipsis stands for the higher order terms of the expansion.

In the next section we will compare these results with the results of calculation of the same diagrams performed in the axial gauge.

\section{Calculation of the NLO corrections to the quark TMDPDF operator in the axial gauge\label{sec:ax-calc}}
In this Section we will discuss the result of calculation of the NLO corrections to the quark TMDPDF operator performed using the axial gauge of the gluon propagator. Our goal is to compare this result with the computation of the NLO correction in the Feynman gauge given in the previous Sec. \ref{sec:Feynmancalc}.

The initial expressions of the NLO correction diagrams in Fig. \ref{fig:Fdiag} can be easily obtained from the corresponding equations in Sec. \ref{sec:Feynmancalc} by replacing the gluon propagator in the Feynman gauge with the one in the axial gauge:
\begin{eqnarray}
&&(x|\frac{-ig_{\mu\nu}\delta^{ab}}{p^2+i\epsilon}|y) \to (x|\frac{-id_{\mu\nu}(p)\delta^{ab}}{p^2+i\epsilon}|y)\,,
\end{eqnarray}
where
\begin{eqnarray}
&&d_{\mu\nu}(p)\equiv g_{\mu\nu} - \frac{e_\mu p_\nu + p_\mu e_\nu}{e\cdot p} + \frac{e^2 p_\mu p_\nu}{(e\cdot p)^2}\,.
\label{eq:ax-num}
\end{eqnarray}
The gauge fixing vector $e_\mu$ fixes the gauge of the quantum fields as $e\cdot A = 0$.\footnote{Note that in the background field approach the gauge of the quantum and background fields can be independently chosen.} To simplify the derivation, let us choose the light-like vector $e_\mu = \bar{n}_\mu$.

The calculation of the diagrams in the axial gauge is straightforward and can be done using methods described in Sec. \ref{sec:Feynmancalc}. The only difference is the spurious $1/p^-$ singularity of the axial gauge gluon propagator, which has to be regularized. There are several prescriptions for this divergence, which correspond to different boundary conditions for the gluon field at the spatial infinity~\cite{Chirilli:2015fza,Ji:2002aa}. We pick the following prescription:
\begin{eqnarray}
&&\frac{1}{p^-} \to \frac{1}{p^- + i\epsilon}\,.
\label{eq:presc}
\end{eqnarray}
In our calculation, we also consider positive values of $x$ only,  $x>0$. With that, we calculate the diagrams in Fig. \ref{fig:Fdiag} and get the following results.

For the matrix element (\ref{def:matrix-fourier1}) of the diagrams in Figs. \ref{fig:Fdiag}a and \ref{fig:Fdiag}b we have
\begin{eqnarray}
&&\Phi^{[\gamma^+]}(x, b_\perp)\Big|^{axial}_{Figs.~\ref{fig:Fdiag}a,~\ref{fig:Fdiag}b} =  - \frac{g^2 C_F}{8\pi^2 } \int \dhd^2p_{\perp} e^{ip_{\perp} b_\perp} \int  \dhd^2k_{\perp} \int^1_0 \frac{dz}{z} \frac{ 1 }{ (1-z) p^2_{\perp} + z k^2_{\perp} } \frac{p_k k_k}{k^2_{\perp}}
\nonumber\\
&&\times \int d^2z_\perp  e^{i (k_{\perp} - p_{\perp})z_\perp }  \int dz^- e^{-i \frac{x}{z} p^+ z^- } \langle P, S| \bar{q}( z^-,   z_\perp) \gamma^+ q(0) |P, S\rangle\,.
\end{eqnarray}

Similarly, the related diagrams in Figs. \ref{fig:Fdiag}d and \ref{fig:Fdiag}e yield
\begin{eqnarray}
&&\Phi^{[\gamma^+]}(x, b_\perp)\Big|^{axial}_{Figs.~\ref{fig:Fdiag}d,~\ref{fig:Fdiag}e} = \frac{g^2C_F}{8\pi^2 } \int \dhd^2p_{\perp} e^{ip_{\perp} b_\perp} \int \dhd^2k_{\perp} e^{-ik_{\perp} b_\perp } \Big[ \int^1_0 dz \frac{ 1 }{ (1-z) p^2_{\perp} + z k^2_{\perp} } 
+ \int^1_0 \frac{dz}{z} \frac{ 1 }{ p^2_{\perp} } \Big] \frac{p_{k} k_{k}}{k^2_{\perp}}
 \nonumber\\
 &&\times \int d^2z_\perp e^{i (k_{\perp} - p_{\perp}) z_\perp}  \int dz^- e^{-i x p^+ z^-} \langle P, S| \bar{q}(z^-, z_\perp ) \gamma^+ q(0)|P, S\rangle\,.
 \label{eq:ax-virt-diverge}
\end{eqnarray}
By contrasting this result with the corresponding Feynman gauge calculation, see Eq. (\ref{eq1dfn}), one makes an important observation that the axial gauge calculations contains an IR singularity $\int dz/z$ not present in the Feynman gauge.  In the next section, we demonstrate  that this divergence is unphysical, and must be removed by applying the EoM for the background fields. For diagrams in Figs. \ref{fig:Fdiag}d and~\ref{fig:Fdiag}e this is further elaborated in Appendix \ref{ap:EoM-div}, where we explicitly demonstrate that the spurious IR singularity in Eq. (\ref{eq:ax-virt-diverge}) corresponds to the contribution of EoM and thus should be eliminated. 

For the diagram in Fig. \ref{fig:Fdiag}c, it is easy to check that taking the $\gamma^+$-projection of the quark operator, see Eq. (\ref{eq:gamma-subst}), the only nontrivial contribution comes from a term of the gluon propagator proportional to the metric tensor, see Eq. (\ref{eq:ax-num}). As a result, the expression for this diagram in the axial gauge coincides with the corresponding expression in the Feynman gauge calculated in Sec. \ref{sec:Feynmancalc}:
\begin{eqnarray}
&&\Phi^{[\gamma^+]}(x_B, b_\perp)\Big|^{axial}_{Fig.~\ref{fig:Fdiag}c} = \frac{ g^2 C_F}{8\pi^2} \int \dhd^2 p_{\perp} e^{ ip_{\perp} b_\perp } \int^1_0  \frac{dz}{z} (1-z) \int \dhd^2 k_{\perp} \frac{ p^2_{\perp} }{ \big( (1-z)p^2_{\perp} + z k^2_{\perp} \big)^2 }   
\nonumber\\
&&\times  \int d^2z_\perp e^{i (k -p )_\perp  z_\perp }  
  \int dz^- e^{-i \frac{ x_B }{z} p^+ z^- } \langle P, S| \bar{q}(z^- , z_\perp) \gamma^+ q(0) |P, S\rangle\,.
\end{eqnarray}

Next, there are three diagrams which are trivial in the Feynman gauge, but get nontrivial values with the axial gauge propagator. Those diagrams are in Fig. \ref{fig:Fdiag}h:
\begin{eqnarray}
&&\Phi^{[\gamma^+]}(x, b_\perp)\Big|^{axial}_{Fig.~\ref{fig:Fdiag}h} = 2 \frac{g^2C_F}{8\pi^2} \int \dhd^2p_\perp e^{ip_\perp b_\perp} \frac{1}{ p^2_\perp } \int^1_0 \frac{dz}{z} \frac{1}{1-z} \int dz^- e^{-i \frac{x}{z} p^+ z^- } 
 \langle P, S| \bar{q}(z^-, b_\perp) \gamma^+ q(0) |P, S\rangle\,
\end{eqnarray}
and Figs. \ref{fig:Fdiag}f and \ref{fig:Fdiag}g:
\begin{eqnarray}
&&\Phi^{[\gamma^+]}(x, b_\perp)\Big|^{axial}_{Figs.~\ref{fig:Fdiag}f,~\ref{fig:Fdiag}g} = - \frac{g^2 C_F}{8\pi^2} \int \frac{\dhd^2p_\perp}{ p^2_\perp } \int^1_0 \frac{dz}{z} \frac{1}{1-z} \int dz^- e^{-ix p^+ z^- } \langle P, S| \bar{q}(z^-, b_\perp) \gamma^+  q(0) |P, S\rangle\,.
\label{eq:fig1g-axial}
\end{eqnarray}
{This result also contains an IR singularity $\int dz/z$, however, as we show in Appendix \ref{ap:EoM-div}, this singularity gets canceled in a sum with contributions of the diagrams in Figs.~\ref{fig:Fdiag}d and \ref{fig:Fdiag}e and does not appear in the final result.}

Finally, for diagrams in Figs. \ref{fig:Fdiag}i and \ref{fig:Fdiag}j we have
\begin{eqnarray}
&&\Phi^{[\gamma^+]}(x, b_\perp)\Big|^{axial}_{Fig.~\ref{fig:Fdiag}i,~\ref{fig:Fdiag}j} = \frac{g^2C_F}{16\pi^2} \int \dhd^2p_{\perp} e^{ip_{\perp} b_\perp } \int^1_0 dz \int \dhd^2k_{\perp} e^{-i k_{\perp} b_\perp } \frac{ 1 }{ (1-z) p^2_{\perp} + zk^2_{\perp} }
\nonumber\\
&&\times \int d^2z_\perp e^{i(k_{\perp} - p_{\perp} ) z_\perp } \int^\infty_{-\infty} dz^- e^{-ix p^+ z^- } \langle P, S| \bar{q}(z^-, z_\perp) \gamma^+ q(0)  |P, S\rangle \,.
\end{eqnarray}

Taking the sum of all diagrams in Fig. \ref{fig:Fdiag}, we find that the  NLO correction to the quark TMD operator calculated in the axial gauge is
\begin{eqnarray}
&&\Phi^{[\gamma^+]}(x, b_\perp)\Big|^{axial}_{Fig.~\ref{fig:Fdiag}} = \frac{g^2 C_F}{4\pi^2 } \int \dhd^2p_{\perp} e^{ip_{\perp} b_\perp} \int  \dhd^2k_{\perp} \int d^2z_\perp e^{i (k -p )_\perp  z_\perp } \int^1_0 \frac{dz}{z} \Big( \Big( \Big[ \frac{ z}{1-z} \Big]_+ 
+ 1 \Big) \frac{1}{ p^2_\perp } e^{-i k_\perp b_\perp}e^{i p_\perp z_\perp}
\nonumber\\
&&- \Big( \frac{p_k k_k}{k^2_{\perp}} - \frac{1-z}{2} \frac{ p^2_{\perp} }{(1-z)p^2_{\perp} + z k^2_{\perp}} \Big) \frac{ 1 }{ (1-z) p^2_{\perp} + z k^2_{\perp} } \Big) 
 \int dz^- e^{-i \frac{ x }{z} p^+ z^- } \langle P, S| \bar{q}(z^- , z_\perp) \gamma^+ q(0) |P, S\rangle
\nonumber\\
 &&+ \frac{g^2C_F}{4\pi^2} \int \dhd^2p_\perp e^{ip_\perp b_\perp} \int \dhd^2k_\perp \int d^2z_\perp e^{i (k-p)_\perp z_\perp} \int^1_0 dz \Big(  \frac{ z}{1-z} e^{i p_\perp z_\perp}
 -  \frac{1}{1-z} e^{i p_\perp (z-b)_\perp}
-  \frac{1}{z} e^{i p_\perp (z-b)_\perp} + \frac{1}{z} \frac{ p_k k_k }{ k^2_{\perp} }
 \nonumber\\
 &&+ \frac{p^2_\perp }{ ( 1 - z) p^2_{\perp} + z k^2_{\perp} } \frac{ p_k k_k }{ k^2_{\perp} } +  \frac{1}{2}\frac{ p^2_\perp }{ (1-z) p^2_{\perp} + zk^2_{\perp} }
 \Big) \frac{1}{ p^2_\perp } e^{- ik_{\perp} b_\perp} \int dz^- e^{-ix p^+ z^- } \langle P, S| \bar{q}(z^-,  z_\perp) \gamma^+ q(0) |P, S\rangle \,.
\label{eq:fig1-fin-axial}
\end{eqnarray}

Comparing this formula with Eq. (\ref{eq:fig1-plus}), we find that the result of calculation of the NLO corrections in the axial gauge does not, in general, coincide with the result of the naive calculation obtained in the Feynman gauge. However, inspecting the leading term of the collinear expansion
\begin{eqnarray}
&&\int dz^- e^{-ix p^+ z^- } \langle P, S| \bar{q}(z^-,  z_\perp) \gamma^+ q(0) |P, S\rangle = \int dz^- e^{-ix p^+ z^- } \langle P, S| \bar{q}(z^-,  0_\perp) \gamma^+ q(0) |P, S\rangle + \dots
\end{eqnarray}
of Eq. (\ref{eq:fig1-fin-axial}):
\begin{eqnarray}
&&\Phi^{[\gamma^+]}(x, b_\perp)\Big|^{axial}_{Fig.~\ref{fig:Fdiag}} = \frac{g^2 C_F}{4\pi^2 } \int \frac{\dhd^2p_{\perp}}{ p^2_\perp } e^{ip_{\perp} b_\perp} \int^1_0 \frac{dz}{z} \Big( \Big[ \frac{ z}{1-z} \Big]_+ + \frac{1-z}{2} \Big) 
 \int dz^- e^{-i \frac{ x }{z} p^+ z^- } \langle P, S| \bar{q}(z^- , 0_\perp) \gamma^+ q(0) |P, S\rangle
\nonumber\\
 &&+ \frac{g^2C_F}{4\pi^2} \int \frac{\dhd^2p_\perp}{ p^2_\perp } e^{ip_\perp b_\perp} \int^1_0 dz \frac{ z}{1-z} \Big(  1
- e^{-i p_\perp b_\perp} \Big) \int dz^- e^{-ix p^+ z^- } \langle P, S| \bar{q}(z^-,  0_\perp) \gamma^+ q(0) |P, S\rangle + \dots\,,
\label{eq:fig1-col-axial}
\end{eqnarray}
we find, taking into account Eq. (\ref{eq:relplus}), the complete agreement with the collinear expansion of the result obtained in the Feynman gauge, see Eq. (\ref{eq:fig1-col}).

Though we see that two results coincide in the leading term of the collinear expansion, we believe that this is merely a coincidence. For example, we would not have an agreement, even in the collinear limit, if we have chosen a different prescription for the $1/p^-$ pole in Eq. (\ref{eq:presc}), e.g. the principle value prescription. Moreover, the results would not agree in general for an arbitrary choice of the gauge fixing vector $e_\mu$ in Eq. (\ref{eq:ax-num}). At the same time we know that the result of calculation of the NLO corrections to a gauge invariant operator must be gauge independent. In the next section we will investigate why our explicit calculation with two choices of the gauge does not satisfy this requirement and how it should be modified to eliminate the dependence on the gauge choice.

\section{Calculation of Feynman diagrams in the axial gauge. The role of transverse gauge links and equations of motion\label{sec:mapping}}
In this section, we will discuss the general structure of calculation of the perturbative corrections to a gauge invariant operator in the axial gauge. Though our discussion will be completely general, for illustration we will use calculation of the NLO corrections to the quark TMDPDF operator discussed in the previous sections.

Our main goal is to show that the result of calculation in the axial gauge does not depend on the choice of the gauge fixing vector $e_\mu$ in the gluon propagator, see Eq. (\ref{eq:ax-num}). So how is it possible that this dependence, while being explicitly present in the gluon propagator, vanishes in the final result? To answer this question, let us consider the contribution of the $e_\mu$ dependent terms of the gluon propagator in the Feynman diagrams calculation.

 \begin{figure}[tb]
 \begin{center}
\includegraphics[width=0.5\textwidth]{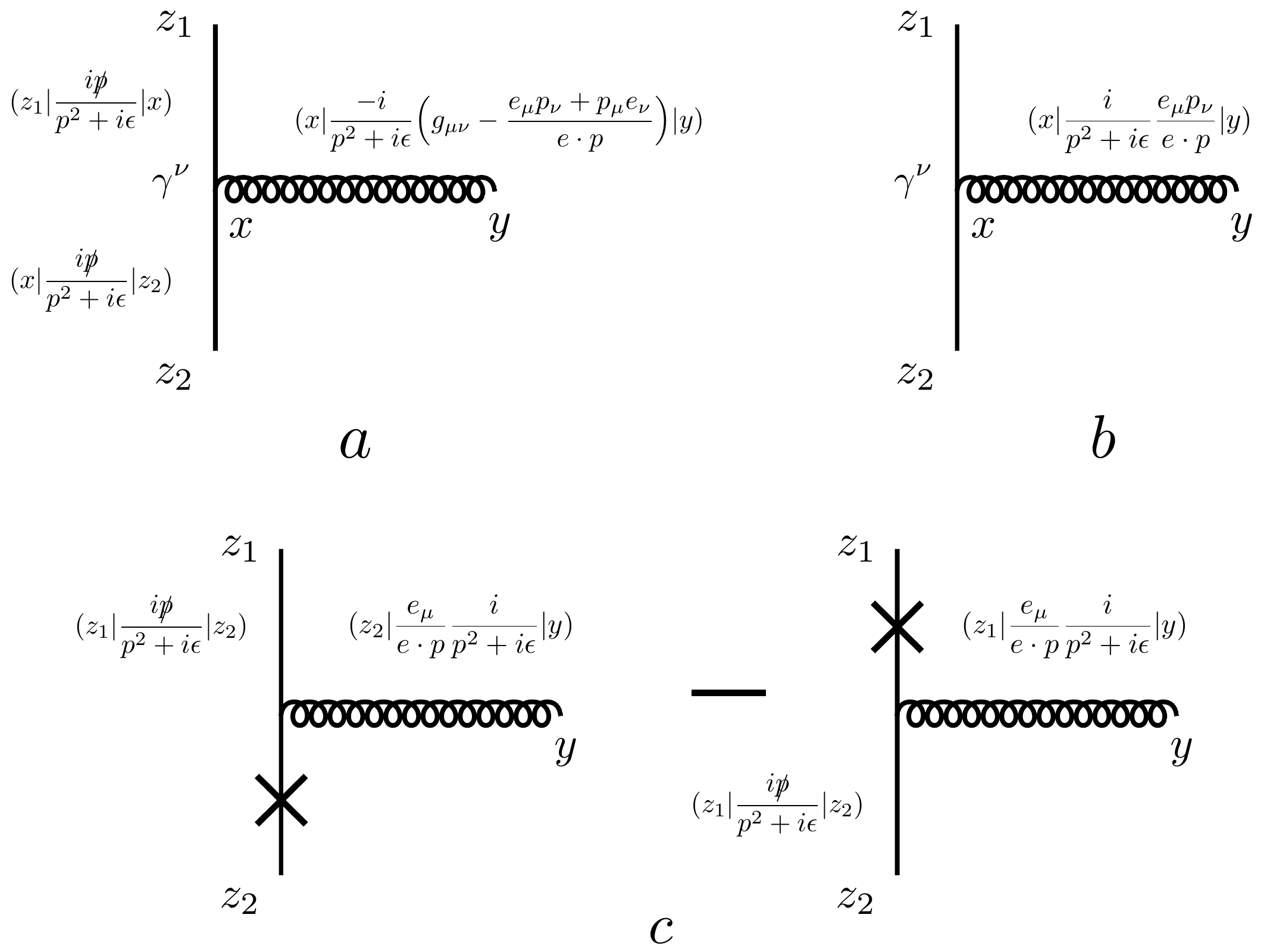}
 \end{center}
\caption{\label{fig:qg_vert}a) The gluon propagator insertion into a quark propagating from $z_2$ to $z_1$, b) The $e_\mu$ dependent term in this insertion, c) The $e_\mu$ dependent term after applying the vertex identity. The cross denotes cancellation of a propagator.}
 \end{figure}
 
Let's consider a basic element in this calculation, which is the gluon insertion into a quark propagator, see Fig. \ref{fig:qg_vert}a. The gluon line in this figure represents a gauge field propagator in the axial gauge
\begin{eqnarray}
&&(x|\frac{-i\delta^{ab}}{p^2+i\epsilon}\Big(g_{\mu\nu} - \frac{e_\mu p_\nu + p_\mu e_\nu}{e\cdot p} + \frac{e^2 p_\mu p_\nu}{(e\cdot p)^2}\Big)|y)
\end{eqnarray}
attaching the quark at position $x$. To simplify our discussion, let us assume that $e_\mu$ is a light-like vector $e^2 = 0$, so the gluon propagator takes a form
\begin{eqnarray}
&&(x|\frac{-i\delta^{ab}}{p^2+i\epsilon}\Big(g_{\mu\nu} - \frac{e_\mu p_\nu + p_\mu e_\nu}{e\cdot p} \Big)|y)\,.
\label{eq:glaxsimp1}
\end{eqnarray}
Our analysis can be easily generalized to the case of $e^2\neq 0$ with no effect on the conclusions.

From Eq. (\ref{eq:glaxsimp1}) we see that the $e_\mu$ dependent terms are proportional to the gluon momentum with Lorentz indexes contracted with the corresponding indexes of $\gamma$-matrices of the quark lines. For example, the $p_\mu e_\nu$ term of the gluon propagator inserted into the quark line, see Fig. \ref{fig:qg_vert}b, has the form\footnote{Here for brevity we suppress the color structure.}:
\begin{eqnarray}
&&ig\int d^4x (z_1|\frac{i\slashed{p}}{p^2+i\epsilon}|x)\gamma^\nu(x|\frac{i\slashed{p}}{p^2+i\epsilon}|z_2) (x|\frac{e_\mu p_\nu }{e\cdot p} \frac{i}{p^2+i\epsilon}|y)\,,
\label{eq:pninscoord}
\end{eqnarray}
that  in the momentum space reduces to 
\begin{eqnarray}
&&ig\frac{i\slashed{k}_1}{k^2_1+i\epsilon}\gamma^\nu \frac{i\slashed{k}_2}{k^2_2+i\epsilon} \frac{e_\mu p_\nu }{e\cdot p} \frac{i}{p^2+i\epsilon}\,,
\label{eq:pninsmom}
\end{eqnarray}
where $k_2$ and $k_1$ are quark momenta before and after interaction with the gluon. Now, let's take into account the momentum conservation $k_1 = k_2 + p$, which corresponds to the integration by parts procedure with respect to $p_\nu$ in Eq. (\ref{eq:pninscoord}), i.e. the vertex identity. As a result, we can rewrite Eq. (\ref{eq:pninsmom}) as
\begin{eqnarray}
&&- g \frac{i\slashed{k}_2}{k^2_2+i\epsilon} \frac{e_\mu }{e\cdot p} \frac{i}{p^2+i\epsilon} + g\frac{i\slashed{k}_1}{k^2_1+i\epsilon} \frac{e_\mu}{e\cdot p} \frac{i}{p^2+i\epsilon}\,,
\end{eqnarray}
where each term corresponds to canceling one of two quark propagators in Eq. (\ref{eq:pninsmom}). In the coordinate representation, see Eq. (\ref{eq:pninscoord}), this equation reads 
\begin{eqnarray}
&&-g (z_1|\frac{i\slashed{p}}{p^2+i\epsilon}|z_2) (z_1| \frac{e_\mu  }{e\cdot p} \frac{i}{p^2+i\epsilon} |y) + g (z_1|\frac{i\slashed{p}}{p^2+i\epsilon}|z_2) (z_2| \frac{e_\mu  }{e\cdot p} \frac{i}{p^2+i\epsilon} |y)\,.
\label{eq:pninscoord-aft}
\end{eqnarray}
This cancellation of propagators adjacent to the point of interaction with the gluon line, i.e. the vertex identity, is schematically represented in Fig. \ref{fig:qg_vert}c.

 \begin{figure}[tb]
 \begin{center}
\includegraphics[width=0.5\textwidth]{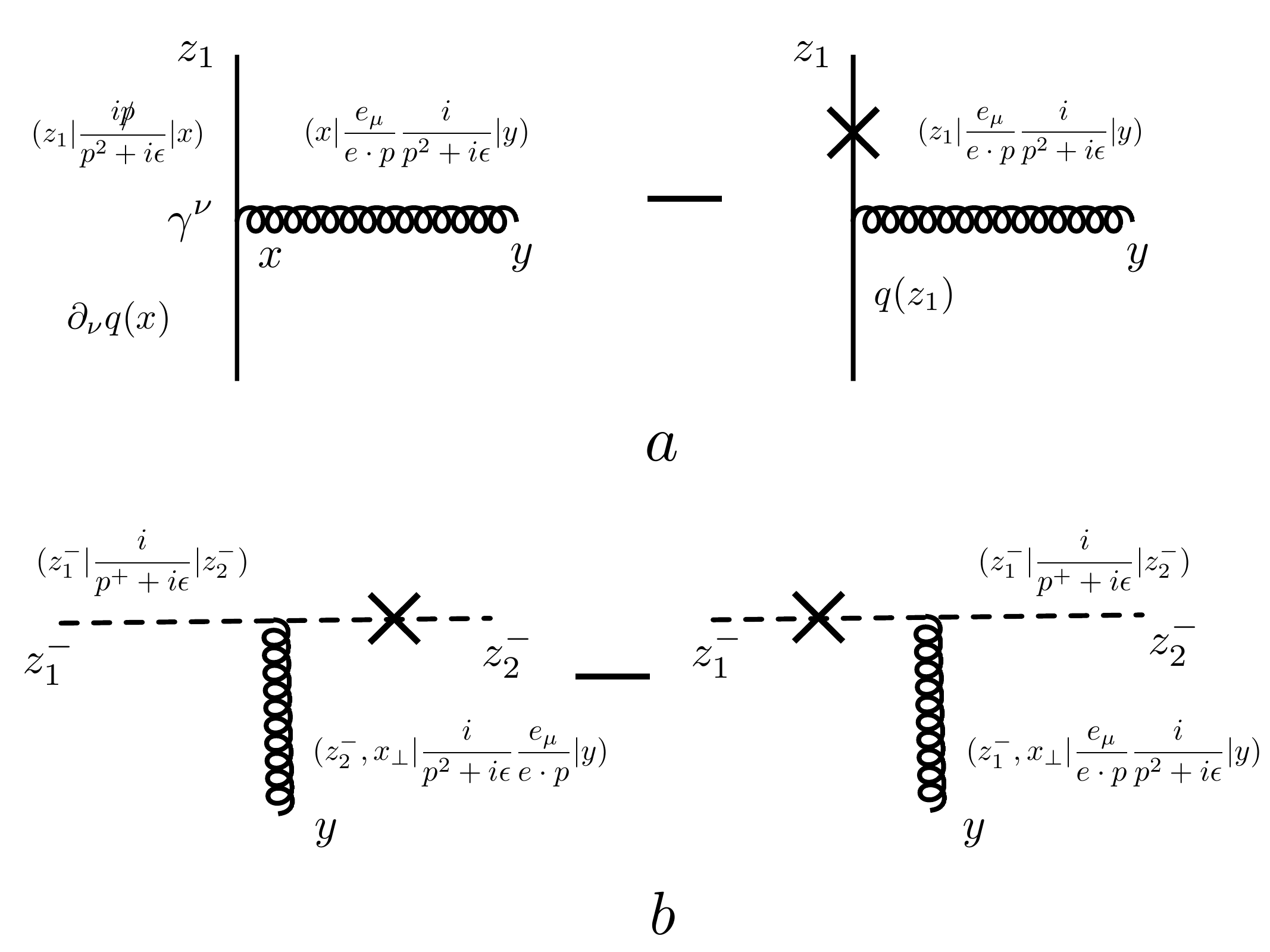}
 \end{center}
\caption{\label{fig:bacqg_wl_vert}a) The vertex identity for the interaction vertex with the background quark field, b) The vertex identity applied to the gluon-Wilson line interaction vertex.}
 \end{figure}

Application of the vertex identity to all $p_\mu$ terms of the gluon propagators is the first step of our analysis of the $e_\mu$ dependence in the Feynman diagrams calculation in the axial gauge. The procedure should be applied to every gluon propagator in all relevant Feynman diagrams, e.g. diagrams in Fig. \ref{fig:Fdiag}. In particular, being applied to the interaction vertex with the background field, see Fig. \ref{fig:bacqg_wl_vert}a,
\begin{eqnarray}
&&ig\int d^4x (z_1|\frac{i\slashed{p}}{p^2+i\epsilon}|x)\gamma^\nu q(x) (x| \frac{e_\mu p_\nu }{e\cdot p} \frac{i}{p^2+i\epsilon} |y)
\label{eq:dil-int-qbg-init}
\end{eqnarray}
the procedure yields
\begin{eqnarray}
&&-g q(z_1) (z_1| \frac{e_\mu }{e\cdot p} \frac{i}{p^2+i\epsilon} |y) + g\int d^4x (z_1|\frac{i\slashed{p}}{p^2+i\epsilon}|x)\gamma^\nu \partial_\nu q(x) (x| \frac{e_\mu }{e\cdot p} \frac{i}{p^2+i\epsilon}  |y)\,,
\label{eq:dil-int-qbg-res}
\end{eqnarray}
where the first term includes cancellation of the quark propagator, which gives a trivial propagation between points $z_1$ and $x$.

Similarly, for the interaction with a Wilson lines\footnote{While we assume that the Wilson line are along vector $n$, the discussion can be generalized to the case of an arbitrary direction.}, see Fig. \ref{fig:bacqg_wl_vert}b, which can be written as
\begin{eqnarray}
&&ig \int dx^- (z^-_1|\frac{i}{p^+ + i\epsilon}|x^-) n^\nu(x^-|\frac{i}{p^+ + i\epsilon}|z^-_2) (x^-, x_\perp| \frac{e_\mu p_\nu }{e\cdot p} \frac{i}{p^2+i\epsilon} |y)\,,
\label{eq:dil-int-wl-init}
\end{eqnarray}
after integrating by parts we obtain
\begin{eqnarray}
&&g (z^-_1|\frac{i}{p^+ + i\epsilon}|z^-_2) (z^-_2, x_\perp| \frac{i}{p^2+i\epsilon} \frac{e_\mu }{e\cdot p} |y) - g (z^-_1|\frac{i}{p^+ + i\epsilon}|z^-_2) (z^-_1, x_\perp| \frac{e_\mu }{e\cdot p} \frac{i}{p^2+i\epsilon}  |y)\,,
\label{eq:dil-int-wl-res}
\end{eqnarray}
where in each term we again see cancellation of the Wilson line propagators.

The next step of our analysis is to observe that after applying the vertex identity to all $e_\mu$ dependent terms of the gluon propagators, there are certain cancellations that appear in the sum of all Feynman diagrams at a given order of the perturbative expansion. Let us demonstrate this cancellation for the diagrams in Fig. \ref{fig:Fdiag}, which contribute to the NLO corrections to the quark TMD operator.

\begin{figure}[tb]
 \begin{center}
\includegraphics[width=0.5\textwidth]{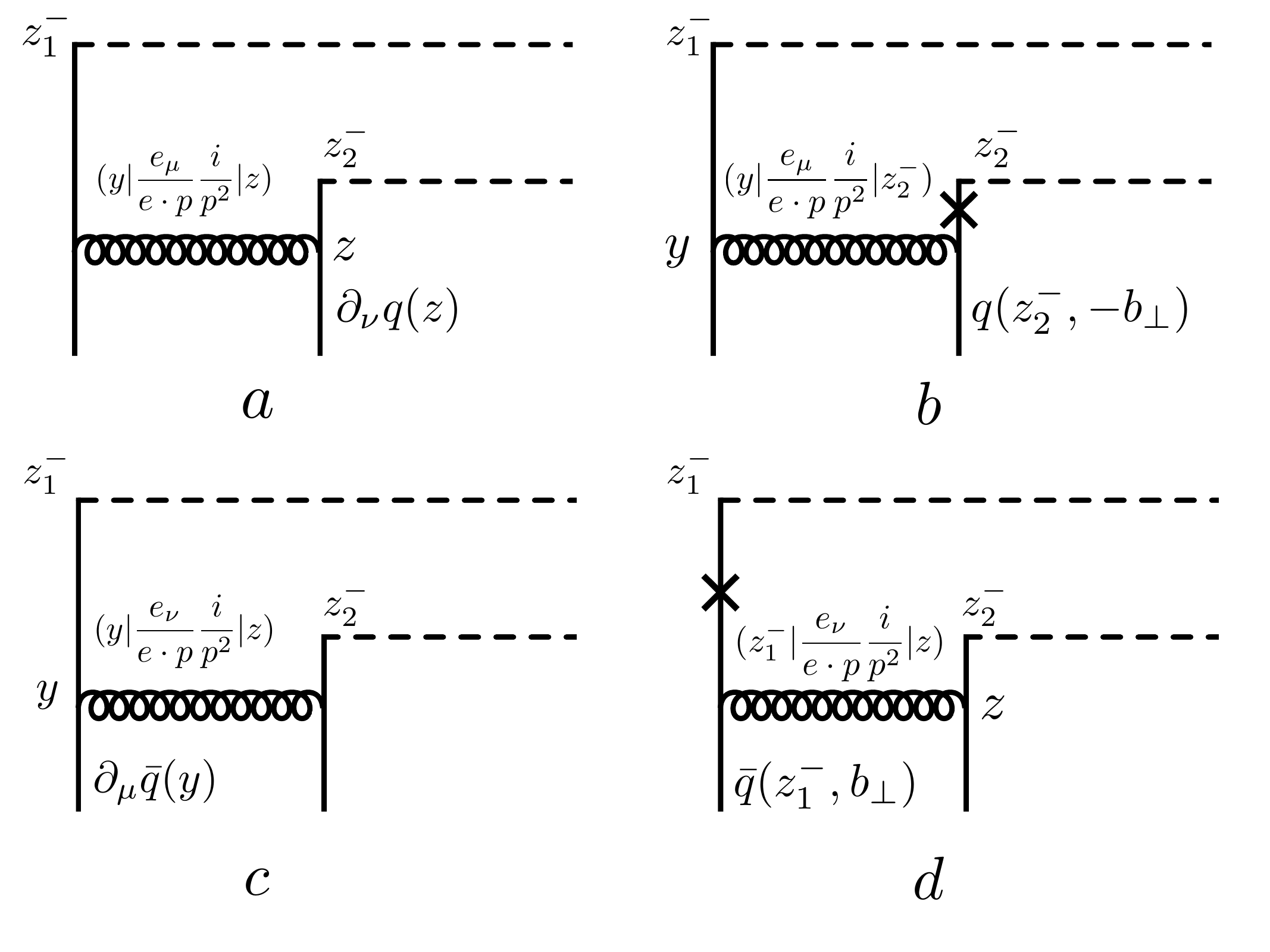}
 \end{center}
\caption{\label{fig:parts1c} The gauge fixing vector $e_\mu$ dependent terms of the diagram in Fig. \ref{fig:Fdiag}c after applying the vertex identity to the $e_\mu$ dependent terms of the gluon propagator in the axial gauge, see Eq. (\ref{eq:1cbyparts}). The cross denotes cancellation of the corresponding propagators.}
 \end{figure}

Starting with a diagram in Fig. \ref{fig:Fdiag}c 
\begin{eqnarray}
&&\mathcal{U}^{[\gamma^+]}(z^-_1, z^-_2, b_\perp)\Big|^{axial}_{Fig.~\ref{fig:Fdiag}c}
\nonumber\\
&&= -g^2\int d^4y \int d^4z \bar{q}(y)\gamma^\mu t^a (y|\frac{i\slashed{p}}{p^2+i\epsilon}|z^-_1, b_\perp) \gamma^+ (z^-_2, -  b_\perp|\frac{i\slashed{p}}{p^2+i\epsilon}|z) \gamma^\nu t^b q(z) (z|\frac{-id_{\mu\nu}(p)\delta^{ab}}{p^2+i\epsilon}|y)\,,
\end{eqnarray}
we subsequently apply the integration by parts procedure and rewrite the expression for the diagram as
\begin{eqnarray}
&&\mathcal{U}^{[\gamma^+]}(z^-_1, z^-_2, b_\perp)\Big|^{axial}_{Fig.~\ref{fig:Fdiag}c}
\nonumber\\
&&= -g^2 C_F \int d^4y \int d^4z \bar{q}(y)\gamma^\mu (y|\frac{i\slashed{p}}{p^2+i\epsilon}|z^-_1, b_\perp) \gamma^+ (z^-_2, -  b_\perp|\frac{i\slashed{p}}{p^2+i\epsilon}|z) \gamma^\nu q(z) (z|\frac{-ig_{\mu\nu}}{p^2+i\epsilon} |y)
\nonumber\\
&&- i g^2 C_F \int d^4y \int d^4z \bar{q}(y)\gamma^\mu (y|\frac{i\slashed{p}}{p^2+i\epsilon}|z^-_1, b_\perp) \gamma^+ (z^-_2, -  b_\perp|\frac{i\slashed{p}}{p^2+i\epsilon}|z) \gamma^\nu \partial_\nu q(z) (z| \frac{e_\mu }{e\cdot p} \frac{-i}{p^2+i\epsilon}   |y)
\nonumber\\
&&+ i g^2 C_F \int d^4y \bar{q}(y)\gamma^\mu (y|\frac{i\slashed{p}}{p^2+i\epsilon}|z^-_1, b_\perp) \gamma^+ q(z^-_2, -  b_\perp) (z^-_2, -  b_\perp| \frac{e_\mu }{e\cdot p} \frac{-i}{p^2+i\epsilon}   |y)
\nonumber\\
&&+i g^2 C_F \int d^4y \int d^4z \partial_\mu \bar{q}(y)\gamma^\mu (y|\frac{i\slashed{p}}{p^2+i\epsilon}|z^-_1, b_\perp) \gamma^+ (z^-_2, -  b_\perp|\frac{i\slashed{p}}{p^2+i\epsilon}|z) \gamma^\nu q(z) (z| \frac{  e_\nu}{e\cdot p} \frac{-i}{p^2+i\epsilon}  |y)
\nonumber\\
&&+i g^2 C_F \int d^4z \bar{q}(z^-_1, b_\perp) \gamma^+ (z^-_2, -  b_\perp|\frac{i\slashed{p}}{p^2+i\epsilon}|z) \gamma^\nu q(z) (z|\frac{  e_\nu}{e\cdot p} \frac{-i}{p^2+i\epsilon}  |z^-_1, b_\perp)\,.
\label{eq:1cbyparts}
\end{eqnarray}
The last four terms of this equation are depicted in Fig.~\ref{fig:parts1c}. Note that the first term in Eq. (\ref{eq:1cbyparts}) is $e_\mu$ independent and coincides, obviously, with the expression for the diagram in the Feynman gauge, see Eq. (\ref{eq:Feynmat1c}) for the matrix element of this contribution. Let us also emphasize that the appearance of terms with derivatives of the background quark fields, e.g. $\gamma^\nu \partial_\nu q$, which will be important in the following discussion.  

\begin{figure}[tb]
 \begin{center}
\includegraphics[width=0.5\textwidth]{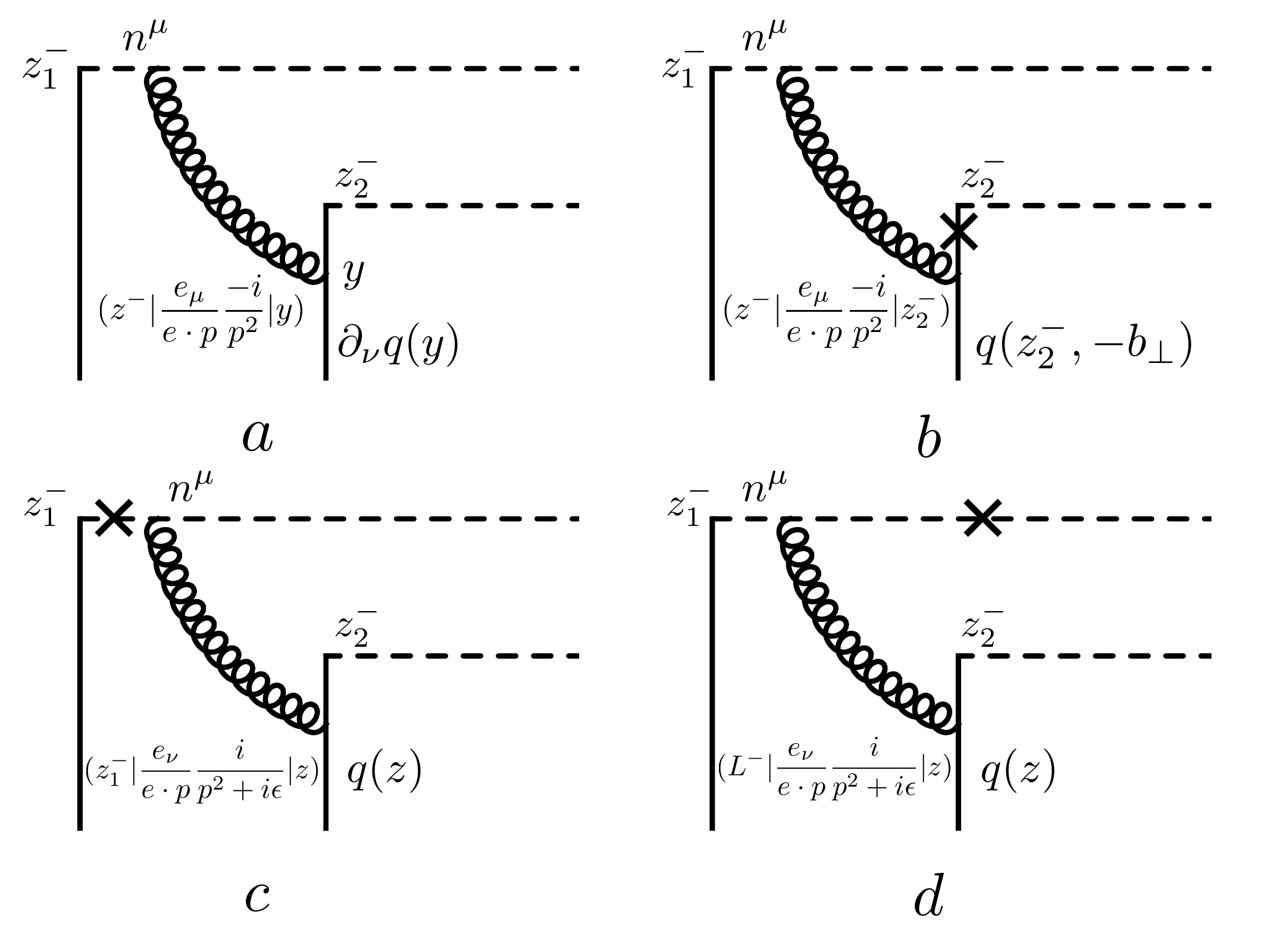}
 \end{center}
\caption{\label{fig:parts1a}
The same as Fig.~\ref{fig:parts1c} but for the diagram in Fig. \ref{fig:Fdiag}a, ,see Eq. (\ref{eq:1abyparts}). }
 \end{figure}

Now, let us perform a similar manipulation with the diagram in Fig. \ref{fig:Fdiag}a:
\begin{eqnarray}
&&\mathcal{U}^{[\gamma^+]}(z^-_1, z^-_2, b_\perp)\Big|^{axial}_{Fig.~\ref{fig:Fdiag}a}
\label{eq:1abyparts}\\
&&= -g^2 C_F \int^{z^-_1}_{\infty}dz^- \int d^4y \bar{q}(z^-_1, b_\perp) \gamma^+ (z^-_2, - b_\perp|\frac{i\slashed{p}}{p^2+i\epsilon}|y) \gamma^\nu q(y) n^\mu (z^-, b_\perp|\frac{-ig_{\mu\nu}}{p^2+i\epsilon}|y)
\nonumber\\
&&+ i g^2 C_F \int_{\infty}^{z^-_1} dz^- n^\mu \int d^4y \bar{q}(z^-_1, b_\perp) \gamma^+ (z^-_2, - b_\perp|\frac{i\slashed{p}}{p^2+i\epsilon}|y) \gamma^\nu \partial_\nu q(y) (z^-, b_\perp| \frac{e_\mu }{e\cdot p} \frac{-i}{p^2+i\epsilon}  |y)
\nonumber\\
&&- i g^2 C_F \int_{\infty}^{z^-_1} dz^- n^\mu \bar{q}(z^-_1, b_\perp) \gamma^+  q(z^-_2, - b_\perp) (z^-, b_\perp| \frac{e_\mu }{e\cdot p} \frac{-i}{p^2+i\epsilon}  |z^-_2, - b_\perp)
\nonumber\\
&&- i g^2 C_F \int d^4z \bar{q}(z^-_1, b_\perp) \gamma^+ (z^-_2, - b_\perp|\frac{i\slashed{p}}{p^2+i\epsilon}|z) \gamma^\nu q(z)  (z| \frac{ e_\nu}{e\cdot p} \frac{-i}{p^2+i\epsilon}  |z^-_1, b_\perp)
\nonumber\\
&&+ i g^2 C_F \int d^4z \bar{q}(z^-_1, b_\perp) \gamma^+ (z^-_2, - b_\perp|\frac{i\slashed{p}}{p^2+i\epsilon}|z) \gamma^\nu q(z)  (z| \frac{ e_\nu}{e\cdot p} \frac{-i}{p^2+i\epsilon}  |L^-, b_\perp)\,,
\nonumber
\end{eqnarray}
where in the last term we introduced a cut-off for the size of the light-cone Wilson lines $L^- \to \infty$ in the definition of the TMDPDF operator. The equation is schematically shown in Fig. \ref{fig:parts1a}.

Comparing Eqs. (\ref{eq:1cbyparts}) and (\ref{eq:1abyparts}), we find that both diagrams contain a term
\begin{eqnarray}
&&i g^2 C_F \int d^4z \bar{q}(z^-_1, b_\perp) \gamma^+ (z^-_2, - b_\perp|\frac{i\slashed{p}}{p^2+i\epsilon}|z) \gamma^\nu q(z)  (z| \frac{ e_\nu}{e\cdot p} \frac{-i}{p^2+i\epsilon}  |z^-_1, b_\perp)
\end{eqnarray}
This term corresponds to contributions in Figs. \ref{fig:parts1c}d and \ref{fig:parts1a}c and involves cancellation of the internal quark or Wilson line propagators. In Eqs. (\ref{eq:1cbyparts}) and (\ref{eq:1abyparts}) the term comes with opposite signs, so in the sum of diagrams this $e_\mu$ dependent term gets canceled. We find that this cancellation is not coincidental and can be observed for other $e_\mu$ dependent terms as well if the sum of all perturbative diagrams is taken.

Indeed, repeating the procedure for all remaining diagrams in Fig. \ref{fig:Fdiag} for the sum of all diagrams we get
\begin{eqnarray}
&&\mathcal{U}^{[\gamma^+]}(z^-_1, z^-_2, b_\perp)\Big|^{axial} = \mathcal{U}^{[\gamma^+]}(z^-_1, z^-_2, b_\perp)\Big|^{Feyn.} + A + B
\label{eq:after-part}
\end{eqnarray}
where $\mathcal{U}^{[\gamma^+]}(z^-_1, z^-_2, b_\perp)|^{Feyn.}$ in the r.h.s. of the equation is the expression for the sum of all diagrams in Fig. \ref{fig:Fdiag} written in the Feynman gauge. From this equation we see that the difference between two gauges is up to two types of terms, i.e. $A$ and $B$, which we discuss below. As a result, we conclude that to ensure the gauge invariance both of these terms have to be trivial. At this point we want to emphasize that our discussion is completely general and applies to an arbitrary choice of the gauge fixing vector $e_\mu$.\footnote{While we discuss a special case of the light-like vectors $e^2 = 0$, our discussion can be easily generalized to the case of arbitrary $e^2\neq 0$.}

Let us now discuss the difference terms in the r.h.s. of Eq. (\ref{eq:after-part}). The explicit form of these terms can be found in Appendix \ref{ap:dif-terms}. Starting with the first group of terms $A$, see Eq. (\ref{eq:termA}), we find that each term in this contribution is defined by behavior of the gluon propagator at the spatial infinity $L^-\to\infty$, and includes an integral of the following form:\footnote{This form is of course specific to our choice of the direction of the Wilson lines, but a similar construction can be obtain for an arbitrary direction of the gauge links.}
\begin{eqnarray}
&&\label{eq:sp-inf-basic-int}(L^-, x_\perp|\frac{-i}{p^2+i\epsilon} \frac{e_\mu}{e\cdot p} |y) = \int \frac{d p^+}{2\pi} e^{-ip^+(L^- - y^-)} \int \frac{d p^-}{2\pi} e^{ip^- y^+} \int \dhd^2p_\perp  e^{ip_\perp(x-y)_\perp}\frac{-i}{2p^+p^- - p^2_\perp+i\epsilon} \frac{e_\mu}{e\cdot p}\,.
\end{eqnarray}

\begin{figure}[tb]
 \begin{center}
\includegraphics[width=0.4\textwidth]{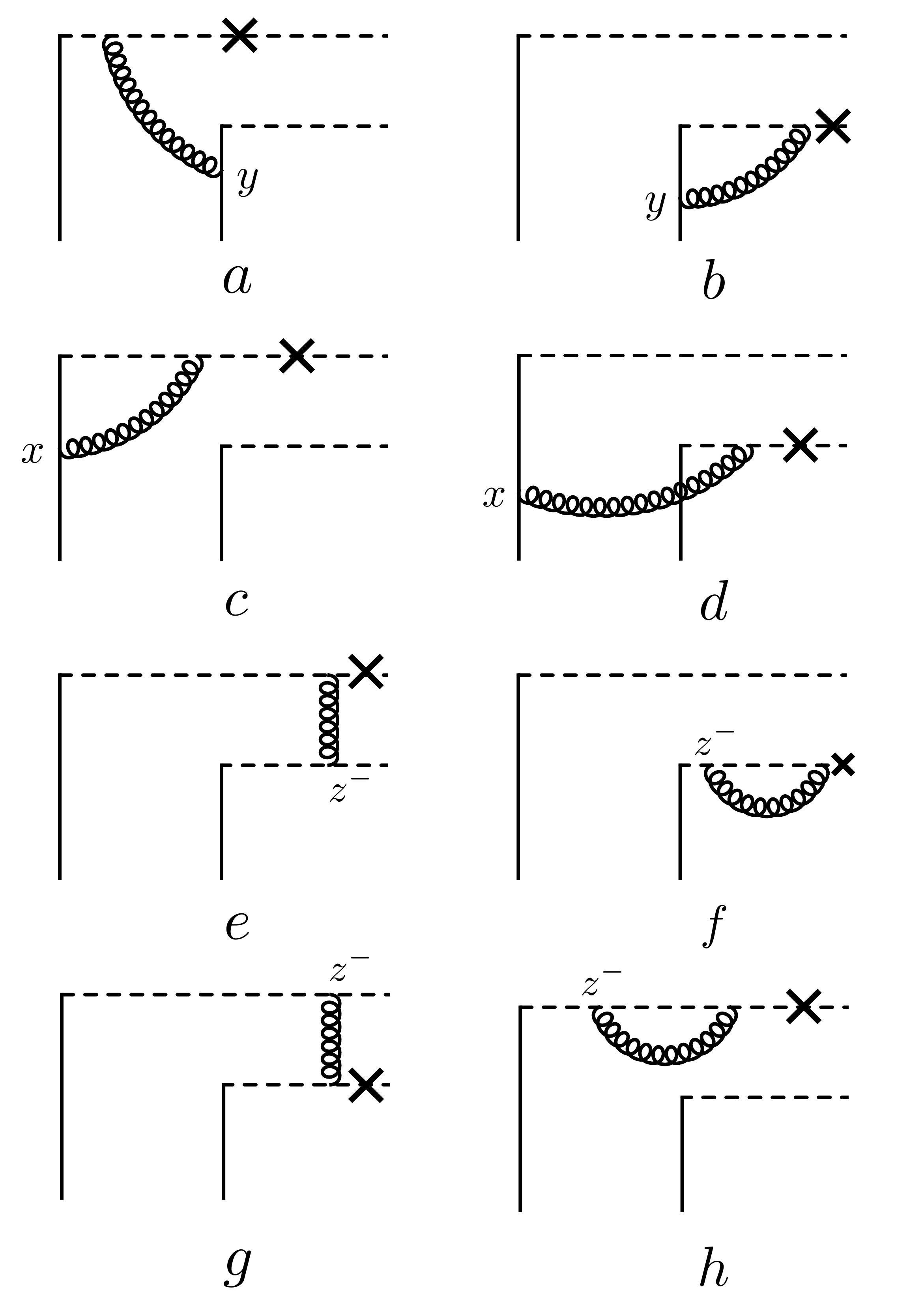}
 \end{center}
\caption{\label{fig:Aterms}The diagrammatic representation of the difference term $A$, see Eq. (\ref{eq:termA}).}
 \end{figure}

It is important to note that all terms in $A$ correspond to cancellation of a Wilson line propagator that goes to the spatial infinity. The diagrammatic representation of these terms is given in Fig. \ref{fig:Aterms}.

Whether the integrals (\ref{eq:sp-inf-basic-int}) in Eq. (\ref{eq:termA}) vanish or are independent the gauge fixing vector $e_\mu$ is subject to the prescription for the axial $1/e\cdot p$ singularity. Nevertheless, regardless of any particular choice, we want to stress that the appears of $A$ terms in Eq. (\ref{eq:after-part}) is fundamental and reflects a flaw in our calculation. 

Indeed, in the end of Sec. \ref{sec:defI} we mentioned that our calculation is ``naive" in a sense that it does not fully account for the gauge invariance of the operator (\ref{def:op}) by neglecting the contribution of the transverse gauge link at the spatial infinity. This can be easily seen by inspecting diagrams in Fig. \ref{fig:Fdiag} that do not contain emission from the transverse Wilson lines at the infinity. However, such emission is essential for ensuring the gauge invariance of the final result of calculation.

\begin{figure}[htb]
 \begin{center}
\includegraphics[width=0.5\textwidth]{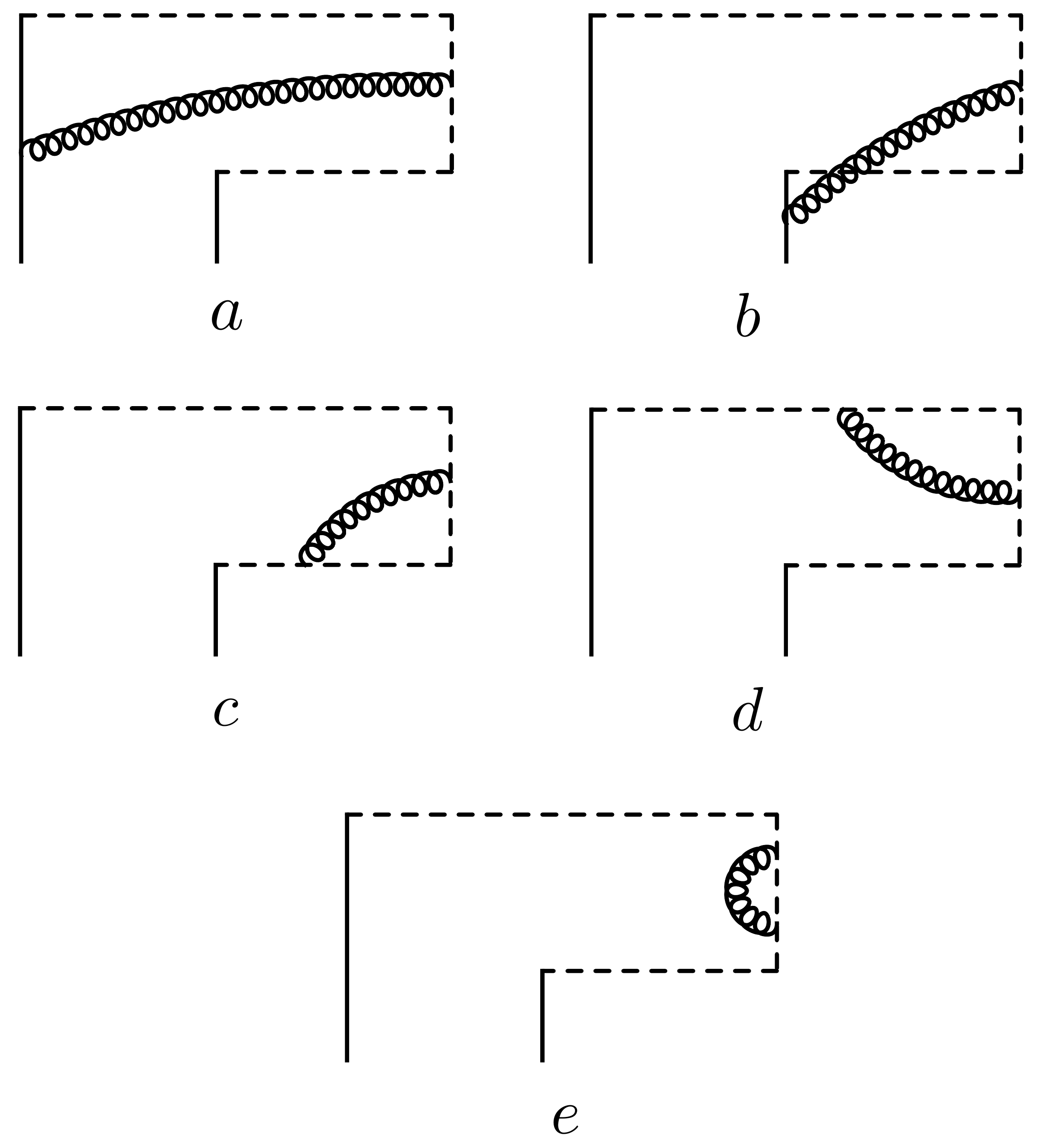}
 \end{center}
\caption{\label{fig:tr_link_emission}The NLO correction diagrams to the quark TMD operator with emission from the transverse gauge link at the spatial infinity.}
 \end{figure}

For the quark TMDPDF operator (\ref{def:op}), the diagrams at the NLO order with emission from the transverse gauge link at the spatial infinity are presented in Fig. \ref{fig:tr_link_emission}. Similarly to our previous analysis, one can apply the vertex identity to these diagrams written in the axial gauge and, in analogy to Eq. (\ref{eq:after-part}), present the result as a sum of the expression for these diagrams in the Feynman gauge and some difference terms. For brevity we do not provide an explicit expression for these difference terms, but their diagrammatic representation can be easily obtained from Fig. \ref{fig:tr_link_emission}.

\begin{figure}[htb]
 \begin{center}
\includegraphics[width=0.7\textwidth]{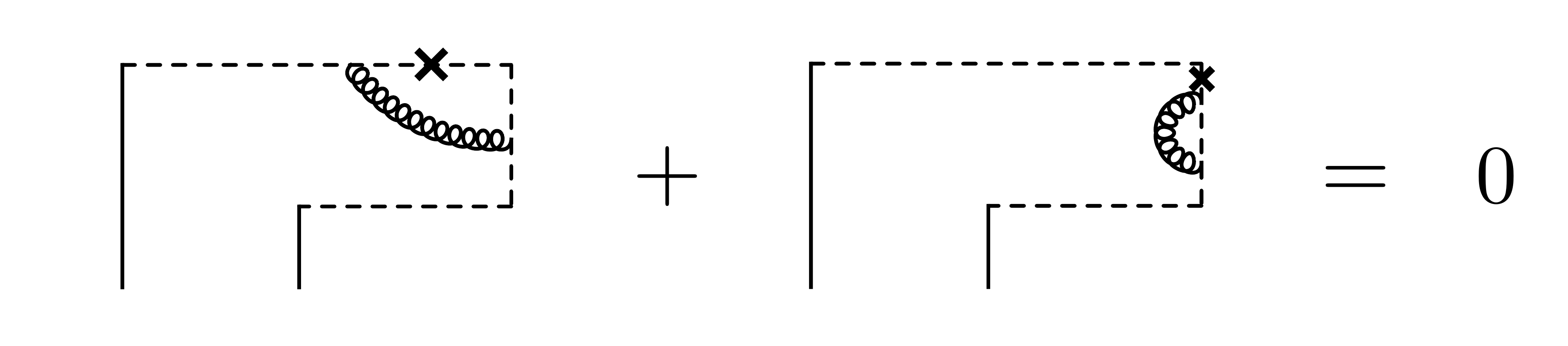}
 \end{center}
\caption{\label{fig:tr_link_int_by parts-some-canc}Cancellation of the difference terms after applying the vertex identity to diagrams in Fig. \ref{fig:tr_link_emission}.}
 \end{figure}

\begin{figure}[htb]
 \begin{center}
\includegraphics[width=0.5\textwidth]{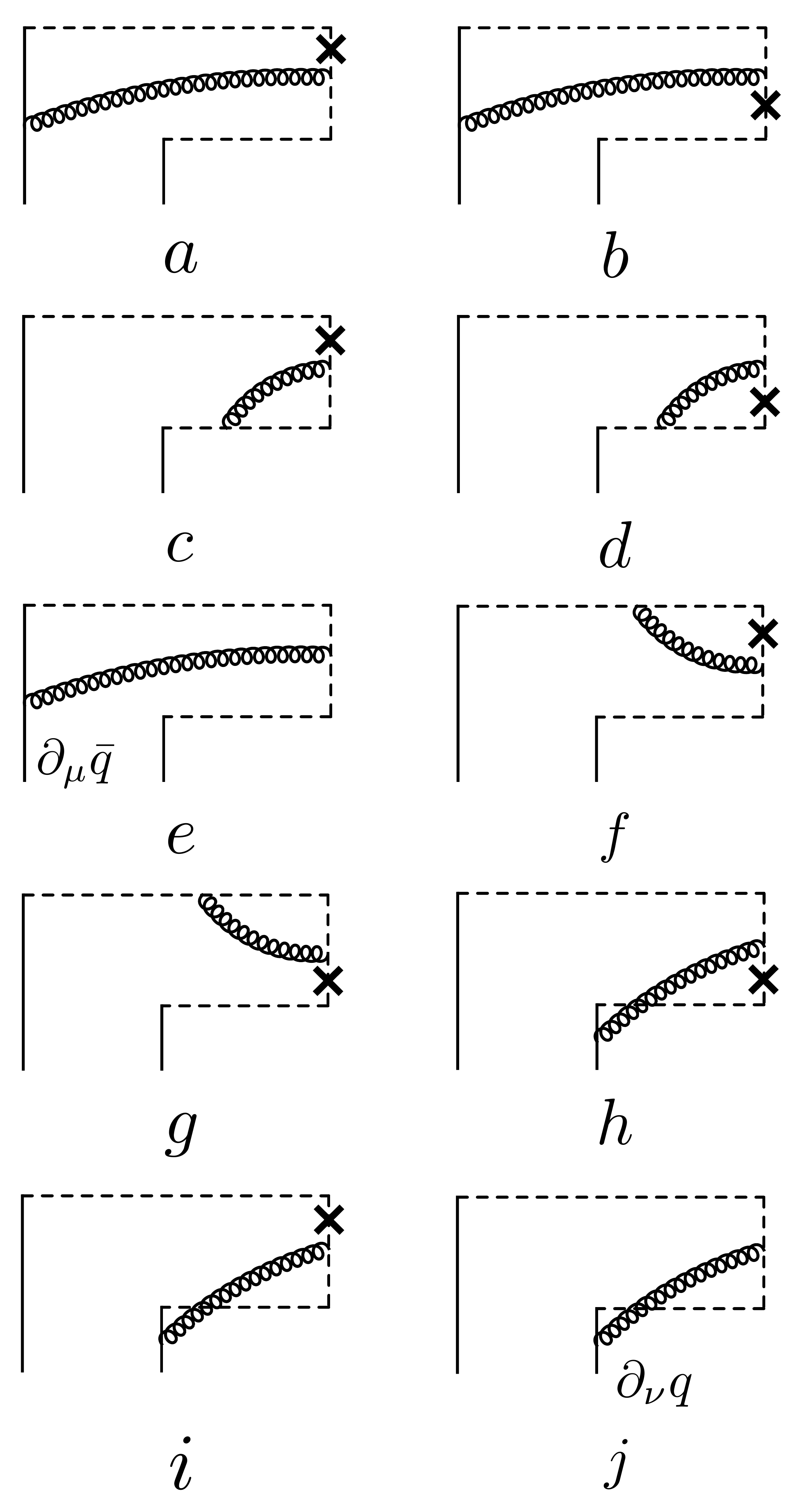}
 \end{center}
\caption{\label{fig:tr_link_int_by parts}The difference terms for the diagrams with emission from the transverse gauge links at the spatial infinity in Fig. \ref{fig:tr_link_emission}. The cross denotes cancellation of a propagator. The $\partial_\mu q$ notation denotes terms with a derivative acting on the background quark field.}
 \end{figure}

Similarly to the analysis of the difference terms for the diagrams in Fig. \ref{fig:Fdiag}, it is easy to find that some of these terms cancel each other, see for example Fig.  \ref{fig:tr_link_int_by parts-some-canc}. Now, remarkably, the remaining terms, see Fig. \ref{fig:tr_link_int_by parts}, apart from the terms with derivatives acting on the background fields, which we will discuss latter, exactly cancel the difference terms $A$ in Eq. (\ref{eq:after-part}), c.f. for example Figs. \ref{fig:Aterms}c and \ref{fig:tr_link_int_by parts}a.

So we conclude that the aftermath of inclusion of the emission from the transverse Wilson lines at the spatial infinity is complete cancellation of the difference terms of the $A$ type in Eq. (\ref{eq:after-part}). As a result we see that the contribution of the transverse Wilson lines cannot be let aside since, as we see from our explicit calculation, neglecting this contribution can easily lead to a wrong result. This is the first important point that we want to make in this paper.

Of course, one can argue that while our general analysis is correct, in practice the contribution of the transverse Wilson lines is not important since it includes emission at the spatial infinity and with an appropriate choice of the gauge fixing vector $e_\mu$ and prescription of the $1/e\cdot p$ singularity it, e.g. all terms in Eq. (\ref{eq:termA}), can be made trivial, so the $A$ terms do not contribute in Eq. (\ref{eq:after-part}), even without the transverse Wilson line in the initial definition of the TMD operator.

\begin{figure}[htb]
 \begin{center}
\includegraphics[width=0.2\textwidth]{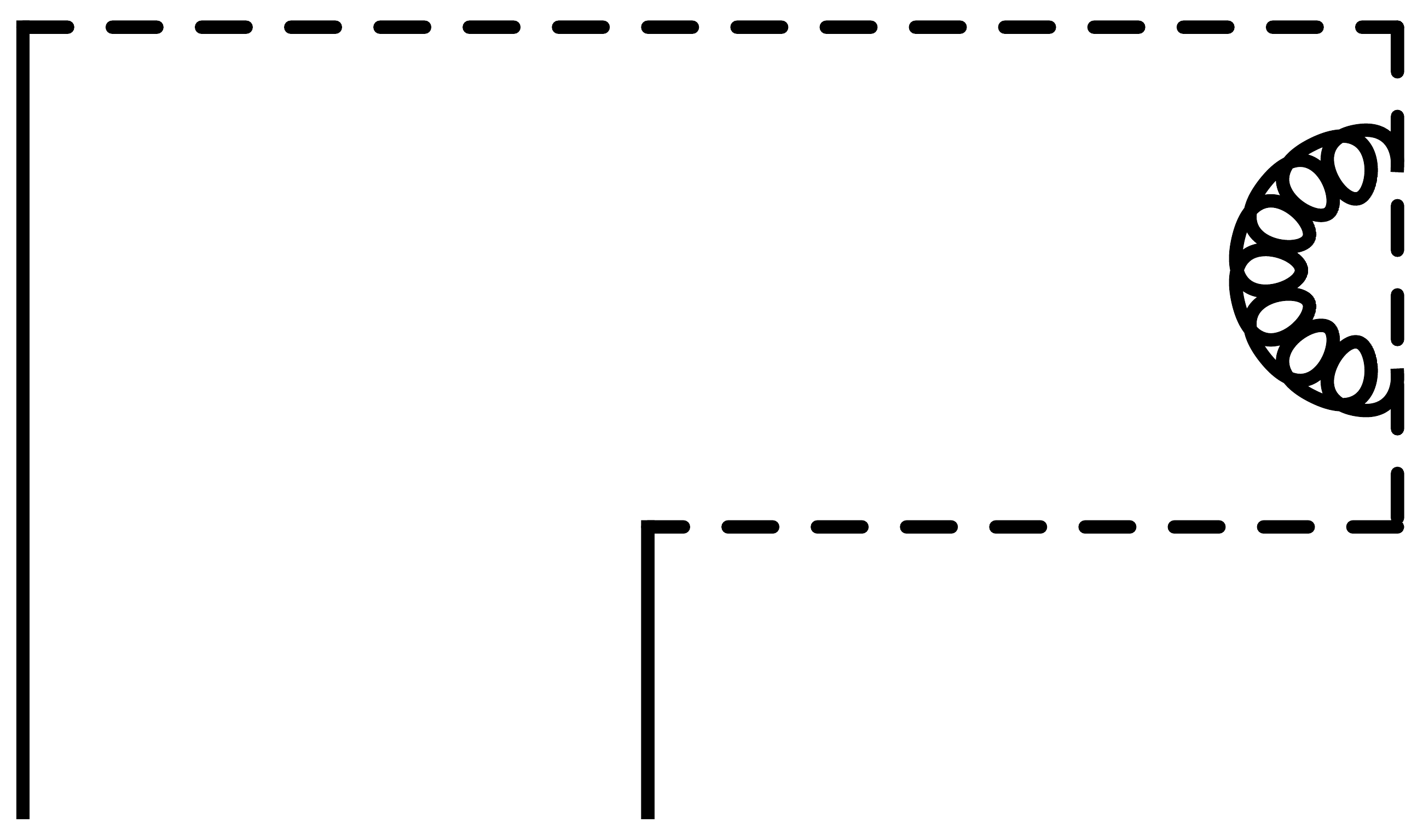}
 \end{center}
\caption{\label{fig:self-en-cor}Self-energy correction to the transverse gauge link at the spatial infinity.}
 \end{figure}

However, one should be careful with this approach since it does not take into account non-trivial self-energy corrections to the transverse gauge link, see Fig. \ref{fig:self-en-cor}. These corrections appear in calculations with an arbitrary gauge fixing vector $e_\mu$, and even in the  Feynman gauge. Subsequently, the analysis of these contributions needs special treatment, which depends on a particular factorization scheme.

Even though we find that contribution of the transverse link at the spatial infinity is necessary to ensure the gauge invariance of the calculation of perturbative corrections to the TMD operator, from inspecting Eq. (\ref{eq:after-part}) we conclude that the gauge invariance of the operator itself is not sufficient to eliminate discrepancies in results calculated in different gauges.

Indeed, in Eq. (\ref{eq:after-part}) the inclusion of the transverse gauge link eliminates only the difference term $A$ and does not change the $B$ contribution. The distinguishing property of the $B$ terms is that all these terms contain derivatives of the background fields, e.g. $\gamma^\mu \partial_\mu q$. From this it becomes obvious that to eliminate the  disagreement between the results of calculation in two gauges we have to assume that the background fields satisfy EoM, which in our dilute approximation are simply
\begin{eqnarray}
&&i\gamma^\mu \partial_\mu q = 0;\ \ \ -i\partial_\mu \bar{q} \gamma^\mu = 0\,.
\label{eq:emo}
\end{eqnarray}

From our analysis we conclude that in the QCD factorization approach the results of perturbative calculations obtained in different gauges agree only up to terms proportional to the EoM for the background partons, which we explicitly see in Eq. (\ref{eq:after-part}).

We believe that the important role of the EoM in the QCD factorization approach has not been sufficiently addressed in the literature. Meanwhile, we find that imposing the EoM for the background fields in not an option but a requirement. We find that while the role of the EoM can be overlooked in some leading order calculations, in general they should be taken into account as  we expect that in the higher orders of the collinear expansion the EoM will start to play the role.

For example, currently the majority of the perturbative calculations of the TMD distributions are done in the collinear limit, which is the small $b_\perp$ approximation for the TMDPDFs. In this approximation the background partons are effectively replaced with the collinear particles that do not carry any transverse momenta and are on the mass-shell. In this approximation the EoM are satisfied and the results of calculation in different gauges do agree, see in particular our results (\ref{eq:fig1-col}) and (\ref{eq:fig1-col-axial}) for the collinear expansion of the quark TMD operator obtained in the Feynman and axial gauges.

However, this approximation is valid only in the region of small $b_\perp \ll \Lambda^{-1}_{\rm QCD}$, while application of the TMD factorization with TMDPDFs describing the non-perturbative structure of the hadron can be justified only in the opposite region of large $b_\perp \lesssim \Lambda^{-1}_{\rm QCD}$. So regardless of whether one tries to calculate the higher order corrections  to the collinear expansion or calculate TMDPDFs in the physical region of large $b_\perp \lesssim \Lambda^{-1}_{\rm QCD}$, as we do in Secs. \ref{sec:Feynmancalc} and \ref{sec:ax-calc}, the EoM for the background fields has to be imposed to ensure the gauge invariance of the calculation.

Another important aspect that we want to emphasize is that since the background fields satisfy the equations of motion, the result of perturbative calculation is not unique. In other words, one can always add a term proportional to the EoM and obtain a physically equivalent result.

In the dilute limit this can be seen as ambiguity in the perturbative coefficients, c.f. for instance Eqs. (\ref{eq:fig1-fin})  and (\ref{eq:fig1-fin-axial}). However, as it follows from our analysis of the problem in the dense limit presented in the next section, the correct way of interpreting this phenomena is ambiguity in the contribution of different non-perturbative matrix elements of the background-field operators to the full result.

In this sense, there is no a ``correct" result for the perturbative calculation. Instead, there is an infinite number of solutions which are different up to the equations of motion. In this sense, what we call a ``correct" form of the final result is a matter of convention.

Note that ``adding" terms proportional to the EoM effectively corresponds to going from one choice of the gauge fixing vector $e_{1\mu}$ to another one $e_{2\mu}$. The limiting case, when the explicit dependence on the $e_{\mu}$ vector is eliminated, in the dilute regime without the background gluons corresponds to the Feynman gauge.

This ambiguity, however, can introduce an unexpected conundrum. Suppose we add to the perturbative result a term, which is proportional to the equation of motion but contains a singularity. This singularity is of course non-physical, because the corresponding term is trivial due to the EoM for the background fields. Yet in practice it is not always trivial to detect this non-physical divergencies and remove them from the final result. 

For example, we observed such a divergence in our ``naive" calculation in the axial gauge, which contains a non-physical IR divergence $\int dz/z$, see Eq. (\ref{eq:ax-virt-diverge}), originating from the contribution of the diagrams in Figs.~\ref{fig:Fdiag}d and~\ref{fig:Fdiag}e. We find that the most efficient way to detect such spurious singularities is to perform the perturbative calculation in a number of distinct gauges and compare the structure of singularities, which has to be the same. If the structure is different, one should apply the EoM to eliminate the unphysical singularities.  For instance, in Appendix \ref{ap:EoM-div} we explicitly show that the IR divergence in Eq. (\ref{eq:ax-virt-diverge}) is related to the contribution of EoM, and for this reason is 
an spurious singularity to be removed from the final result.

Our analysis so far has been done in the dilute limit for the background fields, which is manifested by the form of the corresponding EoM (\ref{eq:emo}). So it is natural to ask whether the analysis and its conclusions can be extended to the dense limit when there are multiple  
interactions with the background parton fields. In the next section we will discuss this limit and necessary modifications to our approach.

\section{Generalization to multi-background field interactions\label{sec:beyondDL}}

So far we limited our discussion to the dilute case, i.e. considering interactions with only two quarks from the background fields. It is, of course, important to understand how our conclusions of the previous section change when multiple interactions with background fields are taken into account.
\begin{figure}[htb]
 \begin{center}
\includegraphics[width=80mm]{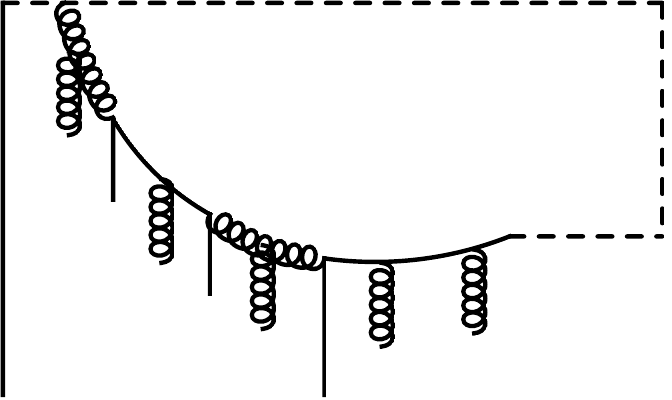}
 \end{center}
\caption{\label{fig:f1-dense}The diagram in Fig. \ref{fig:Fdiag}a generalized to the dense limit with multiple interactions with the background field.}
 \end{figure}

Starting with the dilute diagrams in Fig. \ref{fig:Fdiag}, it is easy to obtain their generalization to the dense limit by replacing all bare quark and gluon propagators with propagators in the background field, which resum multiple interactions with quarks and gluons of the target. For instance, such generalization of Fig. \ref{fig:Fdiag}a can be found in Fig. \ref{fig:f1-dense}. Note that multiple interactions with quarks and gluons of the background mix different types of partons in the quantum loop.

\begin{figure}[htb]
 \begin{center}
\includegraphics[width=80mm]{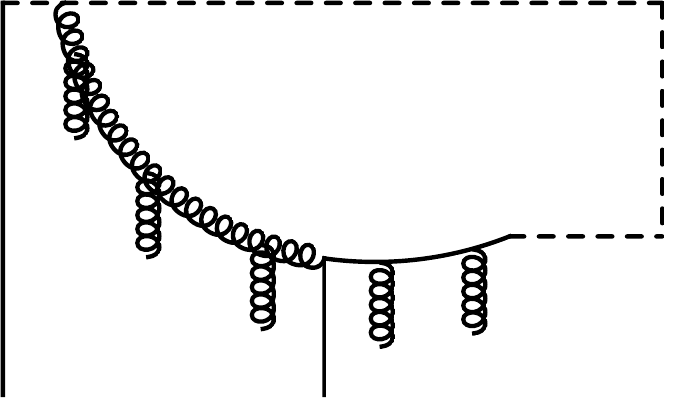}
 \end{center}
\caption{\label{fig:dense-cancellation-internal}Generalization of the diagram in Fig. \ref{fig:Fdiag}a with multiple insertion of the background gluon fields.}
 \end{figure}

Let us, however,  simplify the problem and assume that similarly to diagrams in Fig. \ref{fig:Fdiag} there are only two background quarks, yet the number of gluon insertions is arbitrary, see e.g. Fig. \ref{fig:dense-cancellation-internal}. To compute such diagrams we need to generalize the bare quark and gluon propagators resuming multiple gluon insertions.

\begin{figure}[htb]
 \begin{center}
\includegraphics[width=0.8\textwidth]{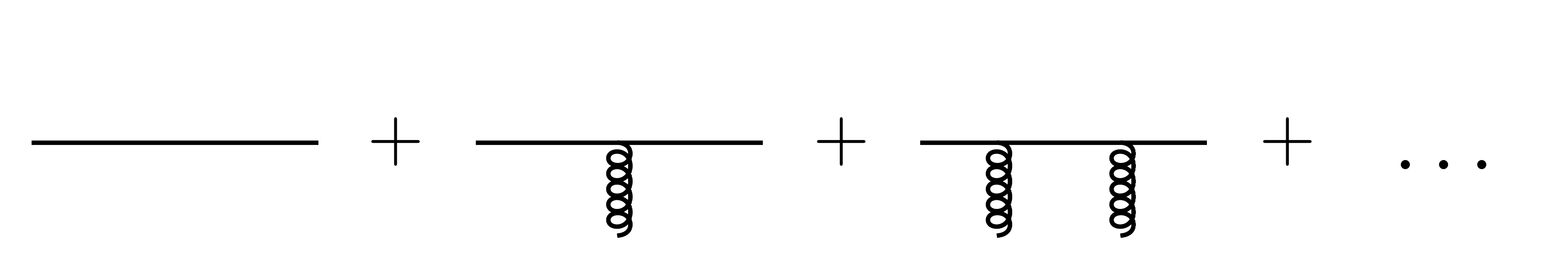}
 \end{center}
\caption{\label{fig:qp-res}Resumation of multiple insertions of the background gluon fields into the quark propagator.}
 \end{figure}

For the quark propagators in Fig. \ref{fig:Fdiag} that implies the following replacement:
\begin{eqnarray}
&&(x|\frac{i}{\slashed{p}}|y) \to (x|\frac{i}{\slashed{p}} + \frac{i}{\slashed{p}}ig\slashed{A}\frac{i}{\slashed{p}} + \frac{i}{\slashed{p}}ig\slashed{A}\frac{i}{\slashed{p}}ig\slashed{A}\frac{i}{\slashed{p}} + \dots |y)\,,
\label{eq:qp-mb}
\end{eqnarray}
where each subsequent term has one extra gluon insertion compared to the previous one. The sum of terms in the right hand side  of Eq. (\ref{eq:qp-mb}) is schematically presented in Fig. \ref{fig:qp-res}. One can formally resum the terms in Eq. (\ref{eq:qp-mb}) and rewrite the generalized propagator as
\begin{eqnarray}
&&(x|\frac{i}{\slashed{p}}|y) \to (x|\frac{i}{\slashed{P} }|y)\,,
\label{eq:qp-mb-sh}
\end{eqnarray}
where $P_\mu \equiv p_\mu + g A_\mu$. Note that the generalized propagator contains a gauge covariant operator $P_\mu$. 

\begin{figure}[htb]
 \begin{center}
\includegraphics[width=0.8\textwidth]{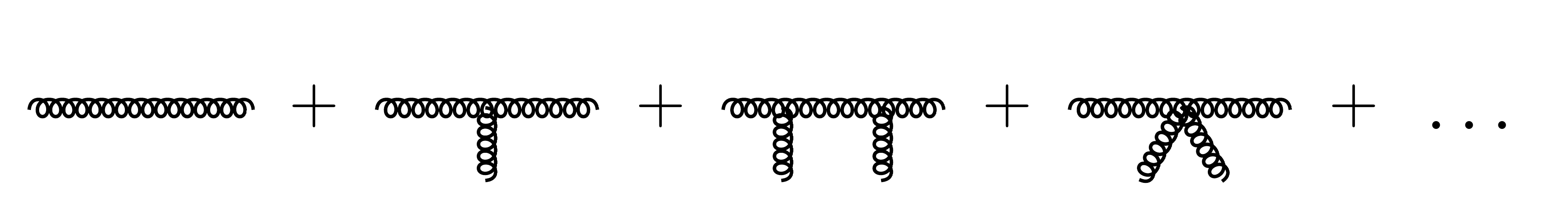}
 \end{center}
\caption{\label{fig:gp-res}Resumation of multiple insertions of the background gluon fields into the gluon propagator.}
 \end{figure}

 Analogous generalization of the gluon propagator is less straightforward. We construct such generalization starting with the bare gluon propagator in the axial gauge by adding multiple background gluon insertions via the standard three- and four- gluon vertexes. Explicitly such construction looks like
 \begin{eqnarray}
&&(x|\frac{-id_{\mu\nu}(p)\delta^{ab}}{p^2} |y) \to (x|\frac{-id_{\mu\nu}(p)\delta^{ab}}{p^2} |y)
\nonumber\\
&&+ (- i g) (x| \frac{-id_{\mu\rho}(p)}{p^2} \Big[ g^{\rho\sigma} \{ p_\alpha,  A^{\alpha} \} + 2 i (\partial^\rho A^\sigma - \partial^\sigma A^\rho) - p^\rho A^\sigma - A^\rho p^\sigma \Big] \frac{-id_{\sigma\nu}(p)}{p^2} |y)^{ab} + \dots\,,
\label{eq:gp-mb}
\end{eqnarray}
where an expression in squared brackets is the three-gluon vertex and ellipsis stands for the higher order terms in the coupling constant. The sum of terms in the right hand side of this equation is presented in Fig. \ref{fig:gp-res}. Here we assume that the gauge of the quantum gluons is fixed with an arbitrary vector $e_\mu$.

Now, similar to the case of the quark propagator, we want to resum terms in the r.h.s. of Eq. (\ref{eq:gp-mb}) and represent the result of this resumation in terms of gauge covariant operators. Though it is not obvious, we find that the following formula offers the necessary solution
\begin{eqnarray}
&&(x|\frac{-id_{\mu\nu}(p)\delta^{ab}}{p^2} |y) \to -i(x|\Big(\delta^\xi_\mu - P_\mu \frac{e^\xi}{P\cdot e}\Big) \Big\{\frac{1}{\square^{\xi\eta}} - \frac{1}{\square^{\xi\sigma}}g\mathcal{O}^{\sigma\rho}\frac{1}{\square^{\rho\eta}} + \dots \Big\}
 \Big(\delta^\eta_\nu - \frac{e^\eta}{P\cdot e} P_\nu\Big)|y)^{ab}\,
 \label{eq:gp-mb-sh}
\end{eqnarray}
where $\square^{\mu\nu} \equiv P^2 g^{\mu\nu} + 2igF^{\mu\nu}$,
\begin{eqnarray}
&&\mathcal{O}^{\sigma\rho} \equiv D_\lambda F^{\lambda\sigma} \frac{e^\rho}{P\cdot e} +  \frac{e^\sigma}{P\cdot e}D_\lambda F^{\lambda\rho} - \frac{e^\sigma}{P\cdot e} P_\beta D_\alpha F^{\alpha\beta} \frac{e^\rho}{P\cdot e}\,,
\label{eq:op-ins}
\end{eqnarray}
and ellipsis stands for terms with a growing number of insertions of the $\mathcal{O}^{\sigma\rho}$ operator. Originally this form of the gluon propagator in the axial gauge was derived in Ref. \cite{Balitsky:1995ub}.

The easiest way to see that Eq. (\ref{eq:gp-mb-sh}) is equivalent to the form of the propagator (\ref{eq:gp-mb}) is to expand both forms in powers of the coupling constant and compare the corresponding terms of the expansion. However, the form of Eq. (\ref{eq:gp-mb-sh}) is advantageous since it contains only gauge covariant operators, and is analogous to Eq. (\ref{eq:qp-mb-sh}). As we will see shortly, this allows straightforward generalization of our previous analysis of the Feynman diagrams in the dilute limit presented in the previous section to the case of multiple insertions of the background fields.

Using Eqs. (\ref{eq:qp-mb-sh}) and (\ref{eq:gp-mb-sh}) we can generalize diagrams in Fig. \ref{fig:Fdiag} to the case of multiple gluon interactions with the background field. For example, the generalization of the diagram in Fig. \ref{fig:Fdiag}a, presented in Fig. \ref{fig:dense-cancellation-internal}, has a formal expression, c.f. Eq. (\ref{eq:f1ainit}),
\begin{eqnarray}
&&\mathcal{U}^{[\gamma^+]}(z^-_1, z^-_2, b_\perp)\Big|^{axial}_{Fig.~\ref{fig:dense-cancellation-internal}}
= -g^2 \int^{z^-_1}_{\infty}dz^- \int d^4y \bar{q}(z^-_1, b_\perp) t^a \gamma^+ (z^-_2, - b_\perp|\frac{i}{\slashed{P}+i\epsilon}|y) \gamma^\nu t^b q(y) n^\mu 
\nonumber\\
&&\times (z^-, b_\perp|(-i)\Big(\delta^\xi_\mu - P_\mu \frac{e^\xi}{P\cdot e}\Big) \Big\{\frac{1}{\square^{\xi\eta}} - \frac{1}{\square^{\xi\sigma}}\mathcal{O}^{\sigma\rho}\frac{1}{\square^{\rho\eta}} + \dots \Big\}
 \Big(\delta^\eta_\nu - \frac{e^\eta}{P\cdot e} P_\nu\Big)|y)^{ab}\,.
\end{eqnarray}
Similar expressions can be easily written for all other diagrams in Fig. \ref{fig:Fdiag}. 

To demonstrate the gauge invariance of the sum of all diagrams we aim to show that the sum does not depend on the gauge fixing vector $e_\mu$. From Eq. (\ref{eq:gp-mb-sh}) we see that the $e_\mu$ dependence in the gluon propagator appears in the boundary factors like
\begin{eqnarray}
&&\delta^\xi_\mu - P_\mu \frac{e^\xi}{P\cdot e}\,,
\label{eq:boundary}
\end{eqnarray}
and the operator insertions $\mathcal{O}^{\sigma\rho}$. The cancellation of the $e_\mu$ dependence in these two structures should be considered independently. 

Let us start with the boundary factors. Similarly to the dilute limit, we expect that the crucial step in proving the cancellation is the vertex identity. So how does this procedure change if we consider a Feynman diagram with a gluon propagator attached to a quark or a Wilson line, when the ``quantum" partons interact multiple times with the background gluon field?

\begin{figure}[htb]
 \begin{center}
\includegraphics[width=0.7\textwidth]{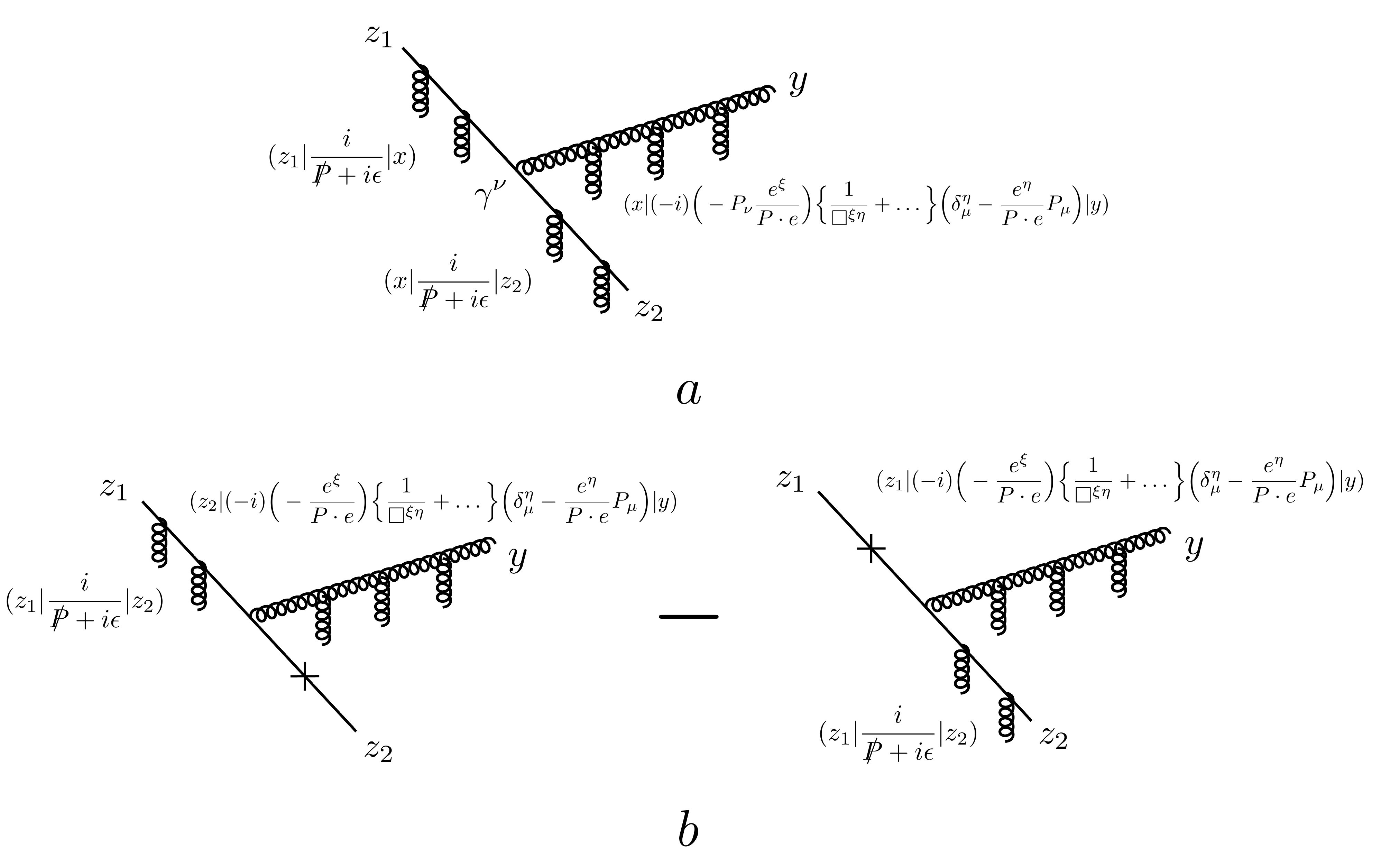}
 \end{center}
\caption{\label{fig:ge-gl-inq}a) Insertion of the gluon propagator into a quark line Eq. (\ref{eq:ql-bg-ins}) and b) the same propagator after the integration by parts procedure Eq. (\ref{eq:ql-bg-ins-byp}).}
 \end{figure}

 To answer this question, similarly to our discussion in Sec. \ref{sec:mapping}, let us consider the generalization of the vertex identity for the $e_\mu$ dependent terms when the gluon propagator is inserted either into a quark line or a Wilson gauge link.
 
Starting with the quark line, we assume that the interaction between the quark and gluon takes place at point $x$. For the $e_\mu$ dependent term at this position we write, see Fig. \ref{fig:ge-gl-inq}a:
\begin{eqnarray}
&&ig\int d^4x (z_1|\frac{i}{\slashed{P}+i\epsilon}|x)\gamma^\nu t^a(x|\frac{i}{\slashed{P}+i\epsilon}|z_2) (x|(-i)\Big( - P_\nu \frac{e^\xi}{P\cdot e}\Big) \Big\{\frac{1}{\square^{\xi\eta}} - \frac{1}{\square^{\xi\sigma}}\mathcal{O}^{\sigma\rho}\frac{1}{\square^{\rho\eta}} + \dots \Big\}
 \Big(\delta^\eta_\mu - \frac{e^\eta}{P\cdot e} P_\mu\Big) |y)^{ab}\,,
\label{eq:ql-bg-ins}
 \nonumber\\
\end{eqnarray}
c.f. Eq. (\ref{eq:pninscoord}). We see that the Lorentz index of the covariant momentum $P_\nu$ is contracted with the index of the $\gamma$-matrix between two quark propagators in the gluon background field. Integrating by parts with respect to this momenta we rewrite Eq. (\ref{eq:ql-bg-ins}) as, c.f. Eq. (\ref{eq:pninscoord-aft}),
\begin{eqnarray}
&&-g t^a(z_1|\frac{i}{\slashed{P}+i\epsilon}|z_2) (z_1|(-i)\Big( -  \frac{e^\xi}{P\cdot e}\Big) \Big\{\frac{1}{\square^{\xi\eta}} - \frac{1}{\square^{\xi\sigma}}\mathcal{O}^{\sigma\rho}\frac{1}{\square^{\rho\eta}} + \dots \Big\}
 \Big(\delta^\eta_\mu - \frac{e^\eta}{P\cdot e} P_\mu\Big) |y)^{ab}
 \nonumber\\
&&+ g  (z_1|\frac{i}{\slashed{P}+i\epsilon}|z_2) t^a (z_2|(-i)\Big( -  \frac{e^\xi}{P\cdot e}\Big) \Big\{\frac{1}{\square^{\xi\eta}} - \frac{1}{\square^{\xi\sigma}}\mathcal{O}^{\sigma\rho}\frac{1}{\square^{\rho\eta}} + \dots \Big\}
 \Big(\delta^\eta_\mu - \frac{e^\eta}{P\cdot e} P_\mu\Big) |y)^{ab} \,.
\label{eq:ql-bg-ins-byp}
\end{eqnarray}
We find, similarly to the dilute case, that the vertex identity leads to cancellation of a quark propagator adjacent to the point of quark-gluon interaction, see Fig. \ref{fig:ge-gl-inq}b.

\begin{figure}[htb]
 \begin{center}
\includegraphics[width=0.7\textwidth]{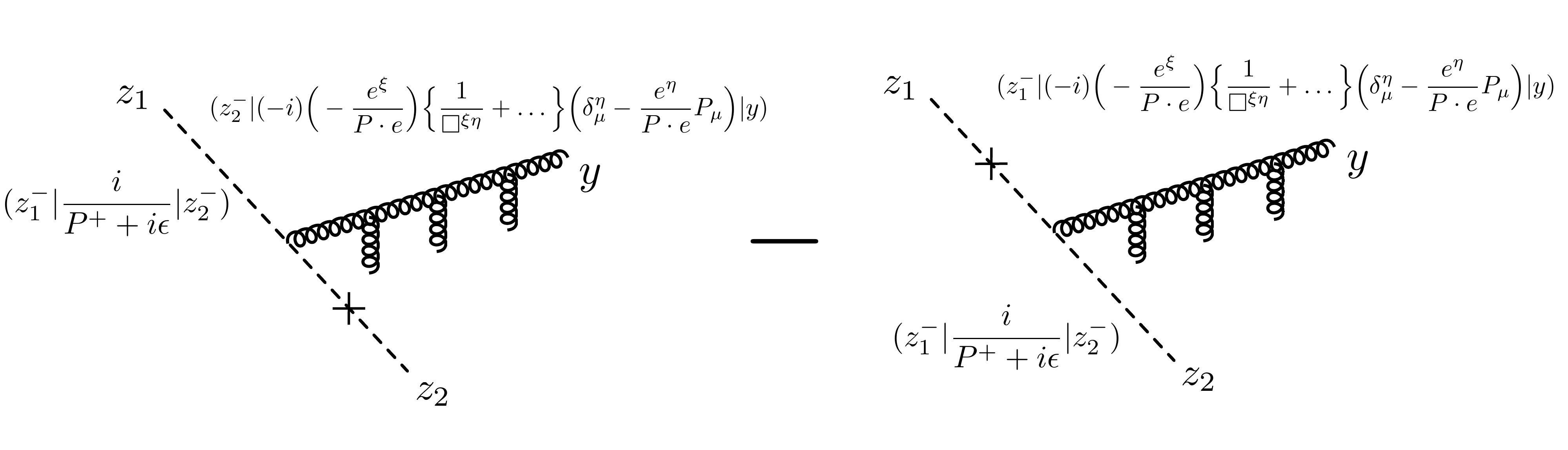}
 \end{center}
\caption{\label{fig:ge-gl-inwl}Insertion of the gluon propagator into a Wilson line after applying the vertex identity, see Eq. (\ref{eq:dense-int-wl-res}).}
 \end{figure}

In full analogy with the dilute case, a similar cancellation takes place when the gluon propagator is inserted into a Wilson line. Starting with a formal expression for the $e_\mu$ dependent term of the gluon propagator in this scenario we have, c.f. Eq. (\ref{eq:dil-int-wl-init}),
\begin{eqnarray}
&&ig \int dx^- (z^-_1|\frac{i}{P^+ + i\epsilon}|x^-) n^\nu t^a (x^-|\frac{i}{P^+ + i\epsilon}|z^-_2) 
\nonumber\\
&&\times (x^-, x_\perp| (-i)\Big( - P_\nu \frac{e^\xi}{P\cdot e}\Big) \Big\{\frac{1}{\square^{\xi\eta}} - \frac{1}{\square^{\xi\sigma}}\mathcal{O}^{\sigma\rho}\frac{1}{\square^{\rho\eta}} + \dots \Big\}
 \Big(\delta^\eta_\mu - \frac{e^\eta}{P\cdot e} P_\mu\Big) |y)^{ab}\,.
\label{eq:dense-int-wl-init}
\end{eqnarray}
Diagrammatic representation of this insertion can be found in Fig. \ref{fig:ge-gl-inwl}.

 Integrating by parts with respect to the covariant momentum $P_\nu$, we can rewrite Eq. (\ref{eq:dense-int-wl-init}) as
\begin{eqnarray}
&&-g t^a (z^-_1|\frac{i}{P^+ + i\epsilon}|z^-_2) (z^-_1, x_\perp| (-i)\Big( -  \frac{e^\xi}{P\cdot e}\Big) \Big\{\frac{1}{\square^{\xi\eta}} - \frac{1}{\square^{\xi\sigma}}\mathcal{O}^{\sigma\rho}\frac{1}{\square^{\rho\eta}} + \dots \Big\}
 \Big(\delta^\eta_\mu - \frac{e^\eta}{P\cdot e} P_\mu\Big) |y)^{ab}
 \nonumber\\
 &&+ g (z^-_1|\frac{i}{P^+ + i\epsilon}|z^-_2) t^a (z^-_2, x_\perp| (-i)\Big( -  \frac{e^\xi}{P\cdot e}\Big) \Big\{\frac{1}{\square^{\xi\eta}} - \frac{1}{\square^{\xi\sigma}}\mathcal{O}^{\sigma\rho}\frac{1}{\square^{\rho\eta}} + \dots \Big\}
 \Big(\delta^\eta_\mu - \frac{e^\eta}{P\cdot e} P_\mu\Big) |y)^{ab}\,,
\label{eq:dense-int-wl-res}
\end{eqnarray}
 with a similar cancellation of the Wilson line propagators, c.f. Eq. (\ref{eq:dil-int-wl-res}) in the dilute approximation.

 \begin{figure}[htb]
 \begin{center}
\includegraphics[width=0.7\textwidth]{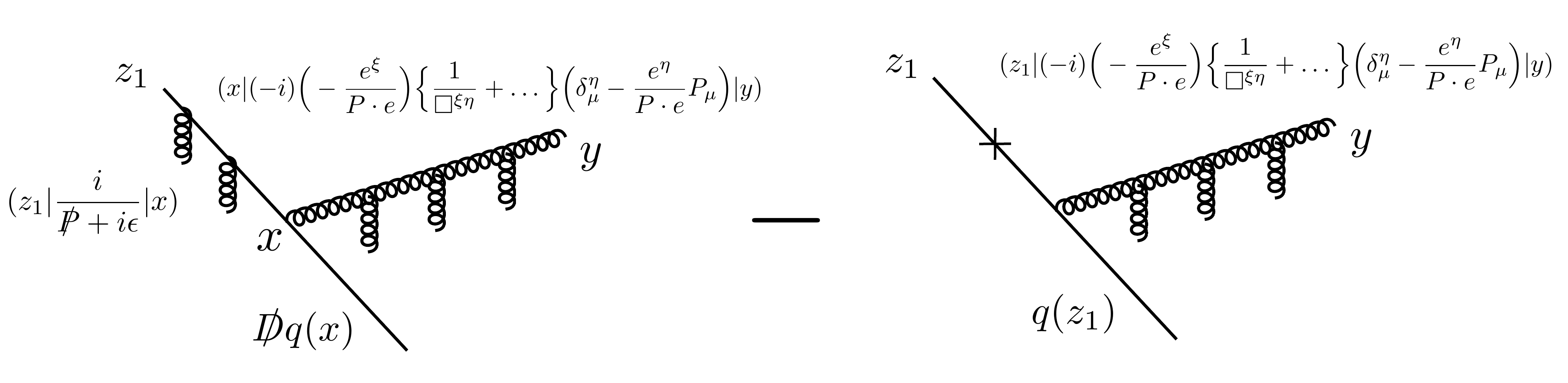}
 \end{center}
\caption{\label{fig:gebq-gl-inwl}The gluon propagator attached to a background quark field after applying the vertex identity, see Eq. (\ref{eq:dense-int-qbg-res}).}
 \end{figure}

Finally, we need to consider a situation when the gluon propagator is attached to a background quark field, see Fig. \ref{fig:gebq-gl-inwl}. In this case we write, c.f. Eq. (\ref{eq:dil-int-qbg-init}),
\begin{eqnarray}
&&ig\int d^4x (z_1|\frac{i}{\slashed{P}+i\epsilon}|x)\gamma^\nu t^a q(x) (x| (-i)\Big( - P_\nu \frac{e^\xi}{P\cdot e}\Big) \Big\{\frac{1}{\square^{\xi\eta}} - \frac{1}{\square^{\xi\sigma}}\mathcal{O}^{\sigma\rho}\frac{1}{\square^{\rho\eta}} + \dots \Big\}
 \Big(\delta^\eta_\mu - \frac{e^\eta}{P\cdot e} P_\mu\Big) |y)^{ab}\,.
\label{eq:dense-int-qbg-init}
\end{eqnarray}

Integrating by parts with respect to $P_\nu$, we get
\begin{eqnarray}
&&-g t^a q(z_1) (z_1| (-i)\Big( - \frac{e^\xi}{P\cdot e}\Big) \Big\{\frac{1}{\square^{\xi\eta}} - \frac{1}{\square^{\xi\sigma}}\mathcal{O}^{\sigma\rho}\frac{1}{\square^{\rho\eta}} + \dots \Big\}
 \Big(\delta^\eta_\mu - \frac{e^\eta}{P\cdot e} P_\mu\Big) |y)^{ab}
 \nonumber\\
 &&+ g\int d^4x (z_1|\frac{i}{\slashed{P}+i\epsilon}|x) t^a \gamma^\nu D_\nu q(x) (x| (-i)\Big( - \frac{e^\xi}{P\cdot e}\Big) \Big\{\frac{1}{\square^{\xi\eta}} - \frac{1}{\square^{\xi\sigma}}\mathcal{O}^{\sigma\rho}\frac{1}{\square^{\rho\eta}} + \dots \Big\}
 \Big(\delta^\eta_\mu - \frac{e^\eta}{P\cdot e} P_\mu\Big) |y)^{ab}\,,
\label{eq:dense-int-qbg-res}
\end{eqnarray}
c.f. Eq. (\ref{eq:dil-int-qbg-res}). We see that, similarly to the dilute case, the vertex identity applied to the $e_\mu$ dependent term of the gluon propagator attached to the background quark field leads to either cancellation of the adjacent quark propagator or appearance of a term with a covariant derivative $D_\mu$ acting on the background quark field $q$, see Fig. \ref{fig:gebq-gl-inwl}.

As a result, we conclude that the vertex identity, being applied to the $e_\mu$ dependent terms of the boundary factors of the gluon propagator (\ref{eq:gp-mb-sh}), has exactly the same structure as in the dilute regime with cancellation of the internal quark and Wilson line propagators, as well as appearance of terms proportional to derivatives of the background quark fields $\slashed{D} q$.

\begin{figure}[htb]
 \begin{center}
\includegraphics[width=0.7\textwidth]{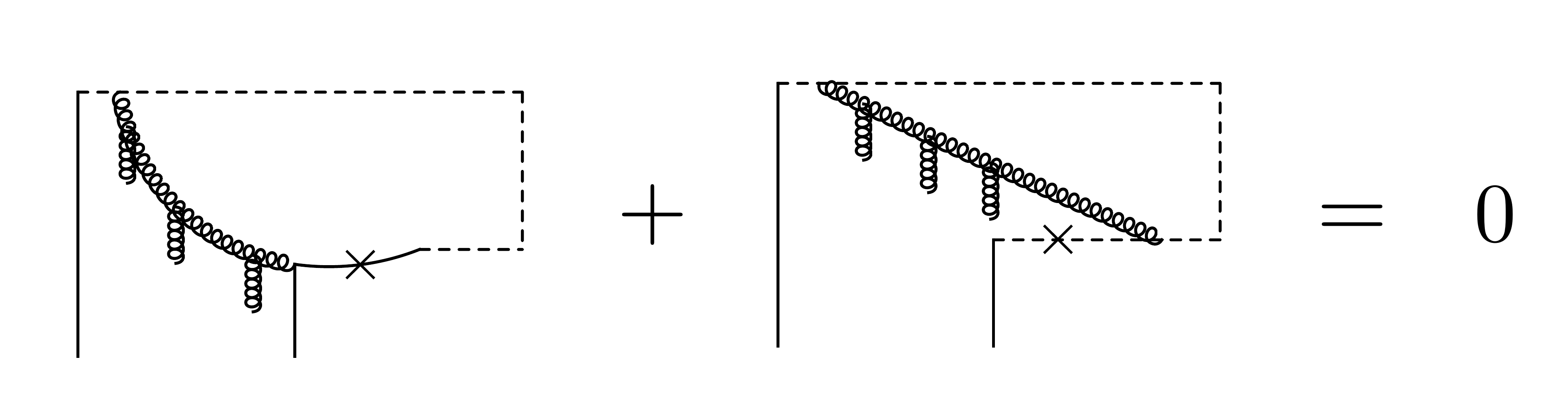}
 \end{center}
\caption{\label{fig:dense-cancellation-internal2}Cancellation of terms after applying the vertex identity with an infinite number of background gluon insertions.}
 \end{figure}

It's easy to see now, following the analysis of Sec. \ref{sec:mapping}, that the dependence on the $e_\mu$ vector in the boundary terms of the gluon propagator cancels in the sum of all Feynman diagrams with all possible insertions of the gluon propagator. This happens due to cancellation of terms with removed (after applying the vertex identity) internal propagators originating in different diagrams, see for example Fig. \ref{fig:dense-cancellation-internal2}.

 However, to have the cancellation of $e_\mu$ dependence, in full analogy to the dilute case, we have to take into account contribution of the transverse gauge link at the spatial infinity and require that the background quark fields satisfy the QCD equations of motion:
\begin{eqnarray}
&&i\gamma^\mu D_\mu q = 0;\ \ \ -i \bar{q} \overleftarrow{D}_\mu\gamma^\mu = 0\,.
\label{eq:emo-full}
\end{eqnarray}
In the dilute approximation with no background gluons these equations obviously reduce to the EoM for free quarks (\ref{eq:emo}).

 Now, after we found that the $e_\mu$ dependence in the boundary factors (\ref{eq:boundary}) of the gluon propagator (\ref{eq:gp-mb-sh}) does not survive in the sum of all Feynman diagrams, let's turn our attention to the $e_\mu$ dependence in the operator insertion $\mathcal{O}^{\sigma\rho}$, see Eq. (\ref{eq:op-ins}).

 To understand cancellation of the $e_\mu$ dependence coming from this operator insertion, we need to extend our previous analysis and take into account multiple insertions of the background quark fields into our diagrams. This generalization can be viewed as insertion of multiple quark ``staples" into ``quantum" gluon lines, see Fig. \ref{fig:f1-dense}.

\begin{figure}[htb]
 \begin{center}
\includegraphics[width=0.5\textwidth]{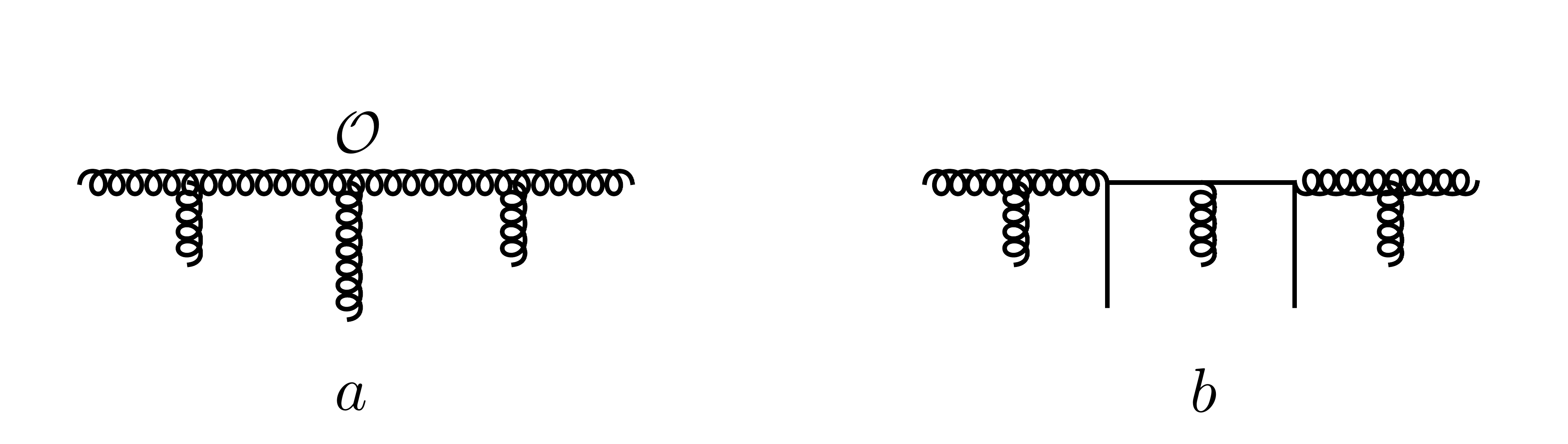}
 \end{center}
\caption{\label{fig:dense-O-and-qst}To see the cancellation of the $e_\mu$ dependence in the gluon propagator (\ref{eq:gp-mb-sh}), we need to combine each operator insertion $\mathcal{O}$ (a) with an insertion of the quark ``staple" (b).}
 \end{figure}

To see the cancellation of the $e_\mu$ dependence in $\mathcal{O}^{\sigma\rho}$ we need to combine each of this operator insertions with an insertion of the quark ``staple", see Fig. \ref{fig:dense-O-and-qst}.

Indeed, each quark insertion of the quark ``staple" by itself depends on the $e_\mu$ vector through boundary factors (\ref{eq:boundary}) of the adjacent gluon propagators, so it has to be combined with the $e_\mu$ dependence of the $\mathcal{O}$ insertion operator, see Eq. (\ref{eq:op-ins}).

\begin{figure}[htb]
 \begin{center}
\includegraphics[width=0.5\textwidth]{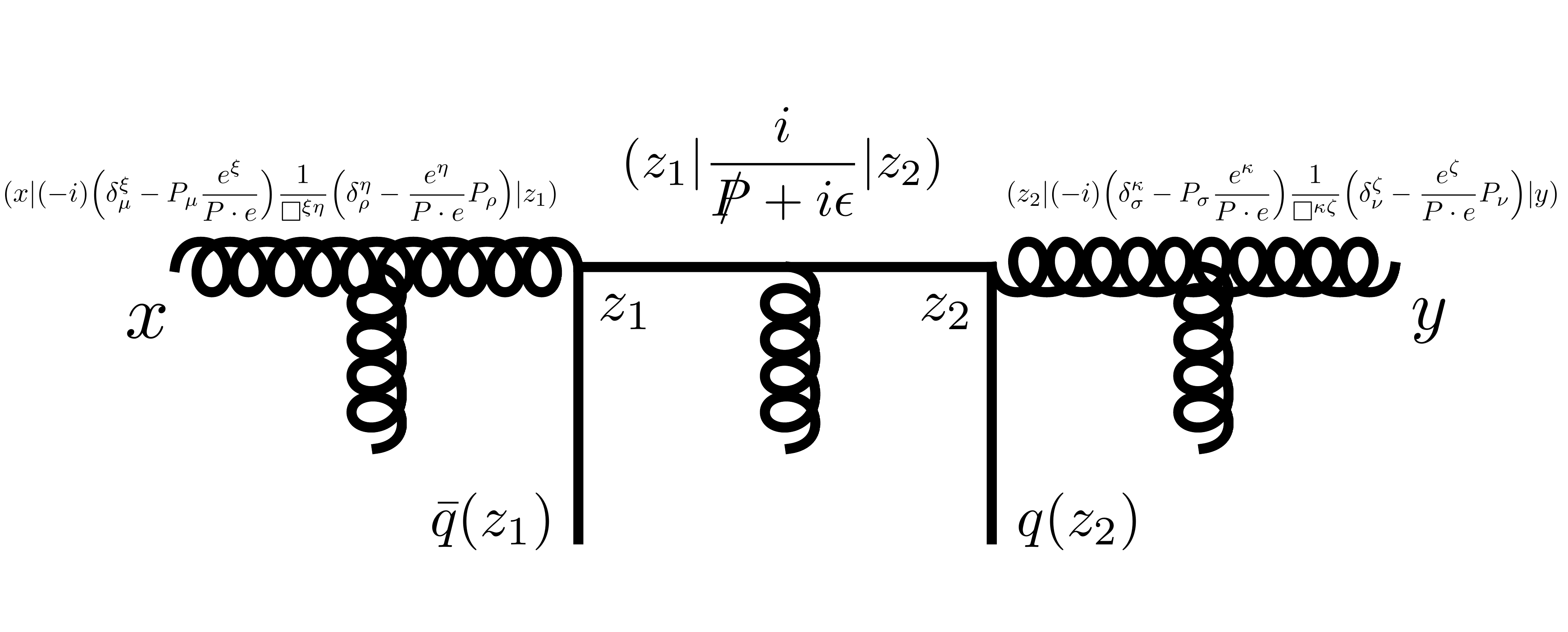}
 \end{center}
\caption{\label{fig:dense-staple-ins}Insertion of the quark ``staple" into a gluon propagator, see the first terms in Eq. (\ref{eq:staple-init}).}
 \end{figure}

For insertion of the quark ``staple" into a gluon propagator, see Fig. \ref{fig:dense-staple-ins}, we write
\begin{eqnarray}
&&-g^2\int d^4 z_1 \int d^4z_2 (x| (-i)\Big(\delta^\xi_\mu - P_\mu \frac{e^\xi}{P\cdot e}\Big) \frac{1}{\square^{\xi\eta}}
 \Big(\delta^\eta_\rho - \frac{e^\eta}{P\cdot e} P_\rho\Big) |z_1)^{ae} 
 \nonumber\\
 &&\times  \Big[\bar{q}(z_1)\gamma^\rho t^e(z_1|\frac{i}{\slashed{P} + i\epsilon}|z_2) t^d\gamma^\sigma q(z_2) +  \bar{q}(z_2)\gamma^\sigma t^d(z_2|\frac{i}{\slashed{P} + i\epsilon}|z_1) t^e \gamma^\rho q(z_1)\Big]
 \nonumber\\
 &&\times (z_2|(-i)\Big(\delta^\kappa_\sigma - P_\sigma \frac{e^\kappa}{P\cdot e}\Big) \frac{1}{\square^{\kappa\zeta}}
 \Big(\delta^\zeta_\nu - \frac{e^\zeta}{P\cdot e} P_\nu\Big)|y)^{db}\,.
 \label{eq:staple-init}
\end{eqnarray}
Note that since we aim to combine this contribution with an $\mathcal{O}$ insertion into the gluon propagator, the gluon propagators around the ``quark" staple are simply $1/\square^{\xi\eta}$ and do not themselves contain any $\mathcal{O}$ insertions. If we add extra $\mathcal{O}$ insertions to the gluon lines, we would need to combine them with extra insertions of the quark ``staples". In other words, one $\mathcal{O}$ insertion into the gluon propagator should be combined with one insertion of the ``quark" staple, which we consider in Eq. (\ref{eq:staple-init}). 

Integrating by parts with respect to $P_\rho$ and $P_\sigma$ we get
\begin{eqnarray}
&&-g^2\int d^4 z_1 \int d^4z_2 \Big\{ (x| \Big(\delta^\xi_\mu - P_\mu \frac{e^\xi}{P\cdot e}\Big) \frac{-i}{\square^{\xi\rho}}|z_1)^{ae} 
 \Big[\bar{q}(z_1)\gamma^\rho t^e (z_1|\frac{i}{\slashed{P} + i\epsilon}|z_2) t^d \gamma^\sigma q(z_2) 
 \nonumber\\
 &&+  \bar{q}(z_2)\gamma^\sigma t^d (z_2|\frac{i}{\slashed{P} + i\epsilon}|z_1) t^e\gamma^\rho q(z_1)\Big] (z_2| \frac{-i}{\square^{\sigma\zeta}}
 \Big(\delta^\zeta_\nu - \frac{e^\zeta}{P\cdot e} P_\nu\Big)|y)^{db}
 \nonumber\\
 &&+ (x| \Big(\delta^\xi_\mu - P_\mu \frac{e^\xi}{P\cdot e}\Big) \frac{-i}{\square^{\xi\rho}}|z_1)^{ae} 
 \Big[ i \bar{q}(z_1)\gamma^\rho [t^e, t^d] q(z_1) \delta^4(z_1 - z_2)
 - \bar{q}(z_1)\gamma^\rho t^e  (z_1|\frac{i}{\slashed{P} + i\epsilon}|z_2) t^d i\slashed{D} q(z_2)
 \nonumber\\
 &&- i \bar{q}(z_2) \overleftarrow{\slashed{D}} t^d (z_2|\frac{i}{\slashed{P} + i\epsilon}|z_1) t^e \gamma^\rho q(z_1)  \Big] (z_2|  \Big( -  \frac{e^\kappa}{P\cdot e}\Big) \frac{-i}{\square^{\kappa\zeta}}
 \Big(\delta^\zeta_\nu - \frac{e^\zeta}{P\cdot e} P_\nu\Big)|y)^{db}
 \nonumber\\
 &&+ (x| \Big(\delta^\xi_\mu - P_\mu \frac{e^\xi}{P\cdot e}\Big) \frac{-i}{\square^{\xi\eta}}
 \Big( - \frac{e^\eta}{P\cdot e} \Big) |z_1)^{ae} 
   \Big[ i \bar{q}(z_1)\overleftarrow{\slashed{D}} t^e (z_1|\frac{i}{\slashed{P} + i\epsilon}|z_2) t^d \gamma^\sigma q(z_2) 
  + i \bar{q}(z_1) [t^e, t^d] \gamma^\sigma q(z_1) \delta^4(z_1-z_2)
  \nonumber\\
  &&+  \bar{q}(z_2)\gamma^\sigma t^d  (z_2|\frac{i}{\slashed{P} + i\epsilon}|z_1) t^e i \slashed{D} q(z_1)\Big] (z_2| \frac{-i}{\square^{\sigma\zeta}}
 \Big(\delta^\zeta_\nu - \frac{e^\zeta}{P\cdot e} P_\nu\Big)|y)^{db}
 \nonumber\\
 &&+ (x| \Big(\delta^\xi_\mu - P_\mu \frac{e^\xi}{P\cdot e}\Big) \frac{-i}{\square^{\xi\eta}}
 \Big( - \frac{e^\eta}{P\cdot e} P_\rho\Big) |z_1)^{ae} 
  \Big[ i \bar{q}(z_1)\gamma^\rho [t^e, t^d] q(z_1) \delta^4(z_1-z_2) - \bar{q}(z_1)\gamma^\rho t^e  (z_1|\frac{i}{\slashed{P} + i\epsilon}|z_2) t^d i \slashed{D} q(z_2) 
  \nonumber\\
  &&-  i \bar{q}(z_2)\overleftarrow{\slashed{D}} t^d (z_2|\frac{i}{\slashed{P} + i\epsilon}|z_1) t^e \gamma^\rho q(z_1) \Big]
 (z_2| \Big( -  \frac{e^\kappa}{P\cdot e}\Big) \frac{-i}{\square^{\kappa\zeta}}
 \Big( \delta^\zeta_\nu - \frac{e^\zeta}{P\cdot e} P_\nu\Big)|y)^{db} \Big\}\,.
 \label{eq:staple-ibp}
\end{eqnarray}
Despite the length, this equation is quite transparent. The first two lines contain terms with no $e_\mu$ dependence in the boundary factors (\ref{eq:boundary}) of the gluon propagators attached to the ``staple". These terms are of no interest to us.

The remaining $e_\mu$ dependence in the boundary factors at points $x$ and $y$ should be treated independently. If these points are attached to the quark lines or gauge links of the initial quark TMD operator,\footnote{Note that our analysis is general and can be applied to any QCD operator.} to eliminate the $e_\mu$ dependence one should repeat the analysis in the beginning of this section. If the gluon line is attached to a quark ``staple", then the procedure that we are developing right now should be applied.
\begin{figure}[htb]
 \begin{center}
\includegraphics[width=0.9\textwidth]{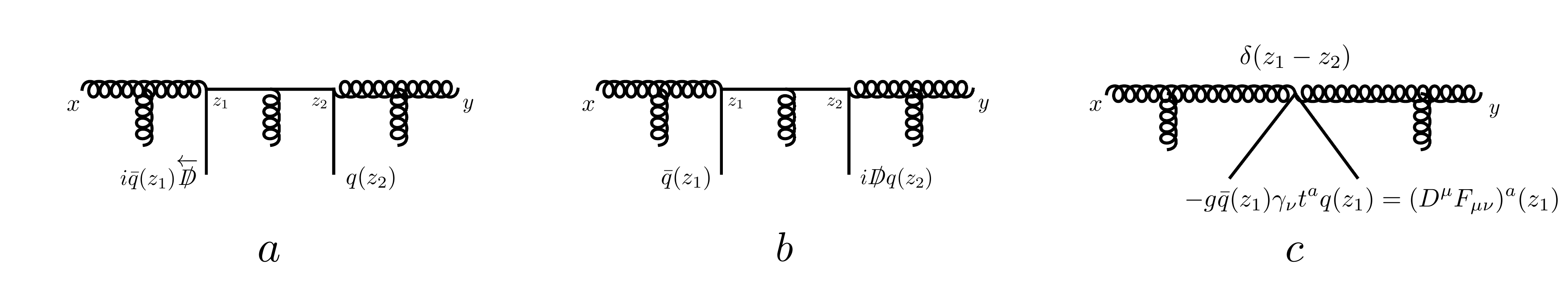}
 \end{center}
\caption{\label{fig:dense-staple-EoMterms}Schematic representation of terms in Eq. (\ref{eq:staple-ibp}) obtained after applying the vertex identity to the quark staple in Fig. \ref{fig:dense-staple-ins}, see Eq. (\ref{eq:staple-init}). Terms ``a" and ``b" can be removed by applying EoM (\ref{eq:emo-full}). Terms ``c" can be related to the background gluon contribution by applying EoM (\ref{eq:emo-gluon}). These terms are canceled by gluonic contributions in Fig. \ref{fig:dense-O-and-qst}a.}
 \end{figure}

The $e_\mu$ dependence associated with the quark ``staple" insertion into a gluon propagator is in the remaining terms of Eq. (\ref{eq:staple-ibp}). The structure of these terms is transparent and similar to what we have seen before. There are $\delta$-functions which manifest cancellation of the internal quark propagator of the ``staple" and terms with covariant derivatives of the background quark fields. {These terms are schematically presented in Fig. \ref{fig:dense-staple-EoMterms}. }

Applying EoM (\ref{eq:emo-full}) we rewrite Eq. (\ref{eq:staple-ibp}), without the first term, as
\begin{eqnarray}
 && \label{eq:staple-ibp-emo}-ig^2\int d^4 z (x| \Big(\delta^\xi_\mu - P_\mu \frac{e^\xi}{P\cdot e}\Big) \frac{-i}{\square^{\xi\rho}}|z)^{ae} \Big[
if^{eds}  \bar{q}(z)\gamma^\rho t^s q(z) \Big] (z|  \Big( -  \frac{e^\kappa}{P\cdot e}\Big) \frac{-i}{\square^{\kappa\zeta}}
 \Big(\delta^\zeta_\nu - \frac{e^\zeta}{P\cdot e} P_\nu\Big)|y)^{db}
\\
 &&-ig^2\int d^4 z (x| \Big(\delta^\xi_\mu - P_\mu \frac{e^\xi}{P\cdot e}\Big) \frac{-i}{\square^{\xi\eta}}
 \Big( - \frac{e^\eta}{P\cdot e} \Big) |z)^{ae} \Big[ if^{eds} \bar{q}(z) t^s \gamma^\sigma q(z) \Big] (z| \frac{-i}{\square^{\sigma\zeta}}
 \Big(\delta^\zeta_\nu - \frac{e^\zeta}{P\cdot e} P_\nu\Big)|y)^{db}
 \nonumber\\
 &&-ig^2\int d^4 z (x| \Big(\delta^\xi_\mu - P_\mu \frac{e^\xi}{P\cdot e}\Big) \frac{-i}{\square^{\xi\eta}}
 \Big( - \frac{e^\eta}{P\cdot e} P_\rho\Big) |z)^{ae} 
  \Big[ if^{eds} \bar{q}(z)\gamma^\rho t^s q(z) \Big] (z| \Big( -  \frac{e^\kappa}{P\cdot e}\Big) \frac{-i}{\square^{\kappa\zeta}}
 \Big( \delta^\zeta_\nu - \frac{e^\zeta}{P\cdot e} P_\nu\Big)|y)^{db}\,,
 \nonumber
\end{eqnarray}
which corresponds to the contribution in Fig. \ref{fig:dense-staple-EoMterms}c.

Now we arrive to a crucial step of our analysis: application of the EoM for the gluon fields. Indeed, since, as we discussed before, the quark fields have to satisfy (\ref{eq:emo-full}), we have to impose the QCD equation of motion for the background gluon fields as well:
\begin{eqnarray}
&&(D^\mu F_{\mu\nu})^a = - g\bar{q}\gamma_\nu t^a q\,.
\label{eq:emo-gluon}
\end{eqnarray}
In fact, as we will see now, this is required to eliminate the dependence on the gauge fixing vector $e_\mu$ and ensure the gauge invariance of our Feynman diagrams calculation.

With the help of Eq. (\ref{eq:emo-gluon}) we rewrite the $e_\mu$ dependent terms in the gluon propagator with a single insertion of the quark ``staple" (\ref{eq:staple-ibp-emo}) as
\begin{eqnarray}
 &&- i g (x|  \Big(\delta^\xi_\mu - P_\mu \frac{e^\xi}{P\cdot e}\Big) \frac{1}{\square^{\xi\sigma}} \Big[  
 D_\lambda F^{\lambda\sigma} \frac{e^\rho}{P\cdot e} + \frac{e^\sigma}{P\cdot e} D_\lambda F^{\lambda\rho} - \frac{e^\sigma}{P\cdot e} P_\beta 
  D_\alpha F^{\alpha\beta} \frac{e^\rho}{P\cdot e}  \Big] \frac{1}{\square^{\rho\eta}} \Big(\delta^\eta_\nu - \frac{e^\eta}{P\cdot e} P_\nu\Big) |y)^{ab}\,.
 \label{eq:staple-ibp-emo-both}
\end{eqnarray}
Comparing this equation with Eq. (\ref{eq:op-ins}) we find that after applying the EoM for the background fields (\ref{eq:emo-full}) and (\ref{eq:emo-gluon}), the $e_\mu$ dependence generated by a quark ``staple" insertion into the gluon propagator reduces to
\begin{eqnarray}
 &&- i g(x|  \Big(\delta^\xi_\mu - P_\mu \frac{e^\xi}{P\cdot e}\Big) \frac{1}{\square^{\xi\sigma}} \mathcal{O}^{\sigma\rho} \frac{1}{\square^{\rho\eta}} \Big(\delta^\eta_\nu - \frac{e^\eta}{P\cdot e} P_\nu\Big) |y)^{ab}\,.
 \label{eq:staple-ibp-emo-both-fin}
\end{eqnarray}

Now, as we promised, we need to combine this contribution with a single $\mathcal{O}$ insertion into the gluon propagator (\ref{eq:gp-mb-sh}):
\begin{eqnarray}
&&+ig(x|\Big(\delta^\xi_\mu - P_\mu \frac{e^\xi}{P\cdot e}\Big) \frac{1}{\square^{\xi\sigma}}\mathcal{O}^{\sigma\rho}\frac{1}{\square^{\rho\eta}} \Big(\delta^\eta_\nu - \frac{e^\eta}{P\cdot e} P_\nu\Big)|y)^{ab}\,.
\end{eqnarray}
We see that two contributions exactly cancel each other which completes our proof of the $e_\mu$ independence of the final result of calculation of the Feynman diagrams in presence of multiple interactions with the background fields.

We find that our general conclusions formulated in the dilute limit are completely valid in the dense regime. To ensure the gauge invariance of the calculation one has to include the contribution of the transverse gauge links at the spatial infinity and enforce the QCD EoM for the background fields.

Meanwhile, the analysis of this section offers some important results on its own. Firstly, we conclude that the gauge invariance of the final result for the sum of all Feynman diagrams in the QCD factorization approach involves an interplay between quark and gluon background-field configurations that is due to the EoM (\ref{eq:emo-full}) and (\ref{eq:emo-gluon}). In other words, there is no meaning in perturbative calculation of a particular operator contribution in isolation from all other possible background-field configurations, see e.g. Fig. \ref{fig:dense-O-and-qst}.

Indeed, one can always add a trivial contribution proportional to the QCD EoM that mix quark and gluon operators. This trivial contribution is going to simultaneously change perturbative contributions corresponding to different configurations of the background fields, i.e. operators. This means that each perturbative contribution to a particular operator is not unique. A change in this contribution, due to the equations of motion, is always compensated by a matching change in other perturbative contributions corresponding to other related background field operators. So only the full set of perturbative contributions of the relevant operators remains the same. {Mixing between different types of operators due to EoM has been extensively studied before, see e.g. Refs. \cite{Mulders:1995dh,Ball:2002ps,Braun:2009vc, Braun:2011dg,Balitsky:2017gis}.  }

\begin{figure}[htb]
 \begin{center}
\includegraphics[width=0.5\textwidth]{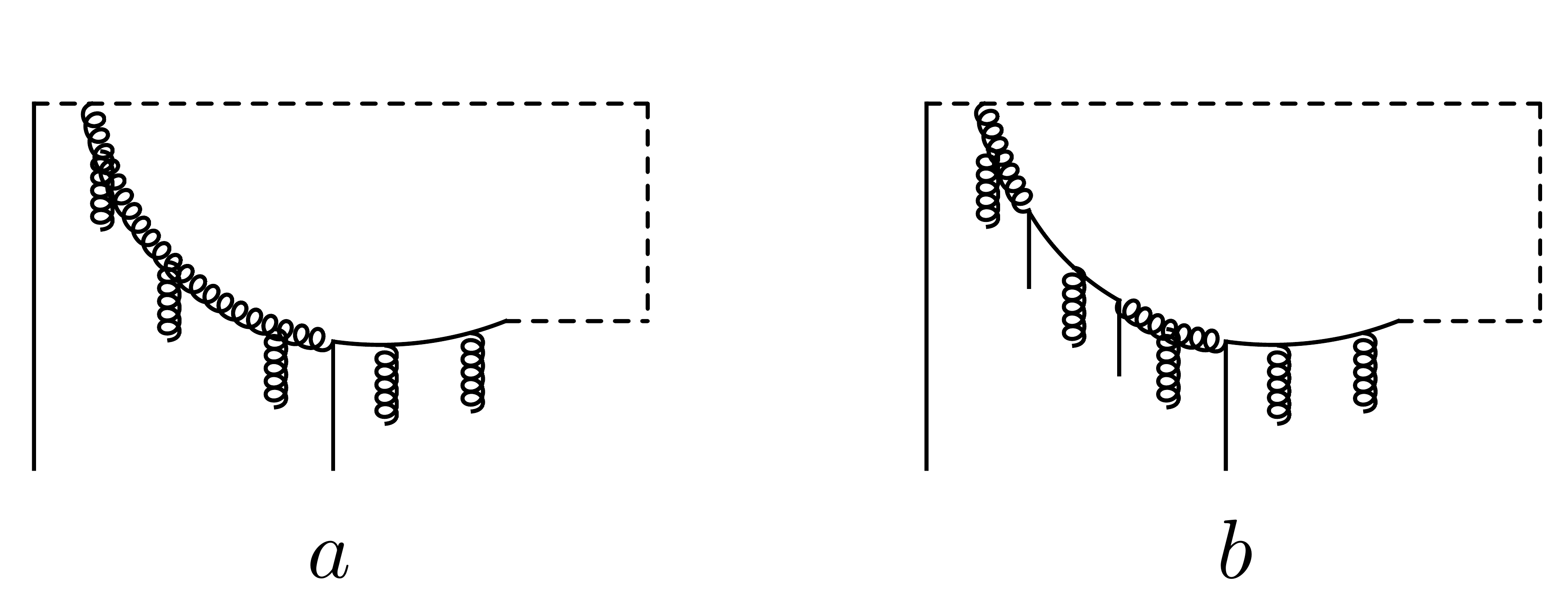}
 \end{center}
\caption{\label{fig:two-qg-channels}Interplay between two different background field configurations.}
 \end{figure}

An illustration of this point can be found in Fig. \ref{fig:two-qg-channels}. Two diagrams in this figure contribute to the perturbative calculation of the quark TMDPDF operator. At the same time, these two diagrams correspond to contribution of two different background field operators, i.e. note two quark fields in the first diagrams and four quark fields in the second one. However, as we showed in this section, two diagrams are related to each other. The contribution of each diagram is not unique due to the QCD EoM for the background fields. Indeed, one can always add a term proportional to the $D^\mu F_{\mu\nu}$ operator in the first diagram, see e.g. Eq. (\ref{eq:staple-ibp-emo-both}), if the second diagram receives a matching contribution proportional to $\bar{q}\gamma_\nu t^a q$, see e.g. Eq. (\ref{eq:staple-ibp-emo}). Though after adding such terms the corresponding perturbative result for each diagram is going to change, the total result for the sum of diagrams remains the same.

Secondly, one can note that since two results of calculation of the perturbative corrections to a gauge invariant operator obtained in two different gauges are the same up to terms proportional to the QCD equations of motion, one can effectively go from one choice of the gauge fixing vector $e_{1\mu}$ to another $e_{2\mu}$ by consistently adding terms proportional to the quark-gluon operators in the QCD EoM (\ref{eq:emo-full}) and (\ref{eq:emo-gluon}).

By doing so, one can in particular eliminate all terms proportional to the gauge fixing vector $e_\mu$. In this case the gluon propagator (\ref{eq:gp-mb-sh}) simplifies to 
\begin{eqnarray}
&& -i(x| \frac{1}{\square^{\mu\nu}} |y)^{ab}\,.
\label{eq:bFprop}
\end{eqnarray}
This form of the gluon propagator is nothing but the gluon propagator in the background-Feynman gauge that is defined by a gauge-fixing term $-\frac{1}{2}(\mathcal{D}^\mu A^a_\mu)^2$.

As a result, we see that in calculation of perturbative corrections to a gauge invariant operator in the background field, the sum of all Feynman diagrams calculated in the axial gauge with an arbitrary gauge-fixing vector $e_\mu$ coincides with the sum of the same diagrams calculated in the background-Feynman gauge if the background fields satisfy the QCD EoM (\ref{eq:emo-full}) and (\ref{eq:emo-gluon}).

Note that the background-Feynman gauge is different from the Feynman gauge which is defined by a gauge fixing term $-\frac{1}{2}(\partial^\mu A^a_\mu)^2$. Two gauges coincide only when the background gluon field is trivial. Indeed, the gluon propagator in the background-Feynman gauge (\ref{eq:bFprop}) reduces to the gluon propagator in the Feynman gauge only when the background field $A_\mu=0$.

However, the converse is not true. Writing the gluon propagator in the background field in the Feynman gauge:
\begin{eqnarray}
&& (x|\frac{-ig_{\mu\nu}\delta^{ab}}{p^2} |y)
\nonumber\\
&&- i g(x| \frac{-ig_{\mu\rho}}{p^2} \Big[ g^{\rho\sigma} \{ p_\alpha,  A^{\alpha} \} + 2 i (\partial^\rho A^\sigma - \partial^\sigma A^\rho) - p^\rho A^\sigma - A^\rho p^\sigma \Big] \frac{-ig_{\sigma\nu}(p)}{p^2} |y)^{ab} + \dots\,,
\label{eq:gp-mb-fg}
\end{eqnarray}
compare with a similar expression for the gluon propagator in the axial gauge (\ref{eq:gp-mb}), it is easy to check that (\ref{eq:gp-mb-fg}) does not coincide with (\ref{eq:bFprop}) for any non-trivial background gluon field.

As a result, we want to make an important remark that an appropriate gauge for calculations in the QCD factorization approach with the dense background fields is either the axial or background-Feynman gauge. While in calculations with trivial background gluon field configurations the results of calculation in the Feynman and background-Feynman gauges coincide, for the reason given above. In general, the Feynman and background-Feynman gauges are different, and in calculations with non-trivial background gluon fields one has to apply the background-Feynman gauge to obrain a gauge invariant result matching the corresponding result calculated in the axial gauge.

\section{Conclusions}

In this paper, we discuss calculation of the NLO correction to the quark TMDPDF operator using the MSTT approach developed in Ref. \cite{Mukherjee:2023snp}. The MSTT approach, designed to bridge the regions of large and small Bjorken-$x_B$, aims to calculate the TMD operators in the region of large $b_\perp \lesssim \Lambda^{-1}_{\rm QCD}$. In this region, the details about the flow of transverse momenta between the perturbative and non-perturbative components of the TMDPDFs become important. Essentially, the MSTT approach provides definition to this flow and allows to extract the IR structure of TMDPDFs in the kinematic region where small-$x_B$ effects can potentially contribute.

This requires calculation of the perturbative corrections in a general background field of the target. By performing such calculation in the dilute limit with only two background quarks, we find an ambiguity in the result. This ambiguity manifests itself in our ``naive" calculation that does not include contribution of the gauge potential at the spatial infinity. We perform this calculation in two gauges, i.e. the Feynman gauge in Sec. \ref{sec:Feynmancalc} and the axial gauge in Sec. \ref{sec:ax-calc}, and find that in our calculation scheme the results explicitly do not agree. At the same time we observe, that in the collinear approximation, when the background quarks do not carry any transverse momenta, the discrepancy in the results do disappear, though this is specific to a particular choice of the gauge fixing vector for the axial gauge and regularization of the corresponding axial divergence.

To understand the source of discrepancy in our calculation scheme we match two results at the order of initial Feynman diagrams. To perform this matching in the dilute limit, in Sec. \ref{sec:mapping} we apply the vertex identity to all quark-gluon interaction vertices. In Sec. \ref{sec:beyondDL} we develop a generalization of this method to the case of the dense field with multiple interactions with the background partons. To our knowledge, such analysis in the dense limit has never been done before and is one of the results of this paper.

Here, we would like to emphasize that our analysis of the gauge invariance is limited in scope to the case of the NLO corrections in ``quantum" fields. For this reason, while we include an infinite number of interactions with the backgorund fields, we do not take into consideration the ``quantum" gluon self-interactions. Though, the analysis of the gauge invariance in presence of the three- and four-gluon vertexes is more involved and goes beyond the scope of this paper, we expect that the main conclusions about the role of the QCD EoM and the gauge invariance of the TMD operator will remain unchanged.

By direct matching of the Feynman diagram contributions we identify two sources of the ambiguity in our ``naive" calculation. The first source is related to the fact that the TMD operator in our ``naive" calculation is not gauge invariant since it does not include contribution of the gauge potential at the spatial infinity. We find that the contribution of this potential is essential in our calculation scheme for the gauge invariance of the obtained results. The role of the transverse link at infinity for the gauge invariance of the TMDPDF operators~\footnote{And any other operators with a transverse direction separation between the light-cone gauge factors.} has of course been known before. Though in any practical calculation its contribution is never included. We argue that one should understand this as an additional scheme dependent prescription. From our calculation it becomes obvious that regardless of the gauge the contribution of the transverse gauge link cannot be formally proved to be zero. While in the axial gauge all types of emission from the transverse link at infinity has to be taken into account, after resummation of all diagrams in both Feynman and axial gauges the contribution of the link reduces to its self-energy corrections. Its not formally obvious that this contribution is zero, though currently there are no signs that this contribution somehow affect the cross section. For this reason, in the TMD physics one usually eliminates these corrections by either appropriately defining the soft function such that the self-energy corrections in the TMDPDF matrix element are completely canceled by a similar corrections in the soft factor~\cite{Collins:2011zzd}, or eliminate them by hand, see e.g. Ref.~\cite{Ebert:2019okf}, which can be understood as a prescription in the definition of the TMDPDF matrix element and the soft factor. In this sense, the self-energy corrections can be understood as artifacts of the TMD factorization formula with operator definitions of its components. So far, it has not been detected that these artifacts lead to any physical effects in the cross sections, for these reason they have to be eliminated in the aforementioned sense.

We believe, the same applies to the small-$x_B$ distributions. There is no an analog of the TMD soft factor for the small-$x_B$ dipole operators. For this reason one has to understand elimination of the self-energy corrections to the transverse links at infinity as a part of the definition of the small-$x_B$ dipole amplitudes. Note that in the axial gauge calculations, to ensure the gauge invariance, one still has to include all possible emissions from the transverse links at infinity. In the sum of all diagrams one has to identify the self-energy contribution, which formally coincides with the self-energy contribution in the Feynman gauge, and eliminate it. Consequently, the result of calculation in the axial gauge will coincide with the corresponding result in the Feynman gauge calculation without the self-energy corrections at infinity.\footnote{The absence of the self-energy corrections at infinity in the Feynman gauge is also a consequence of the quantization procedure in this gauge, which explicitly requires that all fields at infinity should vanish.}

Yet, we find that the gauge potential at the spatial infinity is not the only source of the discrepancy in our calculation scheme. Our background partons are not free particles on the mass-shell. Instead, they are general background fields defined by the MSTT factorization scheme. However, as we show in this paper, the background fields are not arbitrary and has to satisfy the QCD EoM. The EoM for the background fields of the target reflect the gauge invariance of the hadronic state and ensure independence of the hadron properties on the gauge rotations of the background fields. 

It is important to note that the EoM introduce a natural ambiguity into the result of calculations. Indeed, one can always add or subtract a contribution proportional to the EoM and get a physically equivalent result. As we show, this corresponds to going from one choice of the gauge fixing vector in the axial gauge calculations to another one. A limiting case, when the explicit dependence on the gauge fixing vector is eliminated, yields the result of calculation in the background-Feynman gauge, which reduces to the usual Feynman gauge in the dilute approximation.

One concomitant result of our analysis is a clear demonstration that the correct generalization of the Feynman gauge in the dense background field calculations in the QCD factorization approach is the background-Feynman gauge, which for instance was used for calculations in Ref. \cite{Mukherjee:2023snp}.

Since the EoM relate the quark and gluon operators, by adding the equations one can effectively transfer a certain physical content between the quark and gluon sectors in the non-perturbative matrix elements. In the paper, we provide a detailed description of this mechanism.

As a result we see, that quark and gluon sectors can not be considered independently, since only their combination provides a complete physical picture. For this reason, we believe our detailed analysis of the mixing between quark and gluon fields is especially relevant to the small-$x_B$ calculations. In particular, the role of the mixing for spin dependent operators, see e.g. Refs.~\cite{Chirilli:2021lif,Borden:2024bxa}.

While the interplay between quark and gluon sectors due to the EoM does not {\it per se} introduces any troubles. We find, that there still is a potential issue when the terms proportional to the EoM introduces unphysical singularities. We observe such contribution in our explicit result obtained in the axial gauge, which contains rapidity singularities of the IR origin. In our previous calculation for the gluon TMDPDFs, see Ref. \cite{Mukherjee:2023snp}, such singularities generated a logarithmic contribution of the BFKL-type in the IR structure of the gluon TMDPDFs. However, for the gluon case we observe such singularity in the results obtained in both background-Feynman and axial gauge calculations. Which means that the singularity is genuine and is not related to the EoM.

However, in the quark case, presented in this paper, the singularity does not appear in the Feynman gauge calculation. For this reason, we conclude that the observed rapidity singularity is not physical and should be eliminated from the final result.

In practice, however, it is not always obvious whether a particular divergence is related to the EoM or not. To do that, one has to perform a detailed analysis of the operator structure of the result identifying terms which can be recast into the EoM and subsequently eliminated. We believe that methods and techniques developed in this paper provide essential tools for such analysis.

One practical approach, though, is not to rely on a calculation in any given gauge, but instead perform a calculation in different gauges and compare the  results. The genuine divergencies should not be affected by the choice of the gauge. And the difference between results should be up to the EoM. 

From our analysis we conclude that the correct form of the NLO correction to the quark TMDPDF matrix element calculated in the MSTT approach is provided by the calculation in the Feynman gauge and has a form of Eq. (\ref{eq:fig1-fin}). This equation is a bare result for the NLO correction. The divergencies in this result are yet to be regularized and renormalized using the MSTT procedure developped in Ref.~\cite{Mukherjee:2023snp}. We leave this analysis for a subsequent publication.

\section{Acknowledgments}
This material is based upon work supported by The U.S. Department of Energy, Office of Science, Office of Nuclear Physics through Contract Nos.~DE-SC0012704 and DE-SC0020081, and within the framework of Saturated Glue (SURGE) Topical Collaboration in Nuclear Theory. A.T.’s research was also supported by the Center for Frontiers in Nuclear Science at Stony Brook University.

\appendix
\section{Explicit form of the difference terms $A$ and $B$ in Eq. (\ref{eq:after-part})\label{ap:dif-terms}}
In the main part of the paper, we describe the vertex identity which we apply to the $e_\mu$ dependent terms of the gluon propagator in the axial gauge. In the momentum representation this procedure is equivalent to application of the momentum conservation law in each quark-gluon vertex of a Feynman diagram. We find that after performing this operation the sum of diagrams in Fig. \ref{fig:Fdiag} has a form of Eq. (\ref{eq:after-part}) where the difference terms are
\begin{eqnarray}
&&A \equiv - i g^2 C_F \int d^4y \bar{q}(z^-_1, b_\perp) \gamma^+ (z^-_2, - b_\perp|\frac{i\slashed{p}}{p^2+i\epsilon}|y) \gamma^\mu q(y)  (L^-, b_\perp| \frac{-i}{p^2+i\epsilon} \frac{ e_\mu}{e\cdot p} |y)
\nonumber\\
&&+ i g^2 C_F \int d^4y \bar{q}(z^-_1, b_\perp) \gamma^+ (z^-_2, - b_\perp |\frac{i\slashed{p}}{p^2+i\epsilon}|y) \gamma^\mu q(y) (L^-, - b_\perp| \frac{-i}{p^2+i\epsilon} \frac{e_\mu }{e\cdot p} |y)
\nonumber\\
&&- i g^2 C_F \int d^4x \bar{q}(x)\gamma^\mu (x|\frac{i\slashed{p}}{p^2 + i\epsilon}|z^-_1, b_\perp) \gamma^+ q(z^-_2, - b_\perp) (L^-, b_\perp| \frac{-i}{p^2+i\epsilon} \frac{ e_\mu}{e\cdot p} |x)
\nonumber\\
&&+ i g^2 C_F \int d^4x \bar{q}(x)\gamma^\mu  (x|\frac{i\slashed{p}}{p^2 + i\epsilon}|z^-_1, b_\perp) \gamma^+ q(z^-_2, - b_\perp) (L^-, -b_\perp|\frac{-i}{p^2+i\epsilon} \frac{ e_\mu}{e\cdot p} |x)
\nonumber\\
&&- i g^2 C_F \bar{q}(z^-_1, b_\perp) \gamma^+ q(z^-_2, - b_\perp) \int_{z^-_2}^{\infty}dz^- n^\mu  (L^-, b_\perp | \frac{-i}{p^2+i\epsilon} \frac{ e_\mu}{e\cdot p} |z^-, - b_\perp) 
\nonumber\\
&&+ i g^2C_F \bar{q}(z^-_1, b_\perp) \gamma^+ q(z^-_2, - b_\perp) \int_{z^-_2}^{\infty}dz^- n^\mu (L^-, - b_\perp|\frac{-i}{p^2+i\epsilon} \frac{ e_\mu}{e\cdot p} |z^-, - b_\perp)
\nonumber\\
&&+ i g^2 C_F \bar{q}(z^-_1, b_\perp) \gamma^+ q(z^-_2, - b_\perp) \int^{z^-_1}_{\infty}dz^- n^\mu ( L^-, - b_\perp |\frac{-i}{p^2+i\epsilon} \frac{e_\mu }{e\cdot p} |z^-, b_\perp)
\nonumber\\
&&- i g^2C_F\bar{q}(z^-_1, b_\perp) \gamma^+  q(z^-_2, - b_\perp) \int^{z_1}_{\infty}dz^- n^\mu (L^-, b_\perp|\frac{-i}{p^2+i\epsilon} \frac{e_\mu}{e\cdot p} |z^-, b_\perp)
\label{eq:termA}
\end{eqnarray}
and
\begin{eqnarray}
&&B\equiv i g^2 C_F \int_{z^-_2}^{\infty}dz^- \int d^4y \bar{q}(z^-_1, b_\perp) \gamma^+ (z^-_2, - b_\perp |\frac{i\slashed{p}}{p^2+i\epsilon}|y) \gamma^\nu \partial_\nu q(y) n^\mu (z^-, - b_\perp|\frac{-i}{p^2+i\epsilon} \frac{ e_\mu}{e\cdot p} |y)
\nonumber\\
&&+ i g^2 C_F \int_{\infty}^{z^-_1} dz^-  \int d^4y \bar{q}(z^-_1, b_\perp) \gamma^+ (z^-_2, - b_\perp|\frac{i\slashed{p}}{p^2+i\epsilon}|y) \gamma^\nu \partial_\nu q(y) n^\mu (z^-, b_\perp|\frac{-i}{p^2+i\epsilon} \frac{e_\mu }{e\cdot p} |y)
\nonumber\\
&&+ i g^2 C_F \int_\infty^{z^-_1} dz^- \int d^4x \partial_\mu \bar{q}(x)\gamma^\mu (x|\frac{i\slashed{p}}{p^2 + i\epsilon}|z^-_1, b_\perp)    \gamma^+ q(z^-_2, - b_\perp) n^\nu (z^-, b_\perp|\frac{-i}{p^2+i\epsilon} \frac{e_\nu }{e\cdot p} |x)
\nonumber\\
&&+i g^2 C_F \int^\infty_{z^-_2}dz^- \int d^4x \partial_\mu \bar{q}(x)\gamma^\mu  (x|\frac{i\slashed{p}}{p^2 + i\epsilon}|z^-_1, b_\perp)     \gamma^+ q(z^-_2, - b_\perp) n^\nu (z^-, -b_\perp|\frac{-i}{p^2+i\epsilon} \frac{e_\nu }{e\cdot p} |x)
\nonumber\\
&&- i g^2 C_F \int d^4x \int d^4y \bar{q}(x)\gamma^\mu (x|\frac{i\slashed{p}}{p^2+i\epsilon}|z^-_1, b_\perp) \gamma^+ (z^-_2, -  b_\perp|\frac{i\slashed{p}}{p^2+i\epsilon}|y) \gamma^\nu \partial_\nu q(y) (y| \frac{-i}{p^2+i\epsilon}  \frac{e_\mu }{e\cdot p} |x)
\nonumber\\
&&+i g^2 C_F \int d^4x \int d^4y \partial_\mu \bar{q}(x)\gamma^\mu (x|\frac{i\slashed{p}}{p^2+i\epsilon}|z^-_1, b_\perp) \gamma^+ (z^-_2, -  b_\perp|\frac{i\slashed{p}}{p^2+i\epsilon}|y) \gamma^\nu q(y) (y|\frac{-i}{p^2+i\epsilon} \frac{  e_\nu}{e\cdot p} |x)
\nonumber\\
&&+i g^2 C_F \int d^4x \int d^4y \bar{q}(z^-_1, b_\perp)\gamma^+ (z^-_2, - b_\perp |\frac{i\slashed{p}}{p^2+i\epsilon}|x) \gamma^\mu (x|\frac{i\slashed{p}}{p^2+i\epsilon}|y) \gamma^\nu \partial_\nu q(y) (x|\frac{-i }{p^2 + i\epsilon} \frac{e_\mu}{e\cdot p} |y)
\nonumber\\
&&+i g^2 C_F \int d^4x \int d^4y  \partial_\mu \bar{q}(x) \gamma^\mu ( x |\frac{i\slashed{p}}{p^2+i\epsilon}|y) \gamma^\nu (y|\frac{i\slashed{p}}{p^2+i\epsilon}|z^-_1, b_\perp)  \gamma^+ q(z^-_2, - b_\perp) (y|\frac{-i }{p^2 + i\epsilon} \frac{e_\nu  }{e\cdot p} |x)
\label{eq:termB}
\end{eqnarray}

In the main part of the paper we argue that to compensate these terms one has to take into account contribution of the transverse Wilson lines and impose EoM for the background field.

\section{The origin of a spurious IR singularity in Eq. (\ref{eq:ax-virt-diverge}) and the role of EoM\label{ap:EoM-div}}
In this Appendix we discuss calculation of the diagram in Fig.~\ref{fig:Fdiag}e.\footnote{Calculation of a similar diagram in Fig.~\ref{fig:Fdiag}d is almost identical.} Our goal is to show that the IR singularity $\int dz/z$ in the  final result for this diagram (\ref{eq:ax-virt-diverge}) calculated in the axial gauge is related to the contribution of EoM. This indicates that the singularity is unphysical and should be eliminated from the final result. Indeed, such divergence doesn't appear in the final result for this diagram calculated in the Feynman gauge, see Eq. (\ref{eq1dfn}).

Let's start with an initial expression for this diagram in the axial gauge:
\begin{eqnarray}
&&\mathcal{U}^{[\gamma^+]}(z^-_1, z^-_2, b_\perp)\Big|^{axial}_{Fig.~\ref{fig:Fdiag}e} 
\nonumber\\
&&= -g^2 C_F \int_{z^-_2}^{\infty}dz^- \int d^4y \bar{q}(z^-_1, b_\perp) \gamma^+ (z^-_2, - b_\perp |\frac{i\slashed{p}}{p^2+i\epsilon}|y) \gamma^\mu q(y) n^\nu (z^-, - b_\perp|\frac{-id_{\mu\nu}(p)}{p^2+i\epsilon}|y)\,.
\label{eq:Ap1e-init}
\end{eqnarray}

Applying the integration by parts procedure as in Sec. \ref{sec:mapping} we can rewrite this equation as\footnote{Gere we choose the gauge fixing vector $e_\mu = \bar{n}_\mu$.}
\begin{eqnarray}
&&\mathcal{U}^{[\gamma^+]}(z^-_1, z^-_2, b_\perp)\Big|^{axial}_{Fig.~\ref{fig:Fdiag}e}
\nonumber\\
&&= -g^2 C_F \int_{z^-_2}^{\infty}dz^- \int d^4y \bar{q}(z^-_1, b_\perp) \gamma^+ (z^-_2, - b_\perp |\frac{i\slashed{p}}{p^2+i\epsilon}|y) \gamma^+ q(y)  (z^-, - b_\perp|\frac{-i}{p^2+i\epsilon} |y)
\nonumber\\
&&- i g^2 C_F \int d^4y \bar{q}(z^-_1, b_\perp) \gamma^+ (z^-_2, - b_\perp |\frac{i\slashed{p}}{p^2+i\epsilon}|y) \gamma^- q(y) (z^-_2, - b_\perp| \frac{-i}{p^2+i\epsilon} \frac{1 }{p^- + i\epsilon} |y)
\nonumber\\
&&+ i g^2 C_F \int d^4y \bar{q}(z^-_1, b_\perp) \gamma^+ (z^-_2, - b_\perp |\frac{i\slashed{p}}{p^2+i\epsilon}|y) \gamma^- q(y) (L^-, - b_\perp| \frac{-i}{p^2+i\epsilon} \frac{1 }{p^- +i\epsilon} |y)
\nonumber\\
&&+i g^2 C_F \int_{z^-_2}^{\infty}dz^- \int d^4y \bar{q}(z^-_1, b_\perp) \gamma^+ (z^-_2, - b_\perp |\frac{i\slashed{p}}{p^2+i\epsilon}|y) \gamma^\mu \partial_\mu q(y) (z^-, - b_\perp|\frac{-i}{p^2+i\epsilon} \frac{ 1}{p^- + i\epsilon} |y)
\nonumber\\
&&- i g^2 C_F \int_{z^-_2}^{\infty}dz^- \bar{q}(z^-_1, b_\perp) \gamma^+ q(z^-_2, - b_\perp) (z^-, - b_\perp|\frac{-i}{p^2+i\epsilon} \frac{ 1 }{p^- + i\epsilon} |z^-_2, - b_\perp)\,.
\label{eq:Ap-after-parts}
\end{eqnarray}
Here, the last but one term is proportional to $\gamma^\mu \partial_\mu q(y)$, so it can be eliminated by applying the EoM (\ref{eq:emo}). However, we aim to keep this term and show that it leads to the IR divergence in Eq. (\ref{eq:ax-virt-diverge}), which proves that the divergence is unphysical.

Using replacement (\ref{eq:gamma-subst}) and calculating the trace of $\gamma$-matrixes we further simplify the result as
\begin{eqnarray}
&&\mathcal{U}^{[\gamma^+]}(z^-_1, z^-_2, b_\perp)\Big|^{axial}_{Fig.~\ref{fig:Fdiag}e}
\nonumber\\
&&= - 2 g^2 C_F \int_{z^-_2}^{\infty}dz^- \int d^4y (z^-_2, - b_\perp |\frac{ p^+}{p^2+i\epsilon}|y) (z^-, - b_\perp|\frac{1}{p^2+i\epsilon} |y) {\rm tr}\{\bar{q}(z^-_1, b_\perp) \gamma^+ q(y) \}
\nonumber\\
&&+ i g^2 C_F \int_{z^-_2}^{\infty}dz^- \int d^4y (z^-_2, - b_\perp |\frac{ p^k}{p^2+i\epsilon}|y) (z^-, - b_\perp|\frac{1}{p^2+i\epsilon} \frac{ 1 }{ p^- + i\epsilon} |y) {\rm tr}\{\bar{q}(z^-_1, b_\perp) \gamma^+ \partial_k q(y^-, y_\perp) \}
\nonumber\\
&&- g^2 C_F \int_{z^-_2}^{\infty}dz^- (z^-, - b_\perp|\frac{1}{p^2+i\epsilon} \frac{ 1 }{ p^- + i\epsilon } |z^-_2, - b_\perp) {\rm tr}\{ \bar{q}(z^-_1, b_\perp) \gamma^+ q(z^-_2, - b_\perp)\}\,,
\label{eq:Ap-simpl}
\end{eqnarray}
where the second term corresponds to the last but one term proportional to EoM $ \gamma^\mu \partial_\mu q(y)$ in Eq. (\ref{eq:Ap-after-parts}).

For out following discussion it's sufficient to consider only this term
\begin{eqnarray}
&&\mathcal{U}^{[\gamma^+]}(z^-_1, z^-_2, b_\perp)\Big|^{axial}_{Fig.~\ref{fig:Fdiag}e,~ {\rm EoM}}
\nonumber\\
&&= i g^2 C_F \int_{z^-_2}^{\infty}dz^- \int d^4y (z^-_2, - b_\perp |\frac{ p^k}{p^2+i\epsilon}|y) (z^-, - b_\perp|\frac{1}{p^2+i\epsilon} \frac{ 1 }{ p^- + i\epsilon} |y) {\rm tr}\{\bar{q}(z^-_1, b_\perp) \gamma^+ \partial_k q(y^-, y_\perp) \}\,.
\end{eqnarray}

The calculation of this contribution is quite straightforward. Integrating over $y^+$ and $p^-$ component of the internal momenta we obtain
\begin{eqnarray}
&&\mathcal{U}^{[\gamma^+]}(z^-_1, z^-_2, b_\perp)\Big|^{axial}_{Fig.~\ref{fig:Fdiag}e,~ {\rm EoM}} = g^2 C_F \int_{z^-_2}^{\infty}dz^- \int dy^- \int d^2y_\perp \int \dhd^2p_{1\perp} e^{ip_{1\perp}(- b_\perp - y_\perp)} \int \dhd^2p_{2\perp} e^{ip_{2\perp}(- b_\perp - y_\perp)}  
\nonumber\\
&&\times \Big( \int^\infty_0 \dhd p^+_1 \int^\infty_0 \dhd p^+_2 \frac{2p^+_1}{ 2p^+_2 p^2_{1\perp} + 2p^+_1 p^2_{2\perp} } \frac{ p^k_1 }{ p^2_{1\perp} }
+ \int^\infty_0 \dhd p^+_1
 \int^0_{-\infty} \dhd p^+_2 \frac{ p^k_1}{ p^2_{1\perp} } \frac{1}{ p^2_{2\perp} }
\nonumber\\
&&+ \int^0_{-\infty} \dhd p^+_1 \int^0_{-\infty} \dhd p^+_2 \frac{ 2p^+_2 }{ 2p^+_2 p^2_{1\perp} + 2p^+_1 p^2_{2\perp} } \frac{ p^k_1 }{ p^2_{2\perp} } \Big) e^{-ip^+_1 (z^-_2 - y^-)} e^{-ip^+_2 (z^- - y^-)} {\rm tr}\{\bar{q}(z^-_1, b_\perp) \gamma^+ \partial_k q(y^-, y_\perp) \}\,,
\end{eqnarray}
where $p^+_1$ ($p_{1\perp}$) and $p^+_2$ ($p_{2\perp}$) are longitudinal (transverse) components of the quark and gluon propagators in Eq. (\ref{eq:Ap1e-init}).

Now, let's substitute this result into Eq. (\ref{def:matrix-fourier2}) for the matrix element of the TMDPDF. Subsequently integrating over $z^-_2$, $z^-$ and $p^+_2$ for $x>0$ we obtain
\begin{eqnarray}
&&\Phi^{[\gamma^+]}(x, b_\perp)\Big|^{axial}_{Fig.~\ref{fig:Fdiag}e,~ {\rm EoM}} = - \frac{i g^2 C_F}{4\pi \delta(0)} \int d^2y_\perp \int \dhd^2p_{1\perp} e^{ip_{1\perp}(- \frac{1}{2}b_\perp - y_\perp)} \int \dhd^2p_{2\perp} e^{ip_{2\perp}(- \frac{1}{2}b_\perp - y_\perp)} 
\nonumber\\
&&\times \Big( \int^{x p^+}_0 \dhd p^+_1 \frac{1}{x P^+ - p^+_1} \frac{2p^+_1}{ 2(x P^+ - p^+_1) p^2_{1\perp} + 2p^+_1 p^2_{2\perp} } \frac{ p^k_1 }{ p^2_{1\perp} } + \int^\infty_{xp^+} \dhd p^+_1 \frac{1}{ (x P^+ - p^+_1)} \frac{ p^k_1}{ p^2_{1\perp} } \frac{1}{ p^2_{2\perp} } \Big) 
\nonumber\\
&&\times \int dz^-_1 e^{-ix p^+ z^-_1 } \int dy^- e^{i x p^+ y^-} \langle P, S|\bar{q}(z^-_1, \frac{1}{2}b_\perp) \gamma^+ \partial_k q(y^-, y_\perp) |P, S\rangle\,.
\end{eqnarray}

Introducing $z = p^+_1 / xp^+$ in the first term and $z = xp^+/p^+_1$ in the second we get
\begin{eqnarray}
&&\Phi^{[\gamma^+]}(x, b_\perp)\Big|^{axial}_{Fig.~\ref{fig:Fdiag}e,~ {\rm EoM}} = -\frac{i g^2 C_F}{8\pi^2 \delta(0)} \int d^2y_\perp \int \dhd^2p_{1\perp} e^{ip_{1\perp}(- \frac{1}{2}b_\perp - y_\perp)} \int \dhd^2p_{2\perp} e^{ip_{2\perp}(- \frac{1}{2}b_\perp - y_\perp)}  
\nonumber\\
&&\times \Big( \int^1_0 \frac{d z}{1 - z} \frac{z}{ (1 - z) p^2_{1\perp} + z p^2_{2\perp} } \frac{ p^k_1 }{ p^2_{1\perp} } - \int_0^{1} \frac{dz}{z} \frac{1}{ 1 - z } \frac{ p^k_1}{ p^2_{1\perp} } \frac{1}{ p^2_{2\perp} } \Big)
\nonumber\\
&&\times \int dz^-_1 e^{-ix p^+ z^-_1 } \int dy^- e^{i x p^+ y^-} \langle P, S|\bar{q}(z^-_1, \frac{1}{2}b_\perp) \gamma^+ \partial_k q(y^-, y_\perp) |P, S\rangle\,.
\end{eqnarray}

Using translational invariance of the matrix element and changing the notation as $p_{1\perp} = -p_\perp$ and $p_{2\perp} = k_\perp$ we finally obtain
\begin{eqnarray}
&&\Phi^{[\gamma^+]}(x, b_\perp)\Big|^{axial}_{Fig.~\ref{fig:Fdiag}e,~ {\rm EoM}}
= \frac{ g^2 C_F}{8\pi^2} \int \dhd^2p_{\perp} e^{ i p_{\perp} b_\perp} \int \dhd^2k_{\perp} e^{- i k_{\perp} b_\perp}
 \Big( \int^1_0 \frac{d z}{1 - z} \frac{z}{ (1 - z) p^2_{\perp} + z k^2_{\perp} } 
\nonumber\\
&&- \int_0^{1} \frac{dz}{z} \frac{1}{ 1 - z }  \frac{1}{ k^2_{\perp} } \Big) \frac{ p_k ( p_k - k_k ) }{ p^2_{\perp} } \int d^2z_\perp e^{i (k_{\perp} -p_{\perp} ) z_\perp }\int^\infty_{-\infty} dz^- e^{-ix p^+ z^- } \langle P, S|\bar{q}(z^-, z_\perp) \gamma^+ q(0) |P, S\rangle\,.
\label{eq:Ap-fin-EoM}
\end{eqnarray}
This contribution to the matrix element corresponds to the second term in Eq. (\ref{eq:Ap-simpl}) and the last but one term proportional to EoM in Eq. (\ref{eq:Ap-after-parts}). This results contains an infrared singularity $\int dz/z$ which is the infrared singularity in Eqs. (\ref{eq:ax-virt-diverge}) and (\ref{eq:fig1-fin-axial}). However, this singularity is unphysical since contribution (\ref{eq:Ap-fin-EoM}) originates in the term proportional to EoM that should be eliminated.

What about other two terms in Eq. (\ref{eq:Ap-simpl})? They can be calculated in a similar way. For the last term we get
\begin{eqnarray}
&&\frac{g^2 C_F}{8\pi^2} \int \frac{\dhd^2p_{\perp}}{p^2_\perp} \int^1_0 \frac{dz}{z} \frac{1}{1 - z} \int dz^- e^{-ix p^+ z^- } \langle P, S| \bar{q}(z^-, b_\perp) \gamma^+ q(0)|P, S\rangle\,.
\label{eq:Ap-last-g}
\end{eqnarray}
This term contains an IR singularity $\int dz/z$, which is not related to the EoM contribution. However, while this singularity also contribute to Eq. (\ref{eq:ax-virt-diverge}) along with the EoM contribution (\ref{eq:Ap-fin-EoM}), in the sum of all diagrams in Fig.~\ref{fig:Fdiag} this contribution gets canceled by the contribution of Fig.~\ref{fig:Fdiag}g, see Eq. (\ref{eq:fig1g-axial}). So only IR singularity of Eq. (\ref{eq:Ap-fin-EoM}) contributes to the final result (\ref{eq:fig1-fin-axial}), which, as we discussed above, is an unphysical singularity proportional to EoM.

The first term in Eq. (\ref{eq:Ap-simpl}) doesn't contain an IR singularity
\begin{eqnarray}
&& -\frac{ g^2 C_F}{8\pi^2} \int \dhd^2p_{\perp} e^{ i p_{\perp} b_\perp} \int \dhd^2k_{\perp} e^{ - i  k_{\perp} b_\perp} \int^1_0 \frac{d z}{1 - z} \frac{ z}{ (1 - z) p^2_{\perp} +  z k^2_{\perp} } \int d^2z_\perp e^{i (k_{\perp} - p_{\perp} ) z_\perp }
\nonumber\\
&&\times \int dz^- e^{-ix p^+ z^- } \langle P, S|\bar{q}(z^-, z_\perp) \gamma^+ q(0) |P, S\rangle\,.
\label{eq:Ap-firsttm}
\end{eqnarray}

Combining Eqs. (\ref{eq:Ap-fin-EoM}), (\ref{eq:Ap-last-g}), and (\ref{eq:Ap-firsttm}) we obtain
\begin{eqnarray}
&&\Phi^{[\gamma^+]}(x, b_\perp)\Big|^{axial}_{Fig.~\ref{fig:Fdiag}e}
= \frac{ g^2 C_F}{8\pi^2} \int \dhd^2p_{\perp} e^{ i p_{\perp} b_\perp} \int \dhd^2k_{\perp} e^{ - i  k_{\perp} b_\perp} \Big[
 \int^1_0 d z \frac{ 1 }{ (1 - z) p^2_{\perp} + z k^2_{\perp} } + \int^1_0 \frac{d z}{z} \frac{ 1 }{ p^2_{\perp} } \Big] \frac{p_k  k_k}{k^2_\perp}
\nonumber\\
&&\times \int d^2z_\perp e^{i (k_{\perp} -p_{\perp} ) z_\perp } \int dz^- e^{-ix p^+ z^- } \langle P, S|\bar{q}(z^-, z_\perp) \gamma^+ q(0) |P, S\rangle\,,
\end{eqnarray}
which of course coincides with Eq. (\ref{eq:ax-virt-diverge}).

\bibliographystyle{unsrt}
\bibliography{main}

\begin{thebibliography}{10}

\bibitem{Mukherjee:2023snp}
Swagato Mukherjee, Vladimir~V. Skokov, Andrey Tarasov, and Shaswat Tiwari.
\newblock {Unified description of DGLAP, CSS, and BFKL evolution: TMD
  factorization bridging large and small x}.
\newblock {\em Phys. Rev. D}, 109(3):034035, 2024.

\bibitem{Collins:2011zzd}
John Collins.
\newblock {\em {Foundations of perturbative QCD}}, volume~32.
\newblock Cambridge University Press, 11 2013.

\bibitem{Anikin:1978tj}
S.~A. Anikin and O.~I. Zavyalov.
\newblock {Short Distance and Light Cone Expansions for Products of Currents}.
\newblock {\em Annals Phys.}, 116:135--166, 1978.

\bibitem{Shuryak:1981kj}
Edward~V. Shuryak and A.~I. Vainshtein.
\newblock {Theory of Power Corrections to Deep Inelastic Scattering in Quantum
  Chromodynamics. 1. Q**2 Effects}.
\newblock {\em Nucl. Phys. B}, 199:451--481, 1982.

\bibitem{Karchev:1983ai}
N.~i. Karchev.
\newblock {LIGHT CONE EXPANSION FOR A PRODUCT OF TWO GAUGE INVARIANT CURRENTS
  IN QCD}.
\newblock {\em Nucl. Phys. B}, 211:55--76, 1983.

\bibitem{Geyer:1985vw}
B.~Geyer, D.~Robaschik, Michael Bordag, and J.~Horejsi.
\newblock {NONLOCAL LIGHT CONE EXPANSIONS AND EVOLUTION EQUATIONS}.
\newblock {\em Z. Phys. C}, 26:591--600, 1985.

\bibitem{Braunschweig:1985nr}
T.~Braunschweig, B.~Geyer, J.~Horejsi, and D.~Robaschik.
\newblock {Hadron Operators on the Light Cone}.
\newblock {\em Z. Phys. C}, 33:275, 1986.

\bibitem{Balitsky:1987bk}
I.~I. Balitsky and Vladimir~M. Braun.
\newblock {Evolution Equations for QCD String Operators}.
\newblock {\em Nucl. Phys. B}, 311:541--584, 1989.

\bibitem{Dokshitzer:1977sg}
Yuri~L. Dokshitzer.
\newblock {Calculation of the Structure Functions for Deep Inelastic Scattering
  and e+ e- Annihilation by Perturbation Theory in Quantum Chromodynamics.}
\newblock {\em Sov. Phys. JETP}, 46:641--653, 1977.

\bibitem{Gribov:1972ri}
V.~N. Gribov and L.~N. Lipatov.
\newblock {Deep inelastic e p scattering in perturbation theory}.
\newblock {\em Sov. J. Nucl. Phys.}, 15:438--450, 1972.

\bibitem{Altarelli:1977zs}
Guido Altarelli and G.~Parisi.
\newblock {Asymptotic Freedom in Parton Language}.
\newblock {\em Nucl. Phys. B}, 126:298--318, 1977.

\bibitem{Balitsky:1995ub}
I.~Balitsky.
\newblock {Operator expansion for high-energy scattering}.
\newblock {\em Nucl. Phys. B}, 463:99--160, 1996.

\bibitem{Dominguez:2010xd}
Fabio Dominguez, Bo-Wen Xiao, and Feng Yuan.
\newblock {$k_t$-factorization for Hard Processes in Nuclei}.
\newblock {\em Phys. Rev. Lett.}, 106:022301, 2011.

\bibitem{Dominguez:2011wm}
Fabio Dominguez, Cyrille Marquet, Bo-Wen Xiao, and Feng Yuan.
\newblock {Universality of Unintegrated Gluon Distributions at small x}.
\newblock {\em Phys. Rev. D}, 83:105005, 2011.

\bibitem{Xiao:2017yya}
Bo-Wen Xiao, Feng Yuan, and Jian Zhou.
\newblock {Transverse Momentum Dependent Parton Distributions at Small-x}.
\newblock {\em Nucl. Phys. B}, 921:104--126, 2017.

\bibitem{Fadin:1975cb}
Victor~S. Fadin, E.~A. Kuraev, and L.~N. Lipatov.
\newblock {On the Pomeranchuk Singularity in Asymptotically Free Theories}.
\newblock {\em Phys. Lett. B}, 60:50--52, 1975.

\bibitem{Kuraev:1976ge}
E.~A. Kuraev, L.~N. Lipatov, and Victor~S. Fadin.
\newblock {Multi - Reggeon Processes in the Yang-Mills Theory}.
\newblock {\em Sov. Phys. JETP}, 44:443--450, 1976.

\bibitem{Kuraev:1977fs}
E.~A. Kuraev, L.~N. Lipatov, and Victor~S. Fadin.
\newblock {The Pomeranchuk Singularity in Nonabelian Gauge Theories}.
\newblock {\em Sov. Phys. JETP}, 45:199--204, 1977.

\bibitem{Balitsky:1978ic}
I.~I. Balitsky and L.~N. Lipatov.
\newblock {The Pomeranchuk Singularity in Quantum Chromodynamics}.
\newblock {\em Sov. J. Nucl. Phys.}, 28:822--829, 1978.

\bibitem{Ioffe:2010zz}
Boris~Lazarevich Ioffe, Victor~Sergeevich Fadin, and Lev~Nikolaevich Lipatov.
\newblock {\em {Quantum chromodynamics: Perturbative and nonperturbative
  aspects}}.
\newblock Cambridge Univ. Press, 2010.

\bibitem{Balitsky:1997mk}
Ian Balitsky.
\newblock {Operator expansion for diffractive high-energy scattering}.
\newblock {\em AIP Conf. Proc.}, 407(1):953, 1997.

\bibitem{Kovchegov:1999yj}
Yuri~V. Kovchegov.
\newblock {Small x F(2) structure function of a nucleus including multiple
  pomeron exchanges}.
\newblock {\em Phys. Rev. D}, 60:034008, 1999.

\bibitem{Jalilian-Marian:1997qno}
Jamal Jalilian-Marian, Alex Kovner, Andrei Leonidov, and Heribert Weigert.
\newblock {The BFKL equation from the Wilson renormalization group}.
\newblock {\em Nucl. Phys. B}, 504:415--431, 1997.

\bibitem{Jalilian-Marian:1997ubg}
Jamal Jalilian-Marian, Alex Kovner, and Heribert Weigert.
\newblock {The Wilson renormalization group for low x physics: Gluon evolution
  at finite parton density}.
\newblock {\em Phys. Rev. D}, 59:014015, 1998.

\bibitem{Kovner:2000pt}
Alex Kovner, J.~Guilherme Milhano, and Heribert Weigert.
\newblock {Relating different approaches to nonlinear QCD evolution at finite
  gluon density}.
\newblock {\em Phys. Rev. D}, 62:114005, 2000.

\bibitem{Iancu:2000hn}
Edmond Iancu, Andrei Leonidov, and Larry~D. McLerran.
\newblock {Nonlinear gluon evolution in the color glass condensate. 1.}
\newblock {\em Nucl. Phys. A}, 692:583--645, 2001.

\bibitem{Ferreiro:2001qy}
Elena Ferreiro, Edmond Iancu, Andrei Leonidov, and Larry McLerran.
\newblock {Nonlinear gluon evolution in the color glass condensate. 2.}
\newblock {\em Nucl. Phys. A}, 703:489--538, 2002.

\bibitem{Kovner:2013ona}
Alex Kovner, Michael Lublinsky, and Yair Mulian.
\newblock {Jalilian-Marian, Iancu, McLerran, Weigert, Leonidov, Kovner
  evolution at next to leading order}.
\newblock {\em Phys. Rev. D}, 89(6):061704, 2014.

\bibitem{Kovner:2014lca}
Alex Kovner, Michael Lublinsky, and Yair Mulian.
\newblock {NLO JIMWLK evolution unabridged}.
\newblock {\em JHEP}, 08:114, 2014.

\bibitem{Lublinsky:2016meo}
Michael Lublinsky and Yair Mulian.
\newblock {High Energy QCD at NLO: from light-cone wave function to JIMWLK
  evolution}.
\newblock {\em JHEP}, 05:097, 2017.

\bibitem{vanHameren:2025hyo}
Andreas van Hameren and Maxim Nefedov.
\newblock {Hybrid high-energy factorization and evolution at NLO from the
  high-energy limit of collinear factorization}.
\newblock 1 2025.

\bibitem{Kutak:2004ym}
K.~Kutak and A.~M. Stasto.
\newblock {Unintegrated gluon distribution from modified BK equation}.
\newblock {\em Eur. Phys. J. C}, 41:343--351, 2005.

\bibitem{Motyka:2009gi}
Leszek Motyka and Anna~M. Stasto.
\newblock {Exact kinematics in the small x evolution of the color dipole and
  gluon cascade}.
\newblock {\em Phys. Rev. D}, 79:085016, 2009.

\bibitem{Mueller:2013wwa}
A.~H. Mueller, Bo-Wen Xiao, and Feng Yuan.
\newblock {Sudakov double logarithms resummation in hard processes in the
  small-x saturation formalism}.
\newblock {\em Phys. Rev. D}, 88(11):114010, 2013.

\bibitem{Mueller:2012uf}
A.~H. Mueller, Bo-Wen Xiao, and Feng Yuan.
\newblock {Sudakov Resummation in Small-$x$ Saturation Formalism}.
\newblock {\em Phys. Rev. Lett.}, 110(8):082301, 2013.

\bibitem{Balitsky:2015qba}
I.~Balitsky and A.~Tarasov.
\newblock {Rapidity evolution of gluon TMD from low to moderate x}.
\newblock {\em JHEP}, 10:017, 2015.

\bibitem{Iancu:2015vea}
E.~Iancu, J.~D. Madrigal, A.~H. Mueller, G.~Soyez, and D.~N.
  Triantafyllopoulos.
\newblock {Resumming double logarithms in the QCD evolution of color dipoles}.
\newblock {\em Phys. Lett. B}, 744:293--302, 2015.

\bibitem{Iancu:2015joa}
E.~Iancu, J.~D. Madrigal, A.~H. Mueller, G.~Soyez, and D.~N.
  Triantafyllopoulos.
\newblock {Collinearly-improved BK evolution meets the HERA data}.
\newblock {\em Phys. Lett. B}, 750:643--652, 2015.

\bibitem{Balitsky:2016dgz}
I.~Balitsky and A.~Tarasov.
\newblock {Gluon TMD in particle production from low to moderate x}.
\newblock {\em JHEP}, 06:164, 2016.

\bibitem{Boussarie:2020fpb}
Renaud Boussarie and Yacine Mehtar-Tani.
\newblock {A novel formulation of the unintegrated gluon distribution for DIS}.
\newblock {\em Phys. Lett. B}, 831:137125, 2022.

\bibitem{Caucal:2021ent}
Paul Caucal, Farid Salazar, and Raju Venugopalan.
\newblock {Dijet impact factor in DIS at next-to-leading order in the Color
  Glass Condensate}.
\newblock {\em JHEP}, 11:222, 2021.

\bibitem{Taels:2022tza}
Pieter Taels, Tolga Altinoluk, Guillaume Beuf, and Cyrille Marquet.
\newblock {Dijet photoproduction at low x at next-to-leading order and its
  back-to-back limit}.
\newblock {\em JHEP}, 10:184, 2022.

\bibitem{Caucal:2022ulg}
Paul Caucal, Farid Salazar, Bj\"orn Schenke, and Raju Venugopalan.
\newblock {Back-to-back inclusive dijets in DIS at small x: Sudakov suppression
  and gluon saturation at NLO}.
\newblock {\em JHEP}, 11:169, 2022.

\bibitem{Caucal:2023fsf}
Paul Caucal, Farid Salazar, Bj\"orn Schenke, Tomasz Stebel, and Raju
  Venugopalan.
\newblock {Back-to-back inclusive dijets in DIS at small $x$: Complete NLO
  results and predictions}.
\newblock 7 2023.

\bibitem{Altinoluk:2023hfz}
Tolga Altinoluk, N\'estor Armesto, Alexander Kovner, and Michael Lublinsky.
\newblock {Single inclusive particle production at next-to-leading order in
  proton-nucleus collisions at forward rapidities: Hybrid approach meets TMD
  factorization}.
\newblock {\em Phys. Rev. D}, 108(7):074003, 2023.

\bibitem{Duan:2024qck}
Haowu Duan, Alex Kovner, and Michael Lublinsky.
\newblock {Born-Oppenheimer Renormalization group for High Energy Scattering:
  the Setup and the Wave Function}.
\newblock 12 2024.

\bibitem{Duan:2024qev}
Haowu Duan, Alex Kovner, and Michael Lublinsky.
\newblock {Born-Oppenheimer Renormalization group for High Energy Scattering:
  CSS, DGLAP and all that}.
\newblock 12 2024.

\bibitem{Accardi:2012qut}
A.~Accardi et~al.
\newblock {Electron Ion Collider: The Next QCD Frontier}: {Understanding the
  glue that binds us all}.
\newblock {\em Eur. Phys. J. A}, 52(9):268, 2016.

\bibitem{AbdulKhalek:2021gbh}
R.~Abdul~Khalek et~al.
\newblock {Science Requirements and Detector Concepts for the Electron-Ion
  Collider}: {EIC Yellow Report}.
\newblock {\em Nucl. Phys. A}, 1026:122447, 2022.

\bibitem{Aschenauer:2017jsk}
E.~C. Aschenauer, S.~Fazio, J.~H. Lee, H.~Mantysaari, B.~S. Page, B.~Schenke,
  T.~Ullrich, R.~Venugopalan, and P.~Zurita.
\newblock {The electron\textendash{}ion collider: assessing the energy
  dependence of key measurements}.
\newblock {\em Rept. Prog. Phys.}, 82(2):024301, 2019.

\bibitem{Iancu:2003xm}
Edmond Iancu and Raju Venugopalan.
\newblock {\em {The Color glass condensate and high-energy scattering in QCD}},
  pages 249--3363.
\newblock 3 2003.

\bibitem{Weigert:2005us}
Heribert Weigert.
\newblock {Evolution at small x(bj): The Color glass condensate}.
\newblock {\em Prog. Part. Nucl. Phys.}, 55:461--565, 2005.

\bibitem{Gelis:2010nm}
Francois Gelis, Edmond Iancu, Jamal Jalilian-Marian, and Raju Venugopalan.
\newblock {The Color Glass Condensate}.
\newblock {\em Ann. Rev. Nucl. Part. Sci.}, 60:463--489, 2010.

\bibitem{Albacete:2014fwa}
Javier~L. Albacete and Cyrille Marquet.
\newblock {Gluon saturation and initial conditions for relativistic heavy ion
  collisions}.
\newblock {\em Prog. Part. Nucl. Phys.}, 76:1--42, 2014.

\bibitem{Kovchegov:2012mbw}
Yuri~V. Kovchegov and Eugene Levin.
\newblock {\em {Quantum Chromodynamics at High Energy}}, volume~33.
\newblock Oxford University Press, 2013.

\bibitem{Collins:1981uk}
John~C. Collins and Davison~E. Soper.
\newblock {Back-To-Back Jets in QCD}.
\newblock {\em Nucl. Phys. B}, 193:381, 1981.
\newblock [Erratum: Nucl.Phys.B 213, 545 (1983)].

\bibitem{Collins:1984kg}
John~C. Collins, Davison~E. Soper, and George~F. Sterman.
\newblock {Transverse Momentum Distribution in Drell-Yan Pair and W and Z Boson
  Production}.
\newblock {\em Nucl. Phys. B}, 250:199--224, 1985.

\bibitem{Collins:1987pm}
John~C. Collins and Davison~E. Soper.
\newblock {The Theorems of Perturbative QCD}.
\newblock {\em Ann. Rev. Nucl. Part. Sci.}, 37:383--409, 1987.

\bibitem{Collins:1989gx}
John~C. Collins, Davison~E. Soper, and George~F. Sterman.
\newblock {Factorization of Hard Processes in QCD}.
\newblock {\em Adv. Ser. Direct. High Energy Phys.}, 5:1--91, 1989.

\bibitem{Meng:1995yn}
Ruibin Meng, Fredrick~I. Olness, and Davison~E. Soper.
\newblock {Semiinclusive deeply inelastic scattering at small q(T)}.
\newblock {\em Phys. Rev. D}, 54:1919--1935, 1996.

\bibitem{Ji:2004wu}
Xiang-dong Ji, Jian-ping Ma, and Feng Yuan.
\newblock {QCD factorization for semi-inclusive deep-inelastic scattering at
  low transverse momentum}.
\newblock {\em Phys. Rev. D}, 71:034005, 2005.

\bibitem{Ji:2004xq}
Xiang-dong Ji, Jian-Ping Ma, and Feng Yuan.
\newblock {QCD factorization for spin-dependent cross sections in DIS and
  Drell-Yan processes at low transverse momentum}.
\newblock {\em Phys. Lett. B}, 597:299--308, 2004.

\bibitem{Boussarie:2023izj}
Renaud Boussarie et~al.
\newblock {TMD Handbook}.
\newblock 4 2023.

\bibitem{Collins:1981uw}
John~C. Collins and Davison~E. Soper.
\newblock {Parton Distribution and Decay Functions}.
\newblock {\em Nucl. Phys. B}, 194:445--492, 1982.

\bibitem{Kang:2012em}
Zhong-Bo Kang and Jian-Wei Qiu.
\newblock {QCD evolution of naive-time-reversal-odd parton distribution
  functions}.
\newblock {\em Phys. Lett. B}, 713:273--276, 2012.

\bibitem{Sun:2013hua}
Peng Sun and Feng Yuan.
\newblock {Transverse momentum dependent evolution: Matching semi-inclusive
  deep inelastic scattering processes to Drell-Yan and W/Z boson production}.
\newblock {\em Phys. Rev. D}, 88(11):114012, 2013.

\bibitem{Dai:2014ala}
Ling-Yun Dai, Zhong-Bo Kang, Alexei Prokudin, and Ivan Vitev.
\newblock {Next-to-leading order transverse momentum-weighted Sivers asymmetry
  in semi-inclusive deep inelastic scattering: the role of the three-gluon
  correlator}.
\newblock {\em Phys. Rev. D}, 92(11):114024, 2015.

\bibitem{Braun:2009mi}
V.~M. Braun, A.~N. Manashov, and B.~Pirnay.
\newblock {Scale dependence of twist-three contributions to single spin
  asymmetries}.
\newblock {\em Phys. Rev. D}, 80:114002, 2009.
\newblock [Erratum: Phys.Rev.D 86, 119902 (2012)].

\bibitem{Echevarria:2015byo}
Miguel~G. Echevarria, Ignazio Scimemi, and Alexey Vladimirov.
\newblock {Universal transverse momentum dependent soft function at NNLO}.
\newblock {\em Phys. Rev. D}, 93(5):054004, 2016.

\bibitem{Scimemi:2019gge}
Ignazio Scimemi, Andrey Tarasov, and Alexey Vladimirov.
\newblock {Collinear matching for Sivers function at next-to-leading order}.
\newblock {\em JHEP}, 05:125, 2019.

\bibitem{Abbott:1980hw}
L.~F. Abbott.
\newblock {The Background Field Method Beyond One Loop}.
\newblock {\em Nucl. Phys. B}, 185:189--203, 1981.

\bibitem{Abbott:1981ke}
L.~F. Abbott.
\newblock {Introduction to the Background Field Method}.
\newblock {\em Acta Phys. Polon. B}, 13:33, 1982.

\bibitem{Chirilli:2015fza}
Giovanni~A. Chirilli, Yuri~V. Kovchegov, and Douglas~E. Wertepny.
\newblock {Regularization of the Light-Cone Gauge Gluon Propagator
  Singularities Using Sub-Gauge Conditions}.
\newblock {\em JHEP}, 12:138, 2015.

\bibitem{Brodsky:2002ue}
Stanley~J. Brodsky, Paul Hoyer, Nils Marchal, Stephane Peigne, and Francesco
  Sannino.
\newblock {Structure functions are not parton probabilities}.
\newblock {\em Phys. Rev. D}, 65:114025, 2002.

\bibitem{Ji:2002aa}
Xiang-dong Ji and Feng Yuan.
\newblock {Parton distributions in light cone gauge: Where are the final state
  interactions?}
\newblock {\em Phys. Lett. B}, 543:66--72, 2002.

\bibitem{Peskin:1995ev}
Michael~E. Peskin and Daniel~V. Schroeder.
\newblock {\em {An Introduction to quantum field theory}}.
\newblock Addison-Wesley, Reading, USA, 1995.

\bibitem{Ebert:2019okf}
Markus~A. Ebert, Iain~W. Stewart, and Yong Zhao.
\newblock {Towards Quasi-Transverse Momentum Dependent PDFs Computable on the
  Lattice}.
\newblock {\em JHEP}, 09:037, 2019.

\bibitem{Schwinger:1951nm}
Julian~S. Schwinger.
\newblock {On gauge invariance and vacuum polarization}.
\newblock {\em Phys. Rev.}, 82:664--679, 1951.

\bibitem{Mulders:1995dh}
P.~J. Mulders and R.~D. Tangerman.
\newblock {The Complete tree level result up to order 1/Q for polarized deep
  inelastic leptoproduction}.
\newblock {\em Nucl. Phys. B}, 461:197--237, 1996.
\newblock [Erratum: Nucl.Phys.B 484, 538--540 (1997)].

\bibitem{Ball:2002ps}
Patricia Ball, V.~M. Braun, and N.~Kivel.
\newblock {Photon distribution amplitudes in QCD}.
\newblock {\em Nucl. Phys. B}, 649:263--296, 2003.

\bibitem{Braun:2009vc}
V.~M. Braun, A.~N. Manashov, and J.~Rohrwild.
\newblock {Renormalization of Twist-Four Operators in QCD}.
\newblock {\em Nucl. Phys. B}, 826:235--293, 2010.

\bibitem{Braun:2011dg}
V.~M. Braun and A.~N. Manashov.
\newblock {Operator product expansion in QCD in off-forward kinematics:
  Separation of kinematic and dynamical contributions}.
\newblock {\em JHEP}, 01:085, 2012.

\bibitem{Balitsky:2017gis}
I.~Balitsky and A.~Tarasov.
\newblock {Power corrections to TMD factorization for Z-boson production}.
\newblock {\em JHEP}, 05:150, 2018.

\bibitem{Chirilli:2021lif}
Giovanni~Antonio Chirilli.
\newblock {High-energy operator product expansion at sub-eikonal level}.
\newblock {\em JHEP}, 06:096, 2021.

\bibitem{Borden:2024bxa}
Jeremy Borden, Yuri~V. Kovchegov, and Ming Li.
\newblock {Helicity evolution at small x: quark to gluon and gluon to quark
  transition operators}.
\newblock {\em JHEP}, 09:037, 2024.

\end{thebibliography}

\end{document}